\documentclass[review]{elsarticle}  
\usepackage{tikz}
\usepackage{amsmath,bm,algorithm,algorithmic}
\usepackage[mathscr]{euscript}
 \let\mathscr\relax
\usepackage{amssymb}
\usepackage{mathrsfs}
\usepackage{graphicx} 
\usepackage{natbib}
\usepackage{lineno,hyperref}
\usepackage{xcolor}
\usepackage{caption}
\usepackage{subcaption}
\usepackage{booktabs}

\begin{document}

\begin{frontmatter}

\title{Probabilistic modeling of discrete structural response with application to composite plate penetration models} 

\author[add1]{Anindya Bhaduri}
\author[add2]{Christopher S. Meyer}
\author[add2]{John W. Gillespie, Jr.}
\author[add2]{Bazle Z. (Gama) Haque}
\author[add1]{Michael D. Shields}
\author[add1]{Lori Graham-Brady\footnote{Corresponding author. Email address: lori@jhu.edu}}
\address[add1]{Department of Civil Engineering, Johns Hopkins University, Baltimore, MD, USA}
\address[add2]{Center for Composite Materials, University of Delaware, Newark, DE, USA}




\begin{abstract}

Discrete response of structures is often a key probabilistic quantity of interest. For example, one may need to identify the probability of a binary event, such as, whether a structure has buckled or not. In this study, an adaptive domain-based decomposition and classification method, combined with sparse grid sampling, is used to develop an efficient classification surrogate modeling algorithm for such discrete outputs. An assumption of monotonic behaviour of the output with respect to all model parameters, based on the physics of the problem, helps to reduce the number of model evaluations and makes the algorithm more efficient. As an application problem, this paper deals with the development of a computational framework for generation of probabilistic penetration response of S-2 glass/SC-15 epoxy composite plates under ballistic impact. This enables the computationally feasible generation of the probabilistic velocity response (PVR) curve or the $V_0-V_{100}$ curve as a function of the impact velocity, and the ballistic limit velocity prediction as a function of the model parameters. The PVR curve incorporates the variability of the model input parameters and describes the probability of penetration of the plate as a function of impact velocity. 
\end{abstract}


\end{frontmatter}


\section{Introduction}

Physical systems are often described by simulation models with discrete outputs. To understand the effects of the input parameter uncertainties on these discrete outputs, it is essential to locate the surface of separation between regions with different discrete output labels in an input space of many parameters. One such case is the penetration of composite plates under projectile impact.\\
\indent In structural mechanics applications, classification based machine learning approaches are often used to efficiently deal with systems with discrete output response. In the structure health monitoring domain, different classification algorithms are used for damage detection of structures \cite{he2007structural, cury2012pattern, hothu2013damage, salehi2018structural}. For example, Chong et al. \cite{chong2014nonlinear} proposed a multi-class support vector machine-based approach to efficiently detect damage levels for smart structures. In the field of fatigue crack detection, a Continuous k-Nearest Neighbour (CkNN) approach \cite{berry2016consistent, papanikolaou2019spatial} has been used to classify discrete prior deformation amplitudes for uniaxially compressed crystalline thin films. In the field of dynamic brittle material fracture, classification algorithms, such as, Random Forests (RFs), Decision Trees (DTs) and Artifical Neural Networks (ANNs) have been used to predict fracture coalescence using simulation data from a high fidelity Finite-Discrete Element Model \cite{moore2018predictive}. In the metamaterial design domain, a Scalable Variational Gaussian Process (SVGP) \cite{titsias2009variational, hensman2015scalable} classification method has been used to classify coilable (indicating supercompressibility) and noncoilable designs in the pursuit of transforming brittle polymers into supercompressible metamaterials \cite{bessa2019bayesian}. A lot of work has also been done on microstructure classification \cite{decost2015computer, chowdhury2016image, decost2017exploring, ling2017building, bock2019review} using efficient classification techniques. In the context of plate impact analysis, classification algorithms, such as, k-nearest neighbors (KNNs), Support Vector Machines (SVMs), Logistic regression and Decision Trees (DTs) have been used to assess the ballistic performance of perforated plates by predicting three different types of plate damage \cite{bobbili2020machine}. However, most of the classification work in the existing literature uses non-adaptive fixed data generated from simulations or experiments prior to the classification step. If the data generation is expensive, a more efficient way is to adaptively sample more data around the classification boundaries (surfaces of separation between classes) and avoid unnecessary data generation away from the classification boundaries. \\
\indent The current work proposes an efficient systematic approach for generating the probability of obtaining a discrete output, given variations in the input parameters. 
It uses an adaptive domain-based decomposition and classification approach to converge on the surface of separation between the different discrete outputs, and subsequently analyzing the model in regions of the input parameter space dictated by the algorithm. An upper and lower bound of the probability of one of the discrete outputs given a certain input parameter is then estimated analytically using the output label information of samples in the decomposed elements. 
With increase in each level of domain decomposition, it is found that the upper and lower bounds of the probability estimates converge in increments. It is worth mentioning here that the proposed methodology in this work focuses on systems with binary output response but can also be extended to systems with multi-class response.\\
\indent This manuscript is organized as follows: Section \ref{sec:methodology_chap_composite_impact} discusses the proposed methodology in details. In section \ref{sec:num_eg_chap_composite_impact}, the method is applied to a single layer continuum plain weave S-2 glass/SC-15 epoxy composite plate model under ballistic impact by a steel projectile. Section \ref{sec:conclusions_chap_composite_impact} provides conclusions.
\section{Methodology}\label{sec:methodology_chap_composite_impact}
The proposed methodology is a domain decomposition based adaptive classification approach illustrated in figure \ref{fig:Algorithm_schematic} using an arbitrary $2$-dimensional problem where $p_1$ and $p_2$ are the two variable parameters that determine the discrete (binary) output of `+$1$' or `-$1$'. It involves sparse grid \cite{smolyak1963quadrature, klimke2005algorithm, bhaduri2018efficient, bhaduri2018stochastic} sampling to generate samples in an element where the simulation model is run to obtain the corresponding binary outputs. A classification based criterion is then used to decide if the element needs to be further decomposed and if so, along which dimensions it should decompose.
\subsection{Domain decomposition}
Let us consider a $d$-dimensional hypercube input domain, assuming the simulation model is valid everywhere in that domain. If $A_i$ and $B_i$ denote the minimum and maximum bounds of the domain along dimension $i \ [ i \in \{1,2, \dots, d\}]$, the domain can be denoted by $\Xi=[A_1, B_1]\times[A_2, B_2]\times \dots \times[A_d, B_d]$. We further assume the decomposition of the domain takes place orthogonally along the dimensions. In a general orthogonal decomposition, if the domain $\Xi$ is divided along the first $d'\ (1 \leq d' \leq d)$ dimensions and there are $n_j$ divisions along dimension $j \ [ j \in \{1,2, \dots, d'\}]$, then $N$ non-overlapping and space-filling elements are formed where, $N = (n_1)(n_2)\dots (n_{d'})$ and each element is denoted by $\Xi_k$: $\cup_{k=1}^{N} \Xi_k = \Xi$, $\Xi_m \cap \Xi_k = \emptyset$ for $m\neq k$ and $m,k \in [1,2,\dots, N] $. If $a_i^k$ and $b_i^k$ denote the minimum and maximum bounds of element $\Xi_k$ along dimension $i \ [ i \in \{1,2, \dots, d\}]$, $\Xi_k$ is the tensor product given by
\begin{equation}
\Xi_k=[a_1^k,b_1^k)\times[a_2^k,b_2^k)\times \dots \times[a_d^k,b_d^k).
\end{equation}
For example, if a $3$-dimensional hypercube is decomposed along dimensions $1$ and $2$ with $3$ divisions for each dimensions, then $(3)(3) = 9$ elements are formed.\\
\indent If subdivisions are performed along all dimensions, the number of sub-elements formed from each subdivision increases exponentially with increase in dimensionality of the problem and the situation becomes intractable. This is called the `curse of dimensionality' \cite{ernest2003dynamic,ernest1961adaptive}. Tessellation based approaches of subdividing the domain lead to similar problems. In this study, an element is divided into $2$ sub-elements along only one of the dimensions at each iteration. For example, in figure \ref{fig:Algorithm_schematic}, step $2$ denotes subdivision of the original domain along parameter $1$. The criterion to choose the dimension along which the element is subdivided, is discussed in section \ref{sec: Adaptive procedure}. 
\subsection{Domain-based classification}
Classification is a supervised learning method for predicting labels for unseen new samples, based on the class labels of the training samples. For example, in the model of a composite plate under a projectile impact, the labels are penetration (`-1')  or rebound (`+1') of the projectile. Most of the popular classification methods, like k-nearest neighbors (k-NN), support vector machines (SVM), artificial neural networks (ANN), etc., are designed to work with preassigned training data. In this study, data is generated by sampling specific points in the input parameter space and running the simulation model at those parameter combinations. The domain-based classification approach takes advantage of the structured sparse grid sample data in each element in order to classify the entire element space with the same label as the sampling points that define the element. Sparse grid samples are designed to be biased towards the edges of a domain, which, helps increase accuracy of the `same label' assumption for an entire element. The specific combination of the sparse grid sampling and domain decomposition used here enable this type of classification. This domain-based classification can only classify regions with same type of label and is not capable of generating decision surfaces. It is used to define a region of the parameter space that contains the decision surface. In figure \ref{fig:Algorithm_schematic}, step $3$ involves a domain based classification of an element located in the lower right corner of the input domain. Since all the samples in that element have the `-1' (penetration) label, the entire-element is assigned the same label.
\subsection{Adaptive procedure}
\label{sec: Adaptive procedure}

\begin{figure}
\centering
\includegraphics[width=1.0\linewidth, height=10cm]{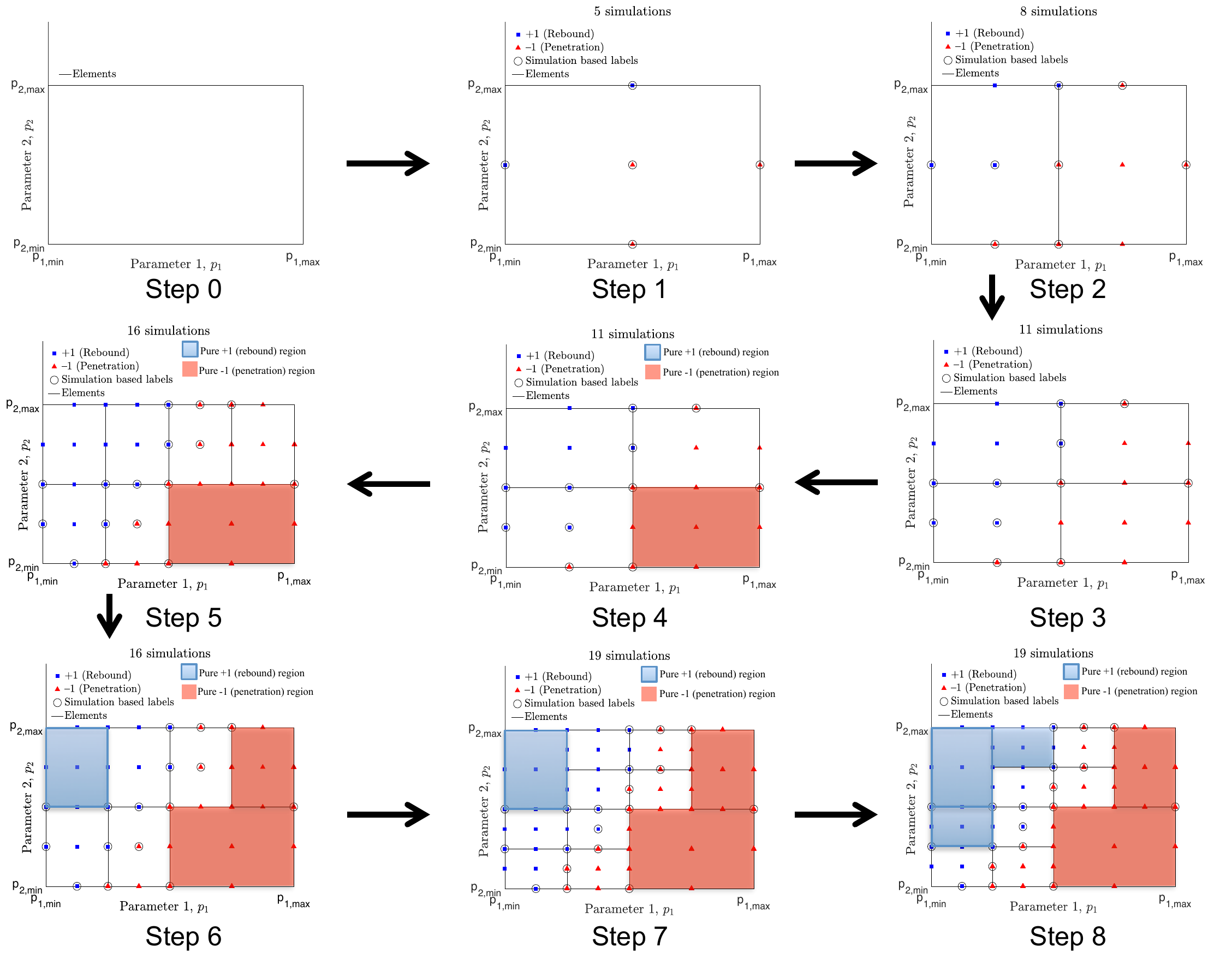}
\caption{Illustration of the adaptive procedure of the proposed algorithm using a $2$-dimensional example: Step $1$ involves generation of sparse grid samples up to level $1$ in the original domain and their labeling using the output (`-1' or `+1') from expensive simulations: Steps $2$, $3$, $5$ and $7$ involve subsequent domain decomposition and further sampling and labeling (either by expensive simulation runs or monotonicity based approximations): Steps $4$, $6$, and $8$ involve domain-based classification, indicated by red or blue regions. Blue squares denote the binary output `+1' for a sample, and red triangles denote the binary output `-1' for a sample. Black circles around these symbols denote samples where expensive simulations have been performed. Samples with labels without the `circle' markers indicate labeling based on monotonicity approximations. Light blue regions are classified as `+1', and red regions classified as `-1'. }
\label{fig:Algorithm_schematic}
\end{figure}
The core objective of the proposed methodology is to direct the sampling towards the region of the boundary of separation between discrete output `-1' (penetration) and `+1' (rebound).
Before explaining the adaptive procedure, some terminology is introduced. 
A sampled element refers to an element where sparse grid samples have been explicitly generated to obtain output labels from model evaluations, in addition to having evaluated samples present in that element from previous iterations. For example, in step $1$ of figure \ref{fig:Algorithm_schematic}, the entire domain can be viewed as a single element which is sampled at the $5$ (up to level $1$) sparse grid points. If a sampled element contains samples with only one type of output label (`-1' or `+1'), then the entire element is classified with that label. This is the domain-based classification step and thus it is assumed that the surface of separation will not pass through that element. For example, in step $4$ of figure \ref{fig:Algorithm_schematic}, the element denoted by a red rectangle contains sampling points with only `-1' (penetration) labels and is thus classified as a penetration region. On the other hand, when a sampled element contains samples with both `-1' and `+1' outputs (for example, penetration and rebound, respectively), the surface of separation passes through that element and thus it needs to be resolved further. The unresolved element is split up equally into two sub-elements along a dimension which corresponds to the maximum edge length of the element. This is done to avoid biasing the samples along particularly sensitive dimensions. If the element edge length is the same along all dimensions, then the subdivision is done along a dimension such that one of the sub-elements has the fewest samples (ideally none) with different output labels. For example, in step $6$ of figure \ref{fig:Algorithm_schematic}, the upper right region is divided into $2$ elements vertically along parameter $1$, such that one of the elements can be classified as a `-1' penetration region.
\begin{figure}
\centering
\begin{subfigure}[b]{0.4\textwidth}
\centering
\includegraphics[width=1.1\linewidth, height=4.5cm]{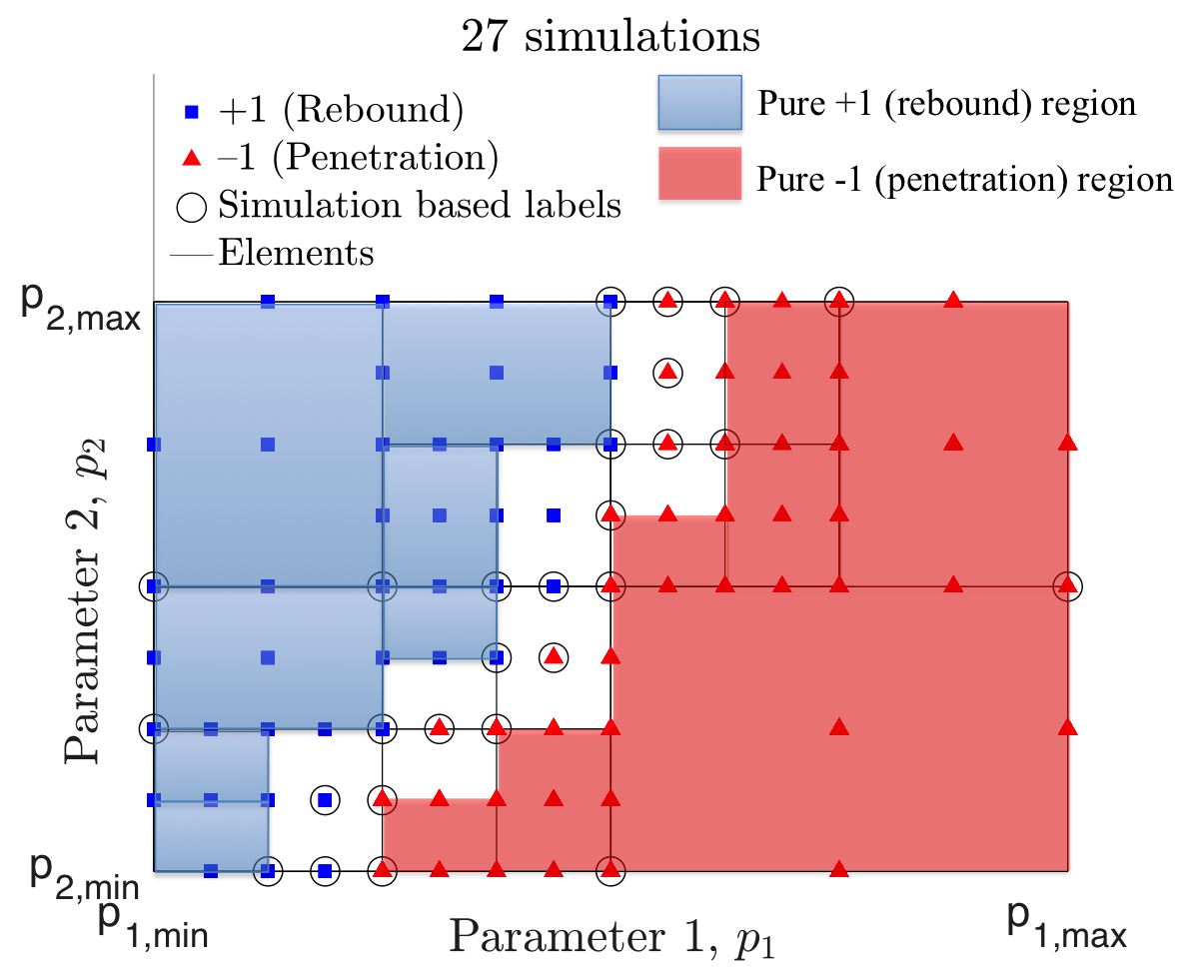}
\caption{}
\label{fig:PVR-BL-schematic_subfig1}
\end{subfigure} \quad \quad  \quad  \quad  \quad  \quad \quad  \quad  \quad  \quad 
\begin{subfigure}[b]{0.4\textwidth}
\centering
\includegraphics[width=1.1\linewidth, height=4.5cm]{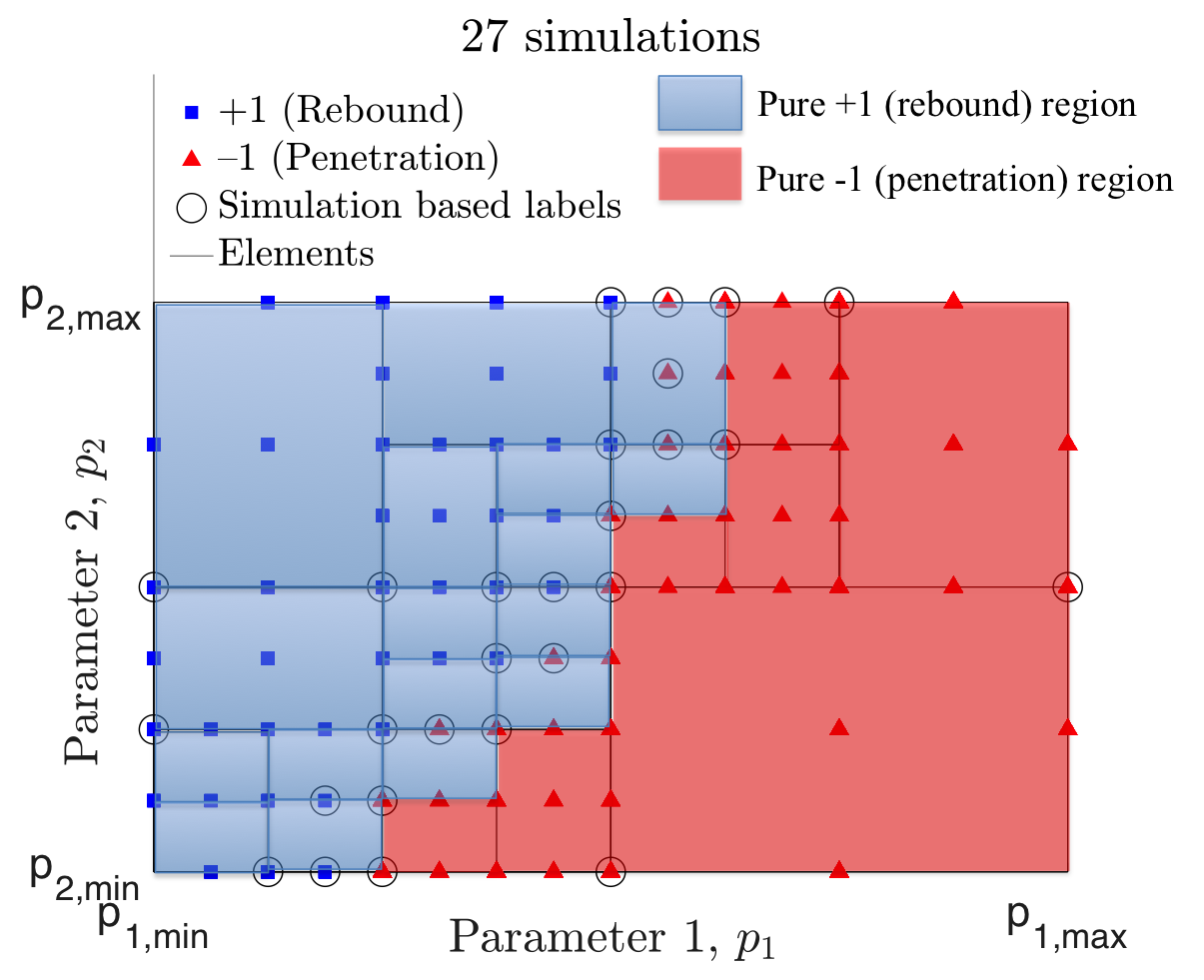}
\caption{}
\label{fig:PVR-BL-schematic_subfig2}
\end{subfigure}  \quad  \quad  \quad 
\begin{subfigure}[b]{0.4\textwidth}
\centering
\includegraphics[width=1.1\linewidth, height=4.5cm]{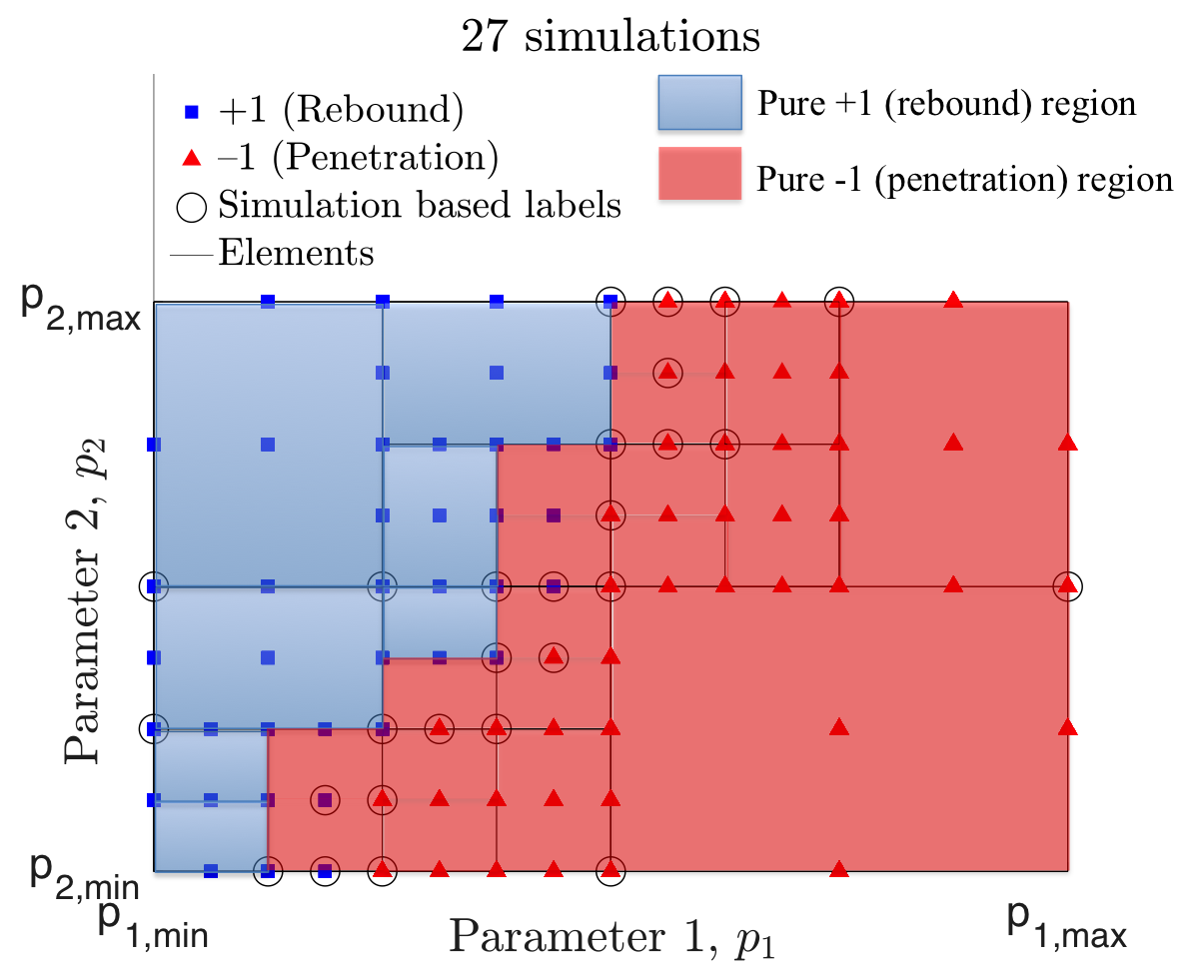}
\caption{}
\label{fig:PVR-BL-schematic_subfig3}
\end{subfigure}\quad 
\begin{subfigure}[b]{0.4\textwidth}
\centering
\includegraphics[width=1.1\linewidth, height=4.5cm]{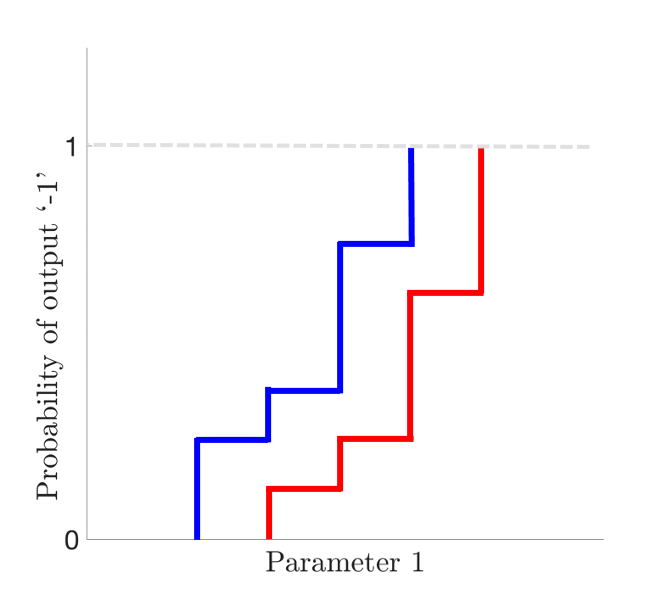}
\caption{}
\label{fig:PVR-BL-schematic_subfig4}
\end{subfigure}
\caption{Schematic demonstration of lower and upper bound probability curve of output response `-1'; (a) shows the pure output '+1' region in blue, the pure output `-1' region in red and the mixed region in between the two in white; (b) shows the mixed region assumed as part of the output `+1' region which leads to the red upper bound probability curve in (d); (c) shows the mixed region assumed as part of the output `-1' region which leads to the blue lower bound PVR curve in (d).}
\label{fig:PVR-BL-schematic}
\end{figure}
\subsection{Probability curve estimation} \label{section: prediction step}
From the adaptive procedure discussed in the previous section, a set of labeled input parameter samples are obtained along with the resolved (classified) and unresolved element boundaries. Using this information, the upper and lower bounds of the probability of obtaining any one of the discrete outputs (`-1' or `+1') for any given value of one of the input parameters. The procedure is illustrated through figure \ref{fig:PVR-BL-schematic} using a $2$-dimensional example where the upper and lower bound probability that output $=$ `-1' for any given value of parameter $1$ is estimated. It is noted that the parameter $2$ can follow any arbitrary marginal probability distribution but in figure \ref{fig:PVR-BL-schematic}, it follows a uniform distribution.

To compute the probability curve, parameter $1$ is fixed at a particular value and the goal is to find out the fraction of points on the projected $(d-1)$-dimensional subdomain that are classified as `-1'.
One approach to generate the probability curve is to estimate, at first, the surface of separation using a suitable classification algorithm with the binary training data. Then, $(d-1)$-dimensional uniformly distributed space filled \cite{mckay1979comparison, halton1960efficiency, sobol1976uniformly, faure1982discrepance, niederreiter1987point, bhaduri2019free} test samples can be generated at a particular value of parameter $1$, and the fraction of the `-1' labels predicted over the total sample set denote the probability that output $=$ `1' at that parameter $1$ value. The process can be repeated for all values in the parameter $1$ range to obtain the probability curve. However, there is classification error as well as sampling error involved in this approach for estimating the probability curve. 

An alternative approach uses a domain-based classification method to analytically compute lower and upper bounds of the penetration probability for each parameter $1$ values of interest. It requires the decomposition of each unresolved element (element with dissimilar output labels) along all its dimensions. This allows generation of sub-elements with uniform output labels (new resolved elements), such that the domain-based classification method can be applied there. Figure \ref{fig:PVR-BL-schematic_subfig1} shows the classification of the $2$-dimensional input parameter space into 3 regions: pure '+1' (rebound) region, the pure `-1' (penetration) region and the mixed region. The true surface (curve) of separation between the two labels lies somewhere in the grey mixed region. The boundaries of the mixed region correspond to edges of the unresolved elements obtained from the proposed adaptive algorithm. The mixed region bounds can thus be used to estimate the lower and upper bounds of the probability curve which will then bound the true probability curve. 
Next, for each parameter $1$ value, all the $d$-dimensional elements that include the value are selected. Let us assume that the total hypervolume of the selected elements in the projected $(d-1)$-dimensional space (at a fixed value of parameter $1$) is denoted by $H_{total}$. Then, the total hypervolume (say, $H_1$) of the projected $(d-1)$-dimensional resolved elements with `-1' labels are calculated. In case of the unresolved elements, the entire element is given a `+1' label to estimate the upper bound, as shown in figure \ref{fig:PVR-BL-schematic_subfig2}, and a `-1' label to estimate the lower bound, as shown in figure \ref{fig:PVR-BL-schematic_subfig3}. If the total hypervolume of the projected $(d-1)$-dimensional unresolved elements is denoted by $H_2$, then the probability that output = `-1' (penetration) has a lower bound value of $H_1/H_{total}$ and a upper bound value of $(H_1 + H_2)/H_{total}$. Following this procedure, the probability curve can be obtained as shown in figure \ref{fig:PVR-BL-schematic_subfig4}. This approach thus avoids the error in sampling in a potentially high-dimensional parameter space. However, an uncertain region exists between the upper and lower probability curves whose area reduces with further refinement of the parameter space.
\subsection{Monotonicity constraints of parameters}
The advantage of dealing with physics-based simulation models is that it is sometimes possible to obtain some initial knowledege about the behavior of the system under study. For the models used in this work, the likelihood of a `-1' outcome can be assumed to have a monotonicity constraint with respect to some parameters of the model. For example, in the composite plate penetration problem, we can assume that an increase in impact velocity of the projectile increases the likelihood of `-1' (penetration) outcome, when all other input parameters are kept constant. This is a case of a monotonicity constraint of increasing type. \\
\indent Monotonic constraint knowledge is expressed in terms of a dominance relation between ordered samples in the input space \cite{cano2019monotonic}. A sample is said to dominate another when each coordinate of the former is equal or greater than the corresponding coordinate of the latter. The class labels are also assumed to be ordered. For example, a student in an exam can get a grade of `A', `B' or `C'. In a binary label output case, either of the labels can be considered of lower value and the other  of higher value. Thus, binary label outputs are ordinal by nature. Monotonicity constraints can be either increasing type or decreasing type. In the case of an increasing monotonicity constraint, the class label assigned to input samples should be equal or higher than the class labels assigned to the samples it dominates. In the case of a decreasing monotonicity constraint, the class label assigned to input samples should be the same or lower than the class labels assigned to the samples it dominates. As an example, consider a monotonicity constraint relating one input attribute and the target class. Keeping other attributes of the sample fixed, if the constraint is of increasing type, a sample with a higher value of the input attribute (dominant sample) should not be associated to a lower class value, and if the constraint is of decreasing type, a dominant sample should not be associated to a higher class value.\\
\indent Following notations and definitions similar to \cite{cano2019monotonic}, we consider $d$ input parameters with $N$ ordered samples $\mathbf{x}^{(i)} \subseteq R^d$ and an associated class label \ $y^{(i)} \ [i=1,\dots,N]$. $C$ ordered labels are considered here such that $y^{(i)} \in \{1,\dots,C\}$. The data set is denoted by $D = \{(x^{(1)}, y^{(1)}),\dots, (x^{(N)}, y^{(N)})\}$. A dominance relation, $\succeq$, is defined as follows:
\begin{equation}
    \mathbf{x} \succeq \mathbf{x}' \iff x_j \geq x^{'}_j, \ \ \forall j \in \{1,\dots,d\},
\end{equation}
where $x_j$ and $x^{'}_j$ are the $j$-th coordinates of samples $\mathbf{x}$ and $\mathbf{x}'$, respectively. This means, $\mathbf{x}$ dominates $\mathbf{x}'$, if each coordinate of $\mathbf{x}$ is not smaller than the corresponding coordinate of $\mathbf{x}'$. Since the samples as well as the corresponding class labels are ordered, the elements in data set $D$ are comparable among each other. Two examples $\mathbf{z}$ and $\mathbf{z}'$ are identical if $z_j = z{'}_j ,\forall j \in \{1,...,d\}$, and they are non-identical if $\exists j$, such that $z_j \neq z^{'}_j$. A pair of elements in $D$, $(\mathbf{x},y)$ and $(\mathbf{x}',y')$ is said to have a monotonic relation of increasing type if,
\begin{align}
&\mathbf{x} \succeq \mathbf{x}' \ \ \wedge \ \ \mathbf{x} \neq \mathbf{x}' \ \ \wedge \ \ y \geq y' \nonumber \\
&\text{or} \nonumber\\
&\mathbf{x} = \mathbf{x}' \ \ \wedge \ \ y = y'.    
\end{align}
and a monotonic relation of decreasing type if,
\begin{align}
&\mathbf{x} \succeq \mathbf{x}' \ \ \wedge \ \ \mathbf{x} \neq \mathbf{x}' \ \ \wedge \ \ y \leq y' \nonumber \\ 
& \text{or} \nonumber \\
&\mathbf{x} = \mathbf{x}' \ \  \wedge \ \ y = y'.    
\end{align}
If all possible pairs of elements in $D$ are either monotone (increasing or decreasing) or incomparable, then data set $D$ is also considered to be monotone. In a general case, a data set can be monotonic of increasing type for a subset of the input parameters, monotonic of decreasing type for another subset of parameters, and non-monotonic in the rest of the parameters. If there are $d_1$ monotonically increasing parameters, $d_2$ monotonically decreasing parameters, and $d_3$ non-monotonic parameters, such that $d=d_1+d_2+d_3$, then for a pair of non-identical elements $(\mathbf{x},y)$ and $(\mathbf{x}',y')$, the condition for a general monotonic relation is given by,
\begin{align} \label{monotonicity criterion}
&\mathbf{x}_{(d_1)} \succeq \mathbf{x}^{'}_{(d_1)} \ \ \wedge \ \  \mathbf{x}_{(d_1)} \neq \mathbf{x}^{'}_{(d_1)} \ \ \wedge \ \ y \geq y' \nonumber\\
&\quad \wedge \nonumber \\ 
&\mathbf{x}_{(d_2)} \succeq \mathbf{x}^{'}_{(d_2)} \ \ \wedge \ \  \mathbf{x}_{(d_2)} \neq \mathbf{x}^{'}_{(d_2)} \ \ \wedge \ \ y \leq y' \nonumber \\
& \quad \wedge \nonumber \\
&\mathbf{x}_{(d_3)} = \mathbf{x}^{'}_{(d_3)}
\end{align} 
where, $\mathbf{x}_{(d_1)}$ is the $d_1$-dimensional subspace containing the increasing monotone parameters, $\mathbf{x}_{(d_2)}$ is the $d_2$-dimensional subspace containing the decreasing monotone parameters, and $\mathbf{x}_{(d_3)}$ is the $d_3$-dimensional subspace containing the non-monotone parameters.
\subsection{Numerical implementation}
\begin{figure}
\centering
\includegraphics[width=0.8\textwidth]{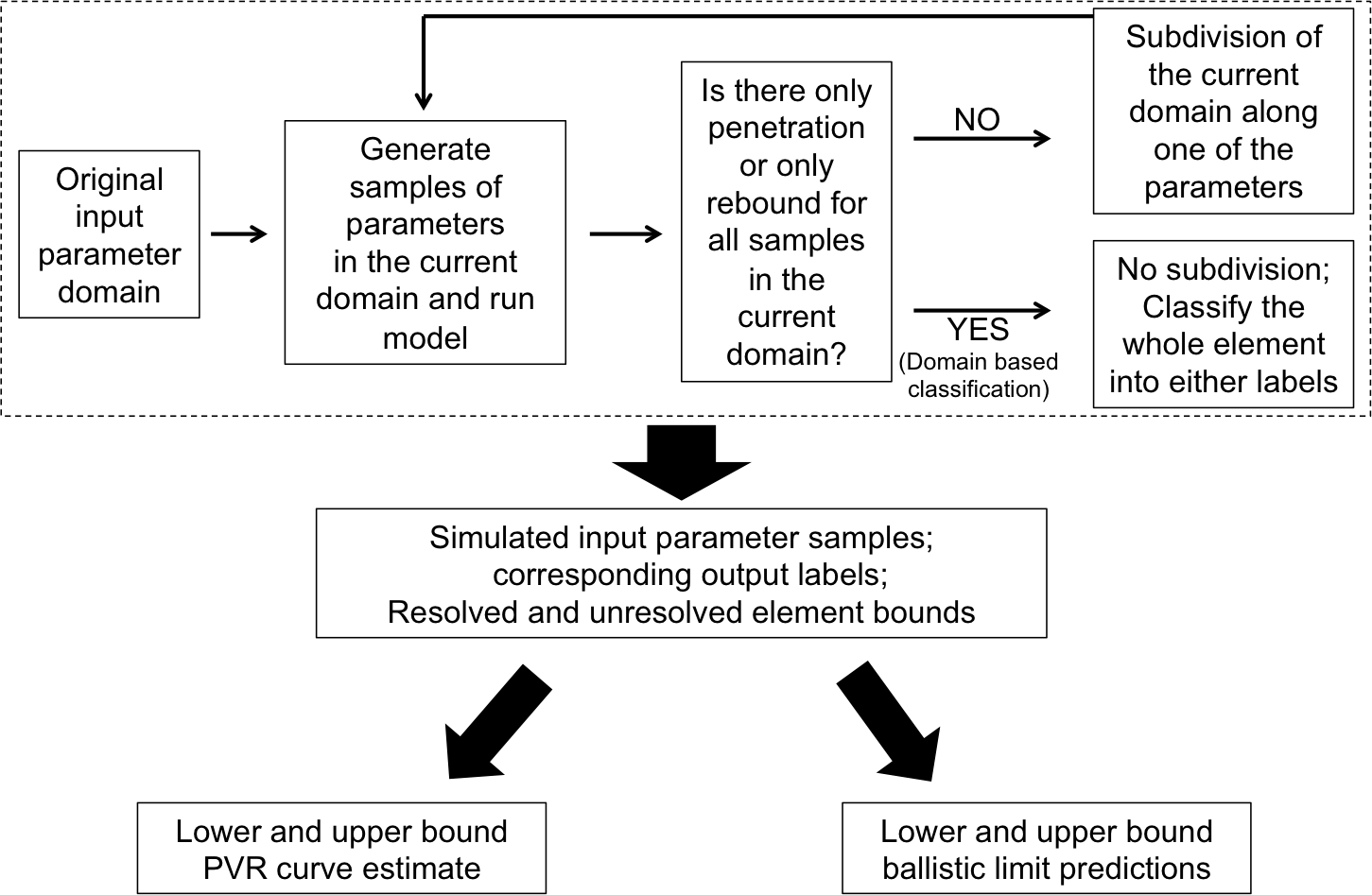}
\caption{Flow chart of the algorithm steps for generating the PVR curve as a function of the impact velocity and predicting ballistic limit velocities as functions of the mode parameters.} '
\label{fig:algorithm_chap_composite_impact}
\end{figure}

The algorithm has the following steps:
\begin{enumerate}
    \item Select the number of variable parameters or input dimensions $d$ and their corresponding ranges of interest.
    \item Set the total number of iterations $N_{iter}$, and the minimum hyper-volume fraction of the non-converged elements $H_{min}$ below which the subdivision into smaller elements is stopped. 
    \item Initialize $iter = 1$. Use sparse grid samples of level $1$ to generate samples in the original input domain (element). Run simulations at the samples and note their output. The output considered here is binary: `+1' or `-1' of the projectile. 
If all the simulations have the same output labels (all `+1' or all `-1'), the element is assumed to be resolved, and go to step \ref{algo: prediction step}. Otherwise, if the output label outcomes are dissimilar (mix of `+1' and `-1'), the element is said to be unresolved.
    \item  Subdivide each unresolved element into two sub-elements along the dimension which corresponds to the maximum element edge length. If the element edge lengths are the same along all dimensions, then the subdivision is done along a dimension such that one of the sub-elements have samples with minimum number of dissimilar output labels. \label{algo: subdivision step}
    \item Generate new samples in each of the new elements using sparse grid sampling of level $1$. Remove duplicate samples. Check if any of these new samples can be assigned a label based on the monotonicity constraint in Eq. (\ref{monotonicity criterion}). Run simulations with the parameter values corresponding to the newly created unlabeled samples. If an element now contains samples with the same type of output labels, it is assumed to have been resolved (domain-based classification). 
    \item At iteration `$iter$', generate upper and lower bounds of the probability curve as a function of any one of the input parameters as described in section \ref{section: prediction step}. \label{algo: prediction step}
    \item Set $iter = iter + 1$. Calculate the cumulative hyper-volume $H$ of all the unresolved sub-elements created and the number of iterations $iter$. Compare with the corresponding critical values $H_{min}$ and $N_{iter}$ respectively to check if either of the two stopping criteria is met. If met, terminate the algorithm. If either of the criteria is not met, go to step \ref{algo: subdivision step}. 
\end{enumerate}
The flow chart for the entire algorithm is shown in figure \ref{fig:algorithm_chap_composite_impact}.
It is noted here that an optional monotonicity enforcement step can be implemented at the end of the algorithm before estimating the quantities of interest. The need of this implementation may arise because the simulation model behavior may become noisy near the classification boundary (surface of separation). This means, if the model is run at input parameter combinations which correspond to regions in input parameter space close to the surface of separation, the outcome of the model may be such that it violates the general monotonic behavior and has the wrong label. A way to enforce monotonicity is to check the label of each sample and see if it violates the general monotonic behavior with respect to the other sample labels. If there is violation, the label is changed. This procedure may be repeated for multiple iterations over the entire sample set to resolve the monotonicity violation issue.
\section{Numerical example: projectile impact of composite plates}\label{sec:num_eg_chap_composite_impact}
In this section, the proposed framework is applied to a single layer plain weave S-2 glass/SC-15 epoxy continuum composite model impacted by a steel projectile. Impact velocity of the projectile is the extrinsic source of variability and the longitudinal tensile strength and punch shear strength of the plate are the two intrinsic sources of variability considered in this study. The first example is a $1$-parameter case which considers only the variation in the impact velocity of the projectile and the strength parameters are fixed at their baseline values. The next example is a $2$-parameter problem which considers the additional variation of the longitudinal tensile strength. Finally a $3$-parameter example is also shown where the impact velocity, the longitudinal tensile strength, and the punch shear strength are all variable parameters for the simulation model. Two types of input distributions for the strength parameters are considered in this study: uniform and normal.
\subsection{Literature review}
\begin{figure}
\centering
\begin{subfigure}[b]{0.75\textwidth}
\centering
\includegraphics[width=1.15\linewidth, height=4.8cm]{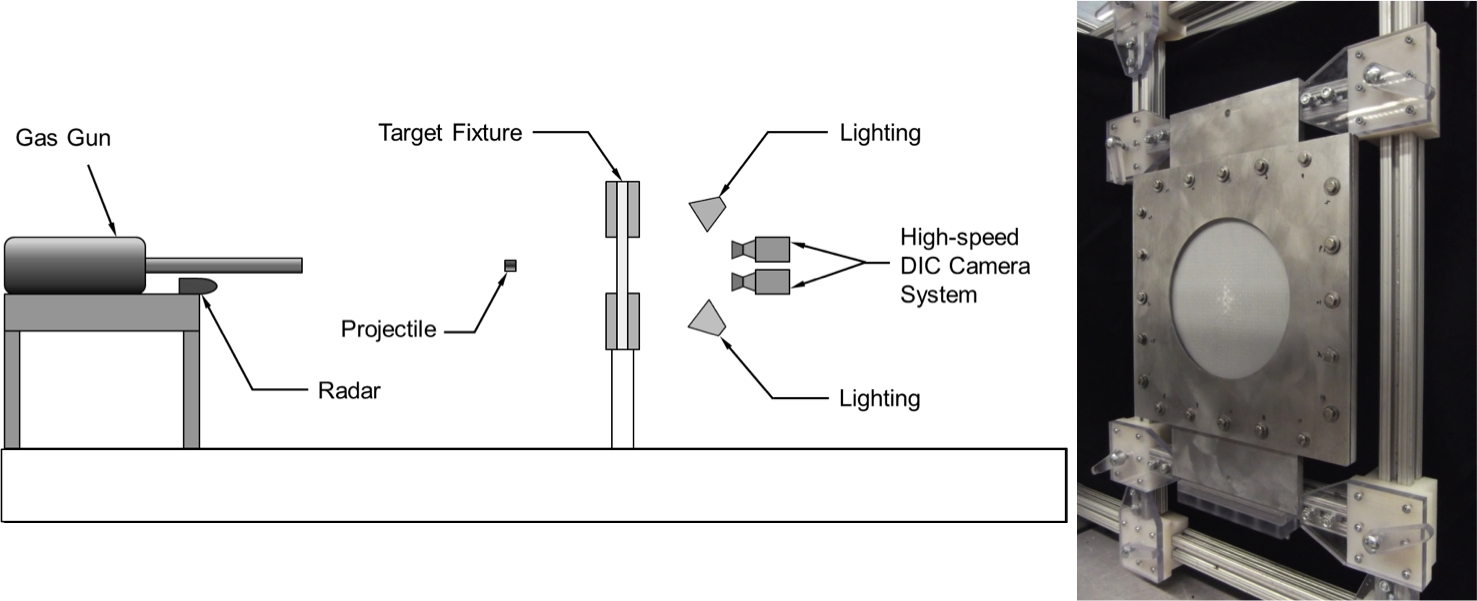}
\caption{}
\label{fig:SOTA_pvr_curve_subfig1}
\end{subfigure} \quad
\begin{subfigure}[b]{0.4\textwidth}
\centering
\includegraphics[width=1.2\linewidth, height=4.0cm]{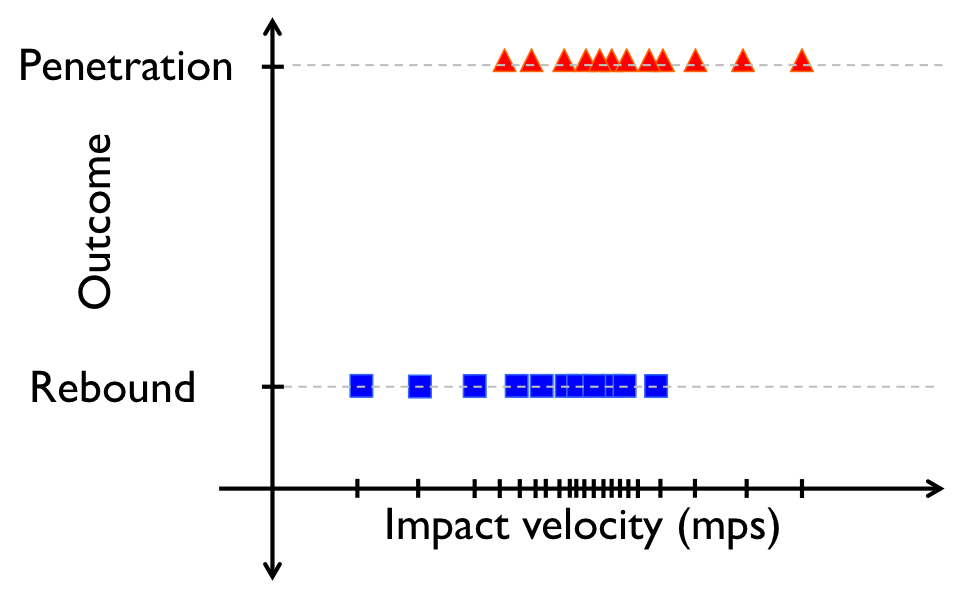}
\caption{}
\label{fig:SOTA_pvr_curve_subfig2}
\end{subfigure} \quad \quad \quad   
\begin{subfigure}[b]{0.4\textwidth}
\centering
\includegraphics[width=1.2\linewidth, height=4.0cm]{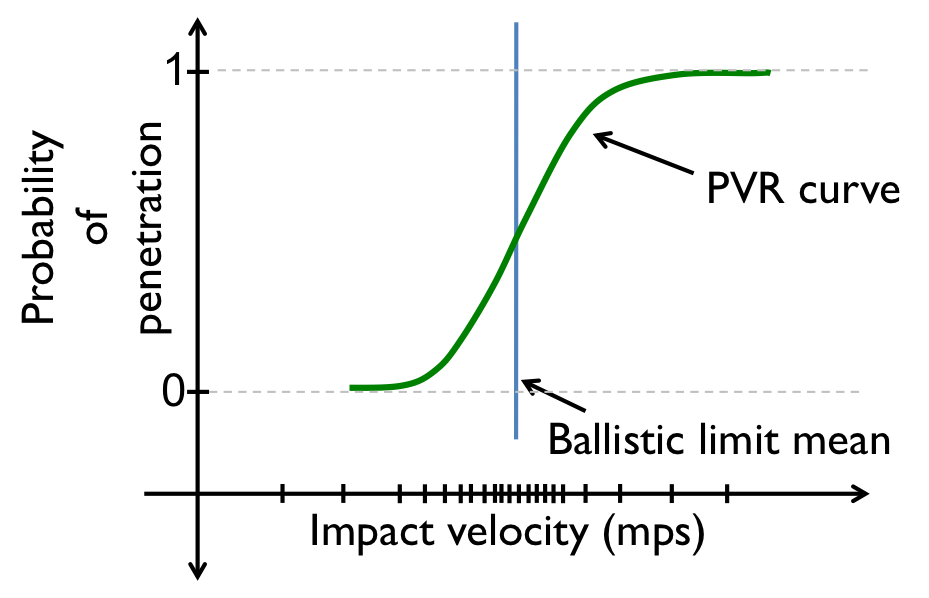}
\caption{}
\label{fig:SOTA_pvr_curve_subfig3}
\end{subfigure}
\caption{(a) A typical experimental setup for a plate under projectile impact; (b) shows the experimental outcomes for a range of impact velocities chosen adaptively: red triangles denote the cases where the projectile penetrates through the plate and blue squares denote the cases where the projectile rebounds from the plate after impact; (c) shows the corresponding PVR curve generated using the ballistic limit mean and variance estimates.}
\label{fig:SOTA_pvr_curve}
\end{figure}

\indent In the field of armor design, performance evaluation of composite plates under ballistic impact in the presence of various sources of statistical variability is an ongoing topic of research. There is a need to explore the high-dimensional conceptual design space comprising material properties and weave architecture; but the cost of prototyping such designs and experimentally characterizing the ballistic impact response is prohibitively high. An alternative approach is virtual armor design and testing, which requires efficient computational frameworks capable of predicting the impact performance. The probabilistic velocity response (PVR) curve, or $V_0-V_{100}$ curve, is of particular interest in this regard. $V_X$ is defined as the velocity at which the probability of penetration is $X\%$. In simulations as well as experiments, the current state-of-the-art approach to generate the PVR curve is to use the Neyer D-Optimal sensitivity test method \cite{neyer1994ad, standard1997v50} as shown in figure \ref{fig:SOTA_pvr_curve}. A typical ballistic experimental setup \cite{meyer2018mesoscale} is shown in figure \ref{fig:SOTA_pvr_curve_subfig1}. In this method, the PVR curve is defined as the cumulative density function (cdf) of the $V_{50}$ impact velocity and is assumed to follow a normal distribution with mean and variance to be determined. The method requires as input, initial guesses for the $V_{50}$ variance and bounds of the $V_{50}$ mean. The method is then used to guide the selection of the projectile impact velocities, such that the outcome of the previous impact velocity test (penetration or rebound) is used to determine the next impact velocity. Figure \ref{fig:SOTA_pvr_curve_subfig2} shows the experimental outcomes for a range of impact velocities chosen adaptively. Once all the impact velocities are obtained, maximum likelihood estimation (MLE) is used to obtain the $V_{50}$ mean and variance estimates, which finally generates the PVR curve as shown in figure \ref{fig:SOTA_pvr_curve_subfig3}. In essence, the whole curve is generated based on $V_{50}$ mean and variance estimates, which are incomplete descriptors, particularly for the tails of the distribution. Also, the generated PVR curve suggests there is a finite (although very small) probability of penetration at zero impact velocity, and also a finite (again very small) probability of rebound at infinite impact velocity, both of which are physically unrealistic.
A lot of work has been done previously in generating the virtual PVR curve by mapping in different intrinsic (for example, yarn tensile strength) and extrinsic (for example, projectile impact location) sources of experimentally characterized variability into fabric finite element (FE) models \cite{nilakantan2010using, nilakantan2014state, wang2016effect, nilakantan2018virtual, nilakantan2018experimentally}.
In some of these studies \cite{nilakantan2018virtual,nilakantan2018experimentally}, the numerical PVR curves have been experimentally validated and are shown to be in good agreement with the experimental PVR curve. It is, however, noted that only $38$ simulations or experiments were performed to generate the PVR curves. While such an approach is understandable given the high costs associated with ballistic experiments, it might not be fully representative of the true experimental PVR curve. In other words, the PVR curve might change if more experiments are performed. As far as the numerical PVR curves are concerned, in \cite{nilakantan2018experimentally}, 
a total of $137$ warp and $137$ fill yarns constituted the FE models and four sources of variability were considered: tensile strength of each yarn ($274$ variables), tensile modulus of each yarn ($274$ variables), inter-yarn friction between each fill and warp yarn ($137^2$ dimensions), and projectile impact location ($1$ variable). This gave rise to a $19318$-dimensional sampling domain from which $38$ samples were selected to generate the PVR curve. In order for $38$ samples to be representative of the model behaviour, the model must not be very sensitive to most of those $19318$ dimensions. Otherwise, if most of the dimensions are sensitive, $38$ samples are never sufficient and the generated PVR curve will be unreliable.
\subsection{Simulation model description}
\begin{figure}
\centering
\begin{subfigure}[b]{0.4\textwidth}
\centering
\includegraphics[width=1.0\linewidth, height=5cm]{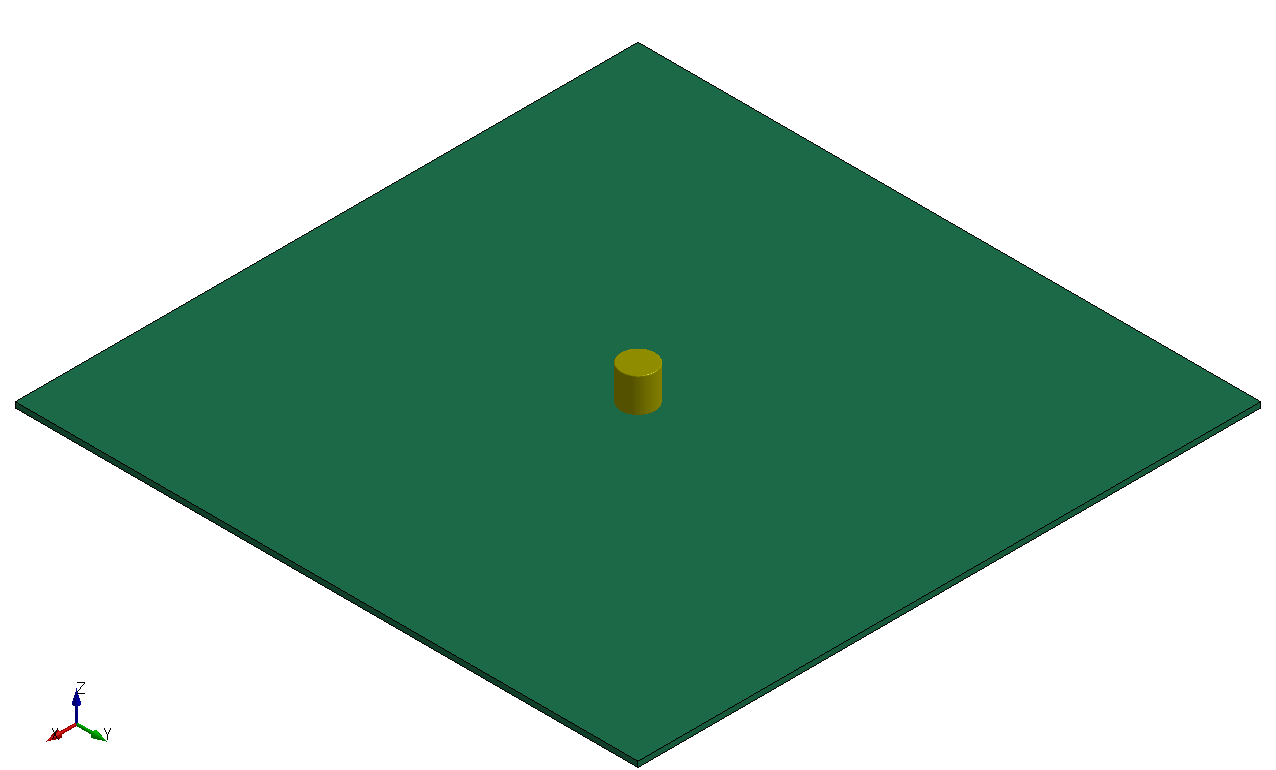}
\caption{}
\label{fig:composite_model_figure_subfig1}
\end{subfigure} \quad 
\begin{subfigure}[b]{0.4\textwidth}
\centering
\includegraphics[width=1.0\linewidth, height=3cm]{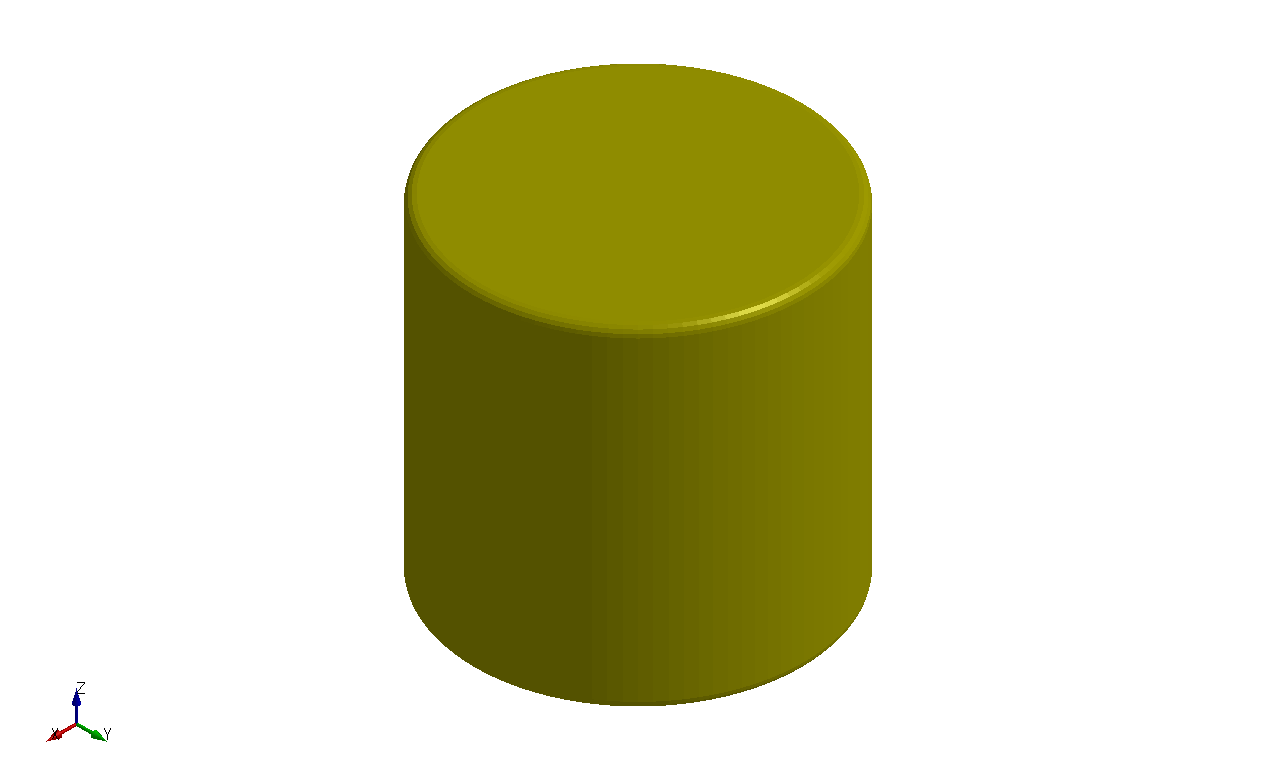}
\caption{}
\label{fig:composite_model_figure_subfig2}
\end{subfigure} 
\begin{subfigure}[b]{0.4\textwidth}
\centering
\includegraphics[width=1.0\linewidth, height=5cm]{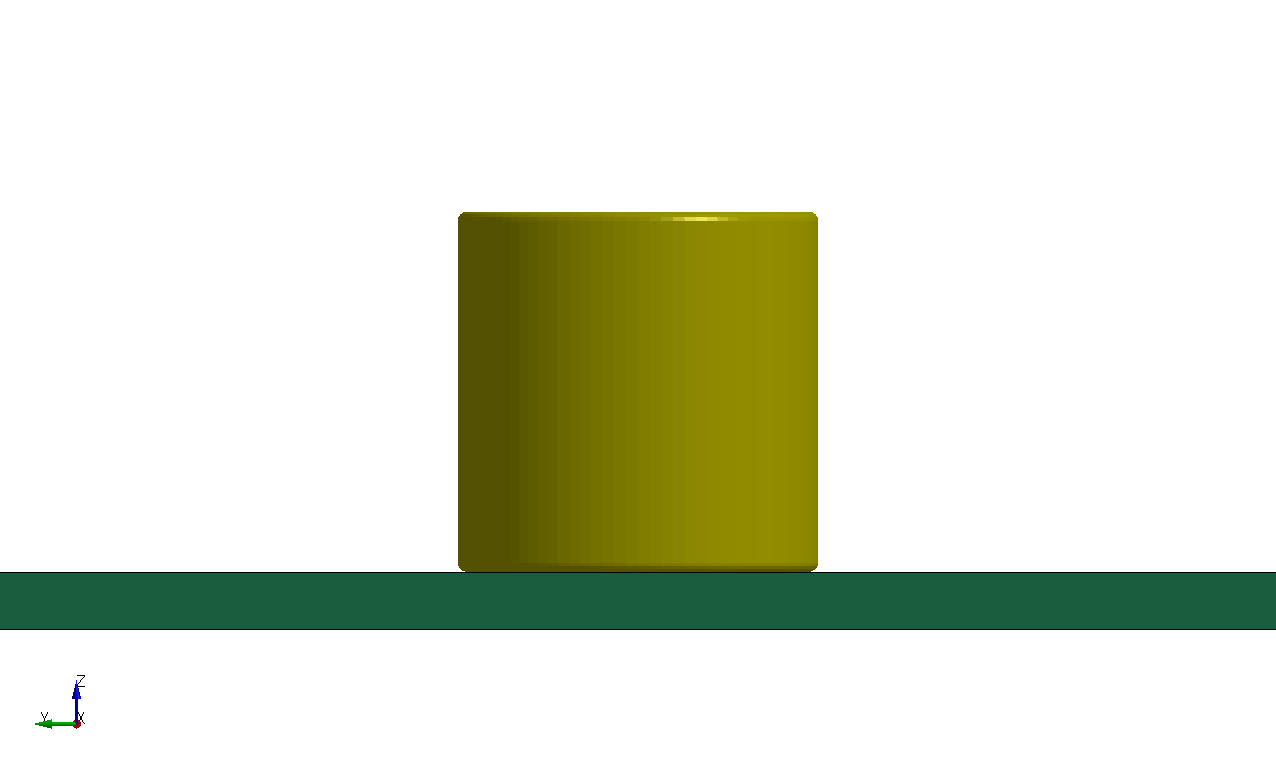}
\caption{}
\label{fig:composite_model_figure_subfig3}
\end{subfigure}
\caption{(a) Isometric view of the $4''\times4''$ composite plate model under impact by a RCC projectile, (b) RCC steel projectile, (c) Magnified front view of the continuum plate model} 
\label{fig:Composite_model_figure}
\end{figure}

The simulation model used in this study is a single layer, continuum plain weave composite plate target as shown in figure \ref{fig:Composite_model_figure}. It is modeled in LS-DYNA \cite{hallquist2014ls} using hexahedral elements. The plate dimensions are $101.6$ mm $\times$ $101.6$ mm $\times$ $0.887$ mm and it is subjected to clamped boundary conditions. The plate model is impacted at the center by a rigid, right-circular, cylindrical elastic steel projectile having a diameter of $5.6$ mm and height $6.1$ mm. A slight fillet ($r\approx0.15$ mm) is applied to projectile edges to reduce stress concentration. 
The damage in the plate is governed by the MAT-162 material model and the modeling approach has been shown to have a good correlation with past ballistic testing \cite{xiao2007progressive, gama2011finite, yen2012ballistic, haque2015progressive}. It is a continuum damage model that models progressive failure in unidirectional and plain weave composite materials subjected to large strain rates and pressures (i.e., impact). The model uses a generalized Hashin failure criteria \cite{hashin1980failure} and the damage progression is characterized by damage evolution laws proposed by Matzenmiller et al. \cite{matzenmiller1995constitutive}. The failure model simulates fiber failure, matrix damage, and delamination under modes I, II, and III loading.  The damage model simulates material softening (degradation of stiffness matrix) following damage initiation (governed by failure model) when loaded beyond the failure strength parameters such as longitudinal tensile strength and punch-shear strength. Thus, varying these strength parameters controls initiation of material softening as well as the total energy dissipation prior to failure under impact loading conditions.
\begin{figure}
\centering
\begin{subfigure}[b]{0.4\textwidth}
\centering
\includegraphics[width=1.15\linewidth, height=5cm]{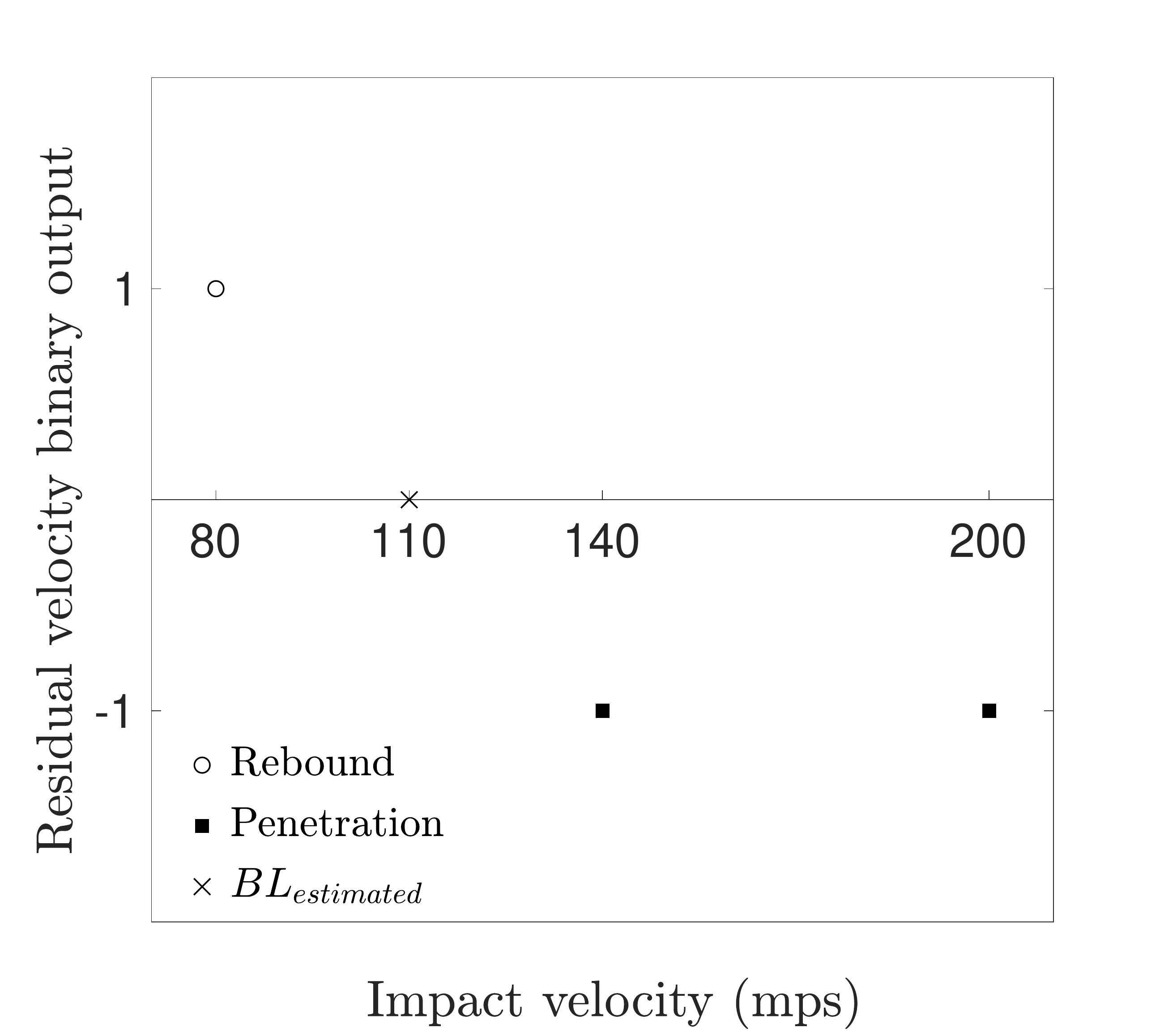}
\caption{}
\label{fig:1-D_ballistic_limit_subfig1}
\end{subfigure}  \quad  \quad \quad \quad 
\begin{subfigure}[b]{0.4\textwidth}
\centering
\includegraphics[width=1.15\linewidth, height=5cm]{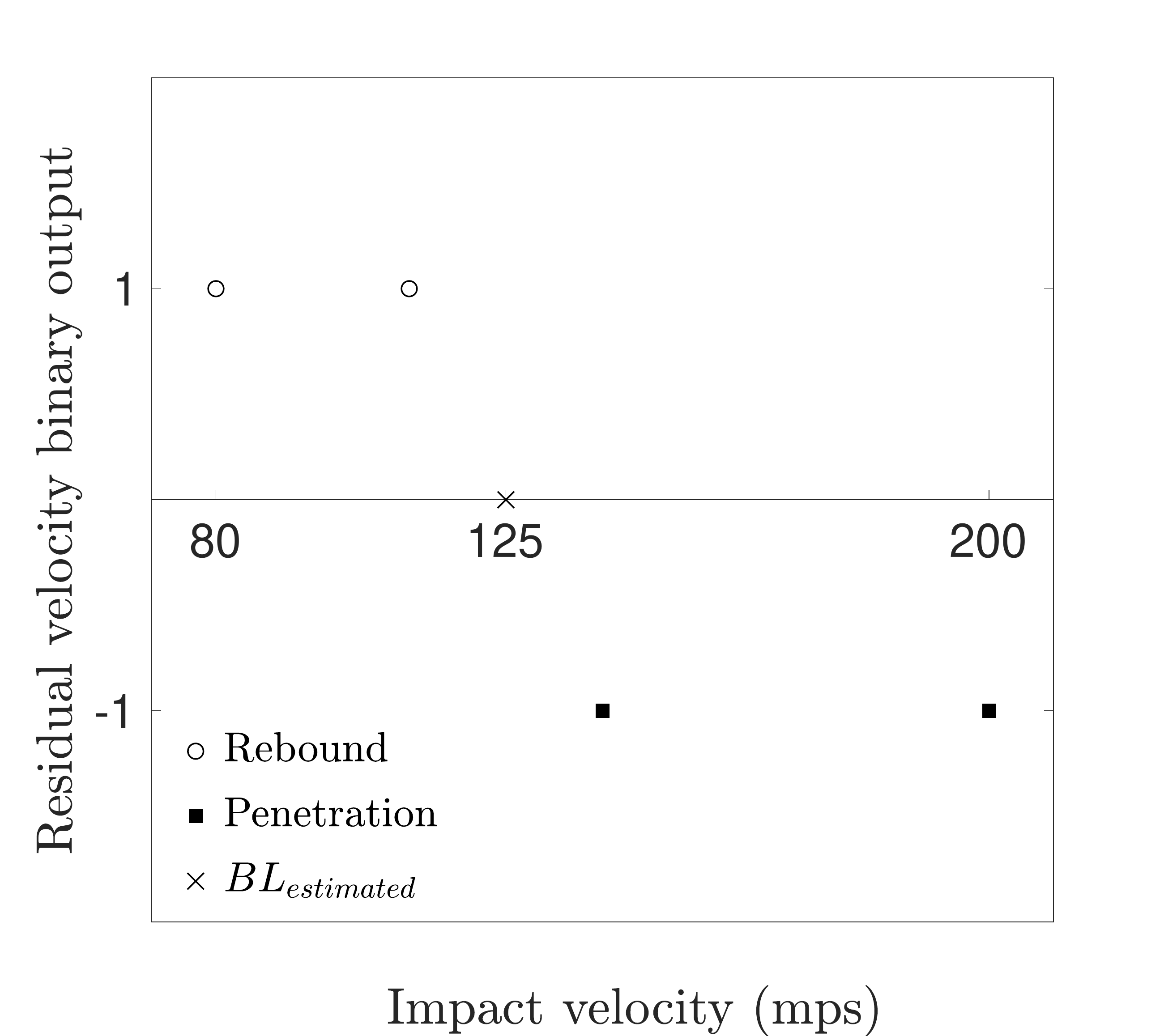}
\caption{}
\label{fig:1-D_ballistic_limit_subfig2}
\end{subfigure} 
\begin{subfigure}[b]{0.4\textwidth}
\centering
\includegraphics[width=1.15\linewidth, height=5cm]{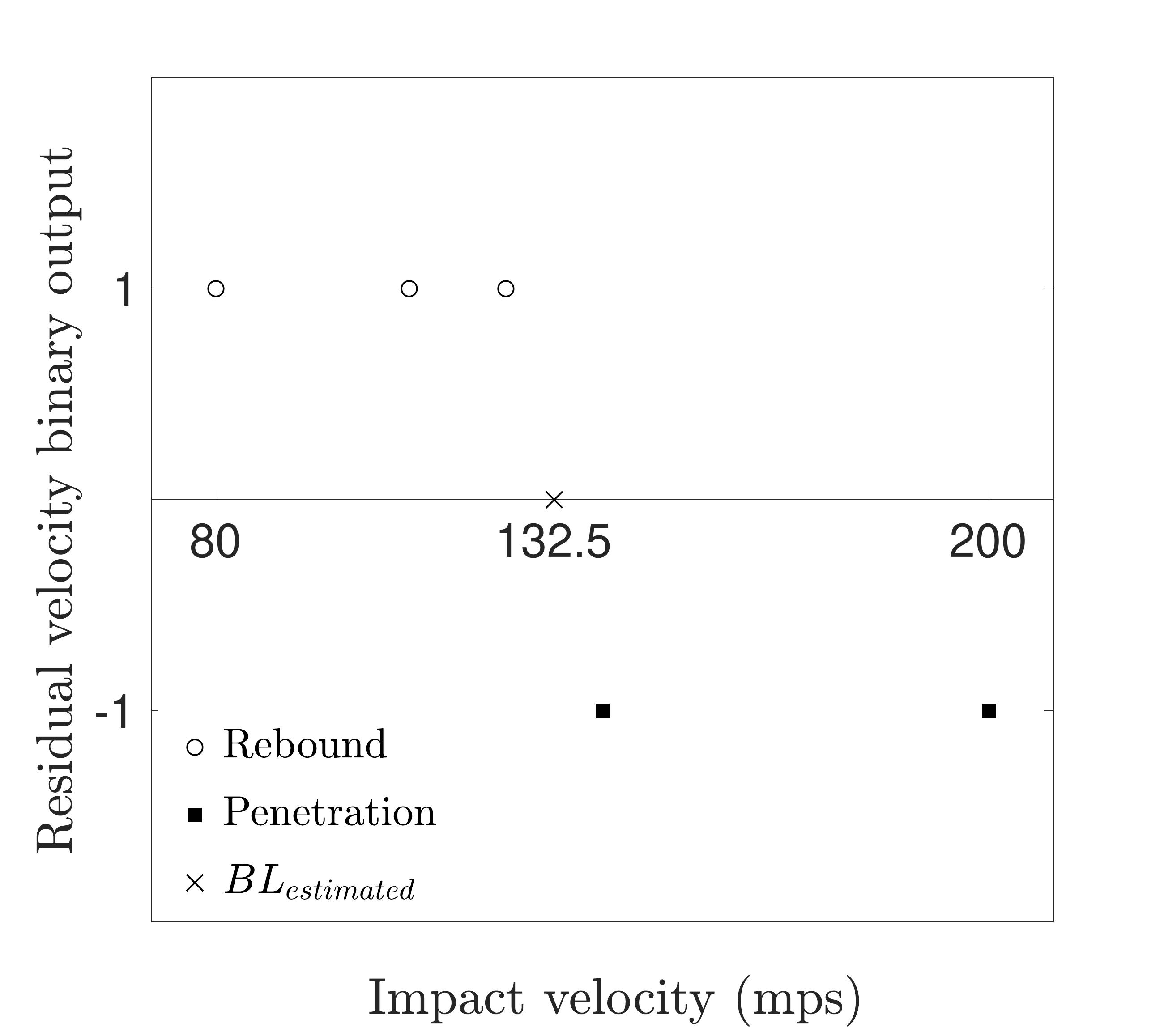}
\caption{}
\label{fig:1-D_ballistic_limit_subfig3}
\end{subfigure}  \quad  \quad \quad \quad 
\begin{subfigure}[b]{0.4\textwidth}
\centering
\includegraphics[width=1.15\linewidth, height=5cm]{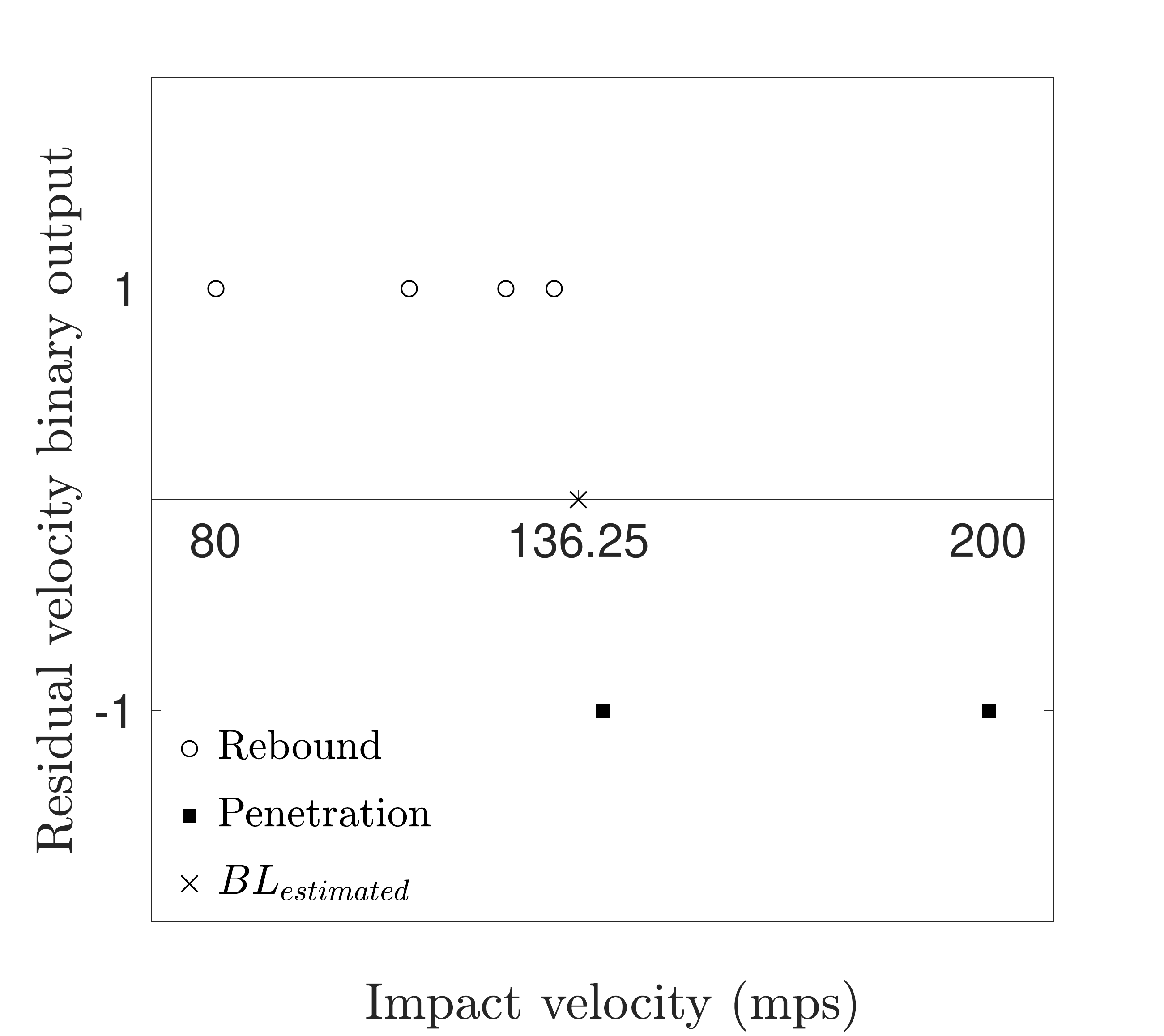}
\caption{}
\label{fig:1-D_ballistic_limit_subfig4}
\end{subfigure}\quad 
\begin{subfigure}[b]{0.4\textwidth}
\centering
\includegraphics[width=1.15\linewidth, height=5cm]{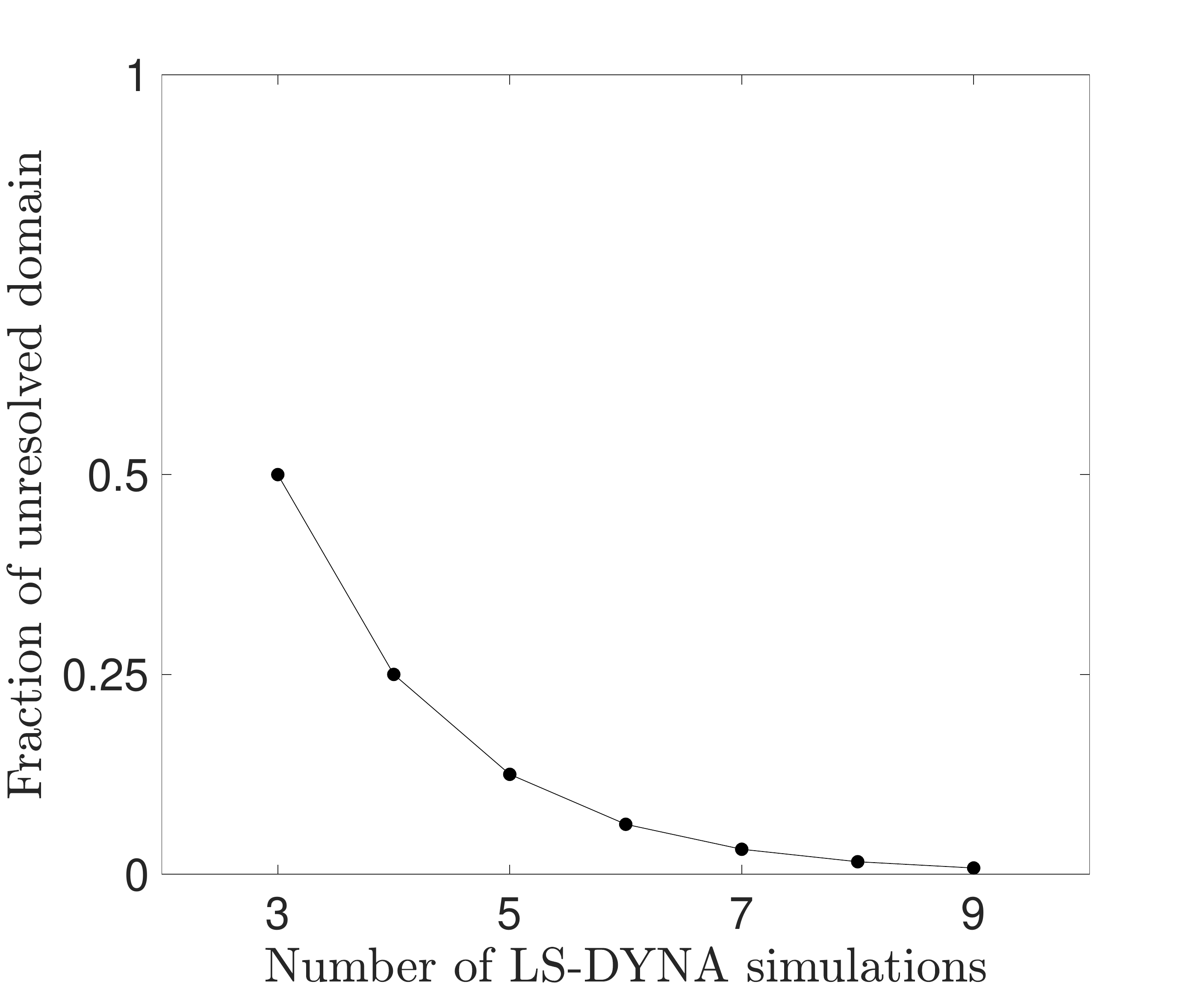}
\caption{}
\label{fig:1-D_ballistic_limit_subfig5}
\end{subfigure}
\caption{(a-d) Ballistic limit prediction with adaptive sampling of impact velocities: samples with rebound (`+1') outcomes are denoted by `circle' markers, samples with penetration (`-1') outcomes are denoted by `square' markers, ballistic limit mean estimate is denoted by `cross' markers; (e) Reduction in the fraction of the unresolved space with increase in the number of LS-DYNA simulations.}
\label{fig:1-D_ballistic_limit}
\end{figure}
\subsection{PVR curve estimation and ballistic limit predictions}
\indent The PVR curve estimation is analogous to the probability curve estimation as discussed in section \ref{section: prediction step} where label `-1' represents the penetration case of the projectile impact and parameter $1$ is the projectile impact velocity.\\
\indent A similar approach is used to obtain lower and upper bound estimates of the ballistic limit velocity predictions as a function of the material parameters. For each $(d-1)$-dimensional input parameter value, the $d$-dimensional unresolved elements that include the value are selected. The minimum lower bound value of the impact velocity among all the unresolved elements corresponds to the lower bound of the ballistic limit velocity and the maximum upper bound value of the impact velocity among those elements corresponds to the upper bound of the ballistic limit velocity.
Thus, the uncertainty of the actual ballistic limit value is bounded by the minimum lower bound and maximum upper bound of the impact velocity values corresponding to the candidate unresolved elements. 
\subsection{Assumptions on monotonicity constraints}
\indent In the numerical example considered, the class labels are binary and hence ordered. The rebound (`+1') label is assumed to be the lower class label and penetration (`-1') label is considered to be the higher class label. There is assumption of monotonicity constraint of increasing type for the impact velocity parameter and monotonicity constraint of decreasing type for the strength parameters (longitudinal tensile strength and punch shear strength). Thus, given all other parameters are fixed, a sample with a higher value of impact velocity should not have a rebound (`+1') label, if the sample which it dominates has a penetration (`-1') label. Similarly, given all other parameters are fixed, a sample with a higher value of longitudinal tensile strength (or punch shear strength) should not have a penetration (`-1') label, if the sample which it dominates has a rebound (`+1') label. When samples are generated in new elements using the adaptive algorithm, the monotonicity criterion in Eq. (\ref{monotonicity criterion}) is implemented to save expensive simulations by cheaper approximation. For example, in step $2$ of figure \ref{fig:Algorithm_schematic}, samples without the `circle' markers are labeled based on the increasing monotonicity assumption of parameter $1$ and decreasing monotonicity assumption of parameter $2$.
\subsection{1-dimensional results}
This section considers the impact velocity variation over the range $[80, 200]$ m/s, with the effective longitudinal tensile strength (LTS) and the punch shear strength (PSS) of the continuum level composite plate set at baseline strengths of $1100$ MPa and $300$ MPa respectively. Since there is no variability in the strength parameters, it is not possible to generate a PVR curve in this case. The only relevant quantity of interest that can be extracted from this scenario is the ballistic limit velocity at the baseline strengths. This case is better viewed as a demonstration of the domain-based decomposition and classification method in $1$-dimension to obtain the ballistic limit velocity at the given strengths. Figures \ref{fig:1-D_ballistic_limit}(a-d) show the evolution of the mean ballistic limit velocity estimate with increase in iterations of the algorithm. After starting with $3$ initial samples as seen in figure \ref{fig:1-D_ballistic_limit_subfig1}, $1$ sample is added at each subsequent iteration of the algorithm as seen in \ref{fig:1-D_ballistic_limit}(b-d). In this $1$-dimensional case, the domain decomposition is very similar to the bisection method. The unresolved domain is the region in the $1$-$d$ impact velocity space bounded by samples with dissimilar labels (rebound (`+1') label on one side and penetration (`-1') label on the other), and the ballistic limit is considered to lie within the unresolved domain. With each new simulation, the fraction of unresolved $1$-$d$ domain gets reduced by $50\%$ as seen in figure \ref{fig:1-D_ballistic_limit_subfig5}. For example, in \ref{fig:1-D_ballistic_limit_subfig1}, the unresolved $1$-$d$ space is between the impact velocity values of $80$ and $140$ $m/s$, which is a half of the entire impact velocity domain. With addition of the $1$ more sample with rebound label at $110$ $m/s$, the unresolved space shifts to values of $110$ and $140$ $m/s$, which is a quarter of the entire impact velocity domain. It is to be noted here that any value of the impact velocity in the unresolved $1$-dimensional space could be a potential candidate for the ballistic limit. For this particular case, the candidate of choice is the mean.
\subsection{2-dimensional results}
\begin{figure}
\centering
\begin{subfigure}[b]{0.3\textwidth}
\centering
\includegraphics[width=\linewidth]{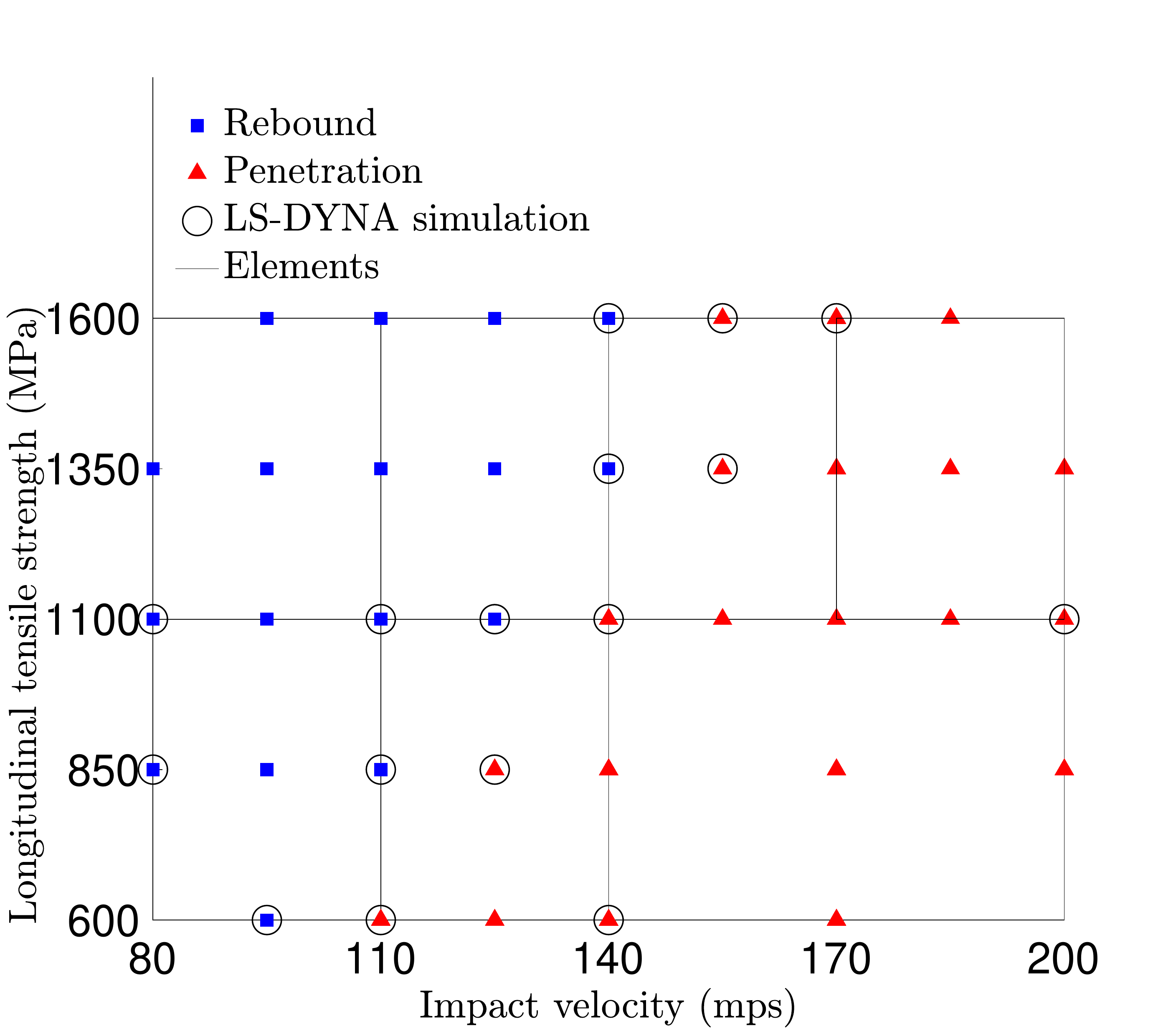}
\caption{}
\label{fig:2-D_uniform_samples_PVR_subfig1}
\end{subfigure} \quad
\begin{subfigure}[b]{0.3\textwidth}
\centering
\includegraphics[width=\linewidth]{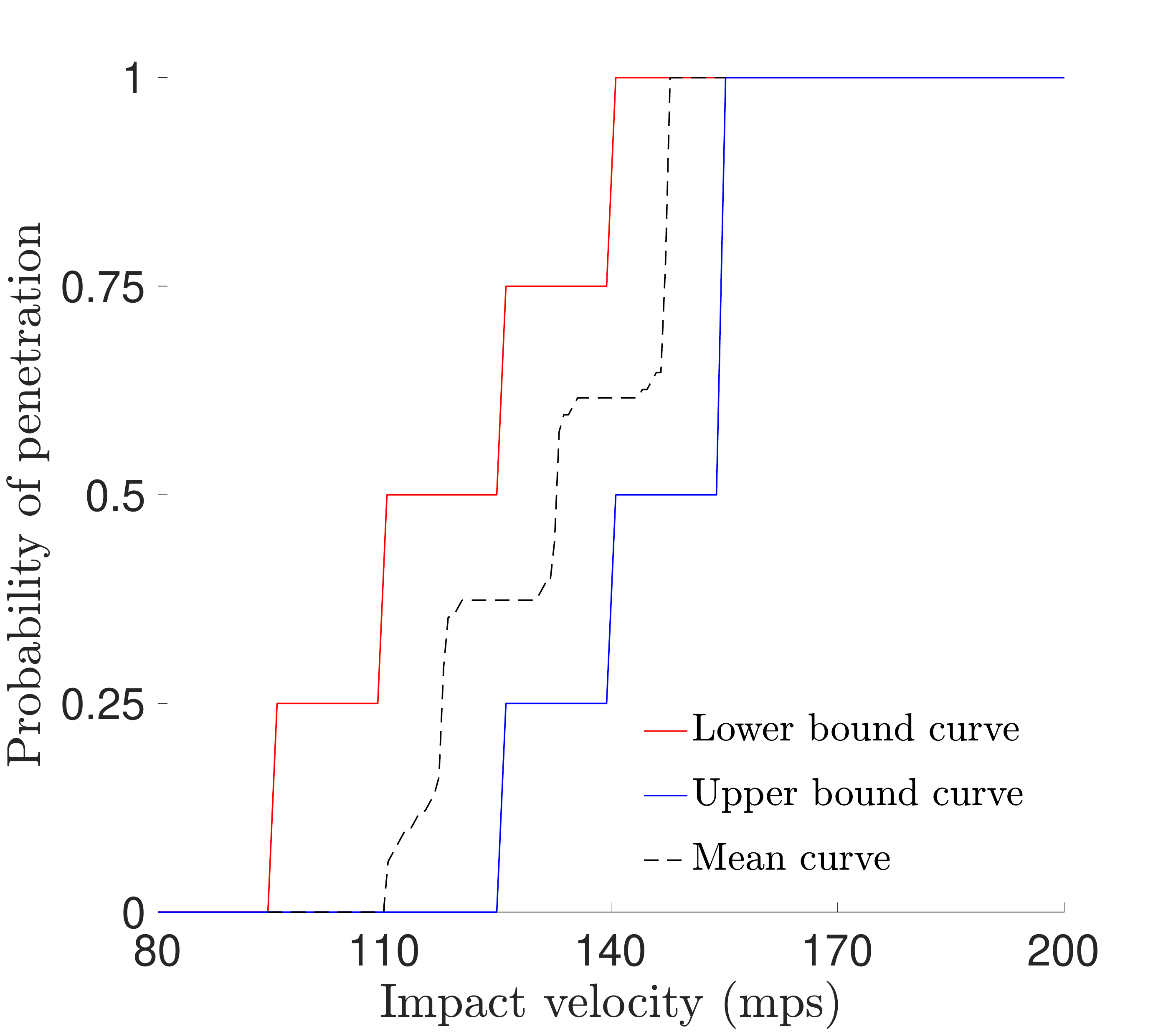}
\caption{}
\label{fig:2-D_uniform_samples_PVR_subfig2}
\end{subfigure}\quad 
\begin{subfigure}[b]{0.3\textwidth}
\centering
\includegraphics[width=\linewidth]{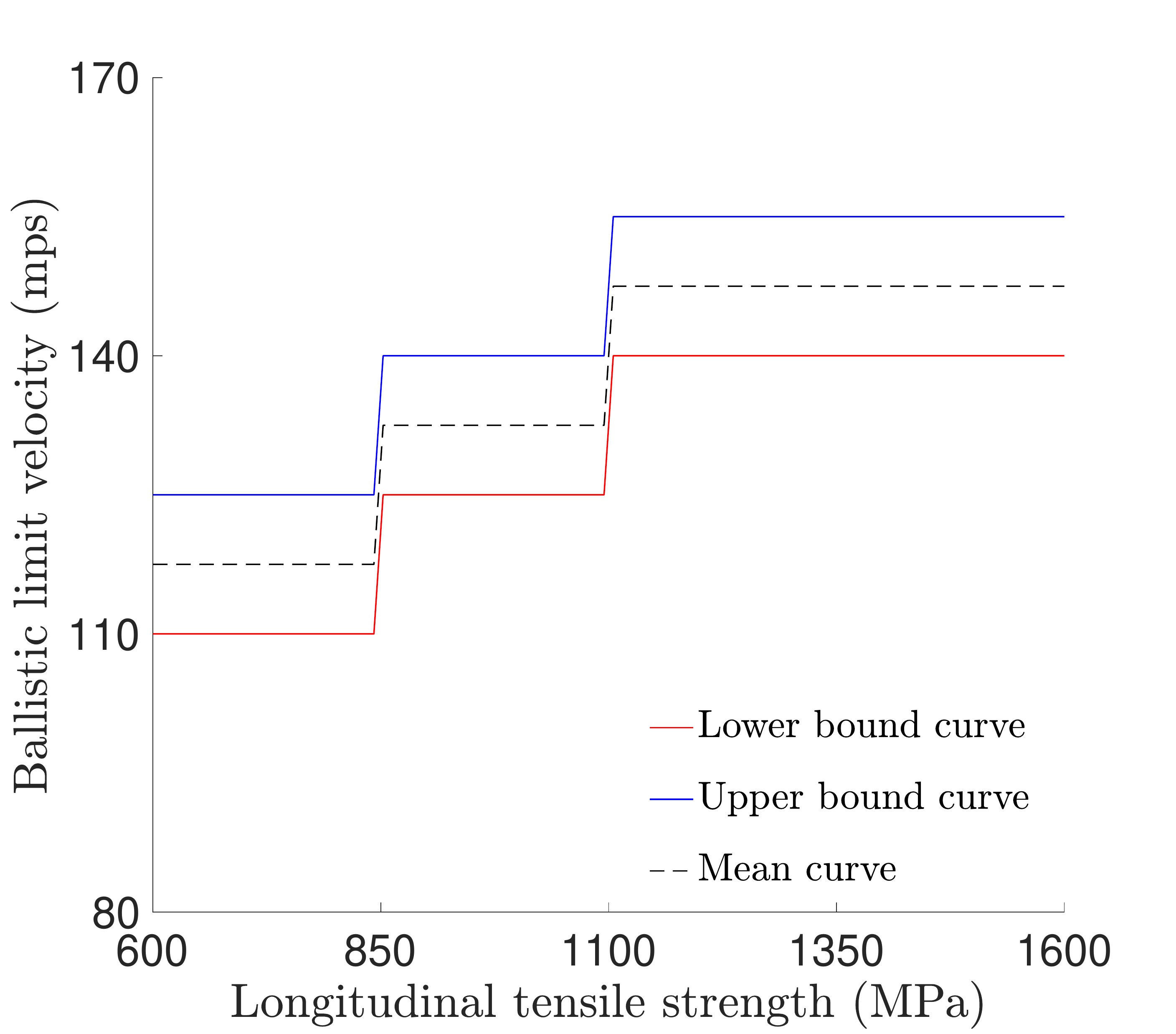}
\caption{}
\label{fig:2-D_uniform_samples_PVR_subfig3}
\end{subfigure}
\begin{subfigure}[b]{0.3\textwidth}
\centering
\includegraphics[width=\linewidth]{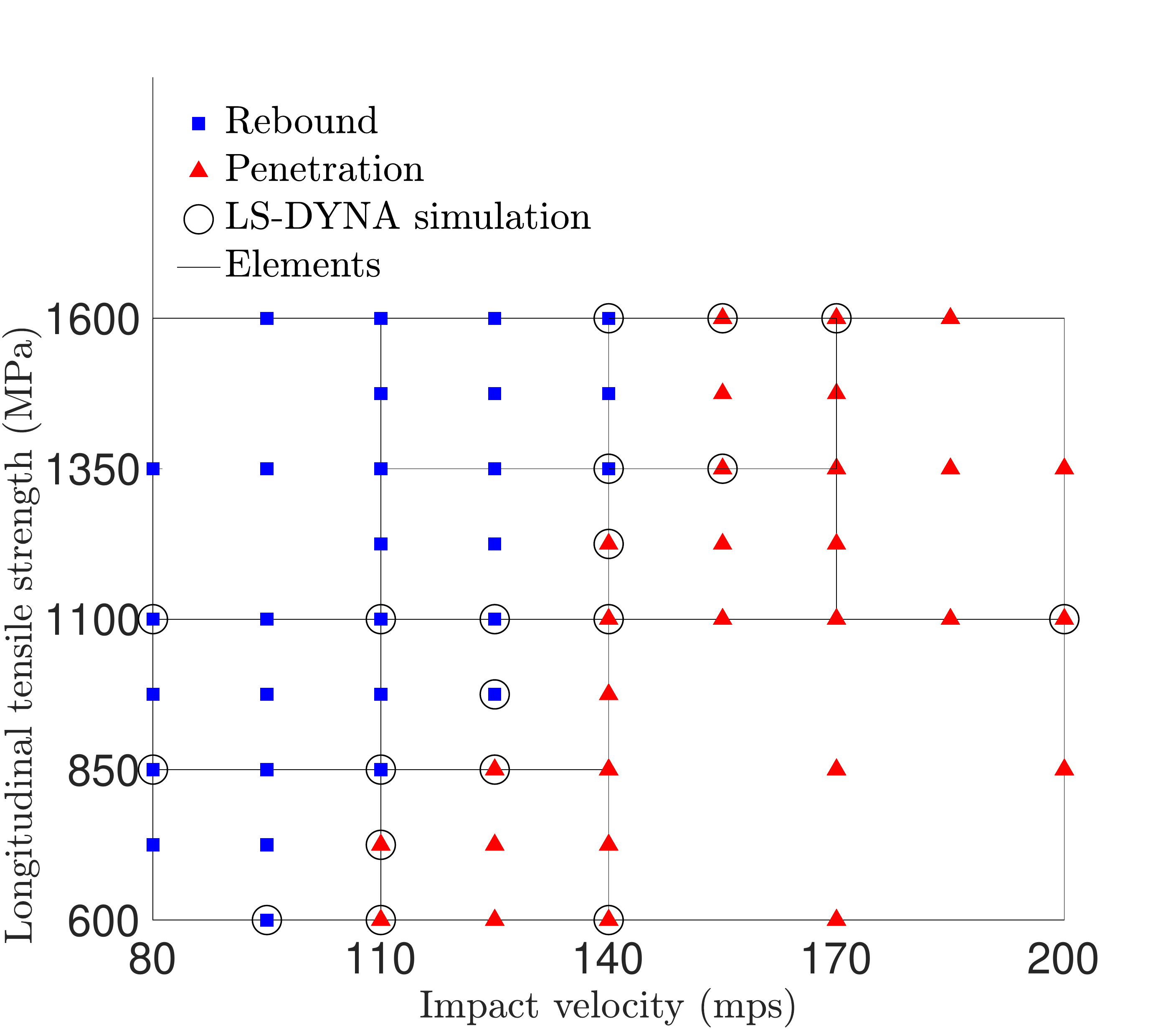}
\caption{}
\label{fig:2-D_uniform_samples_PVR_subfig4}
\end{subfigure}\quad 
\begin{subfigure}[b]{0.3\textwidth}
\centering
\includegraphics[width=\linewidth]{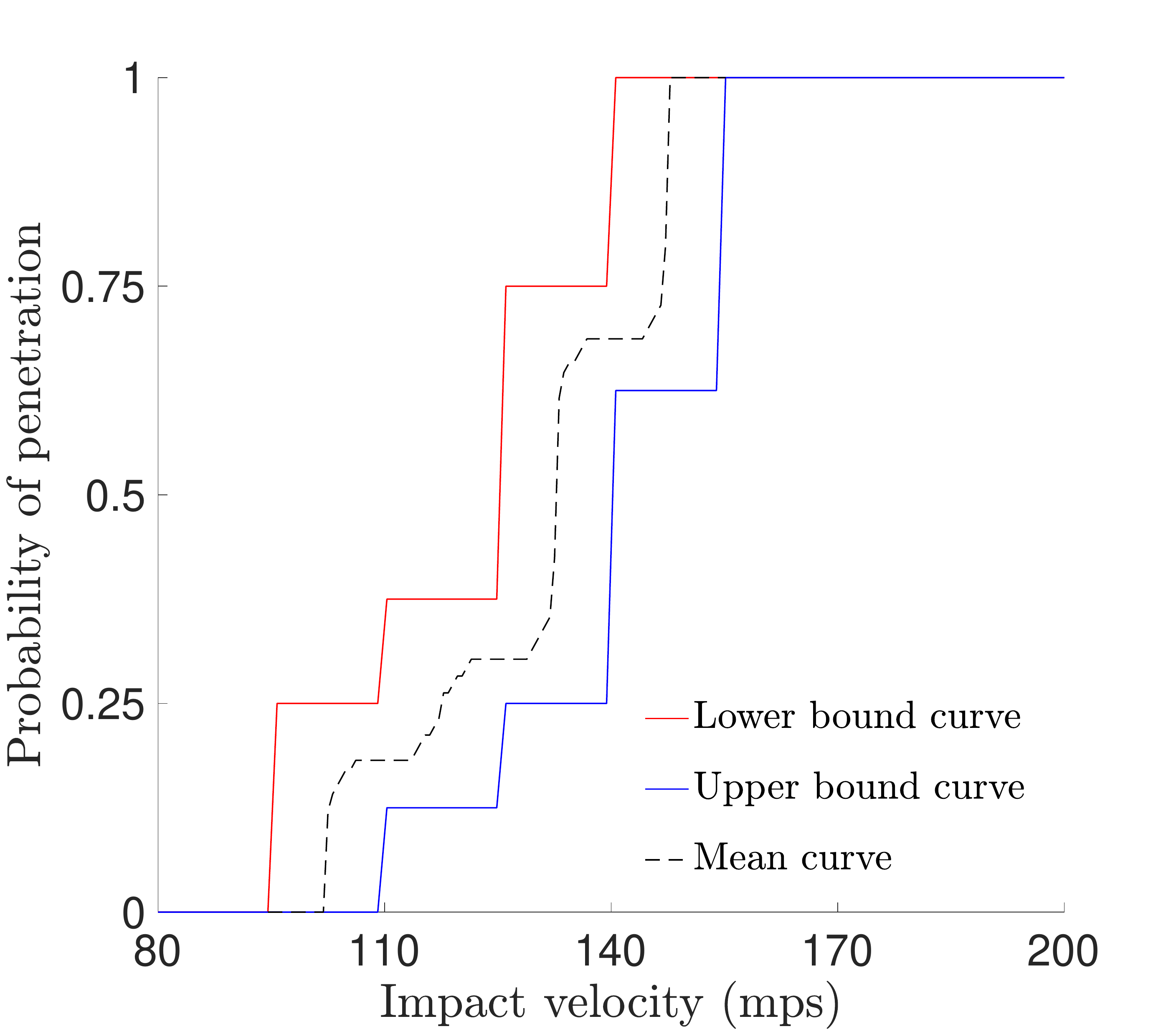}
\caption{}
\label{fig:2-D_uniform_samples_PVR_subfig5}
\end{subfigure}\quad 
\begin{subfigure}[b]{0.3\textwidth}
\centering
\includegraphics[width=\linewidth]{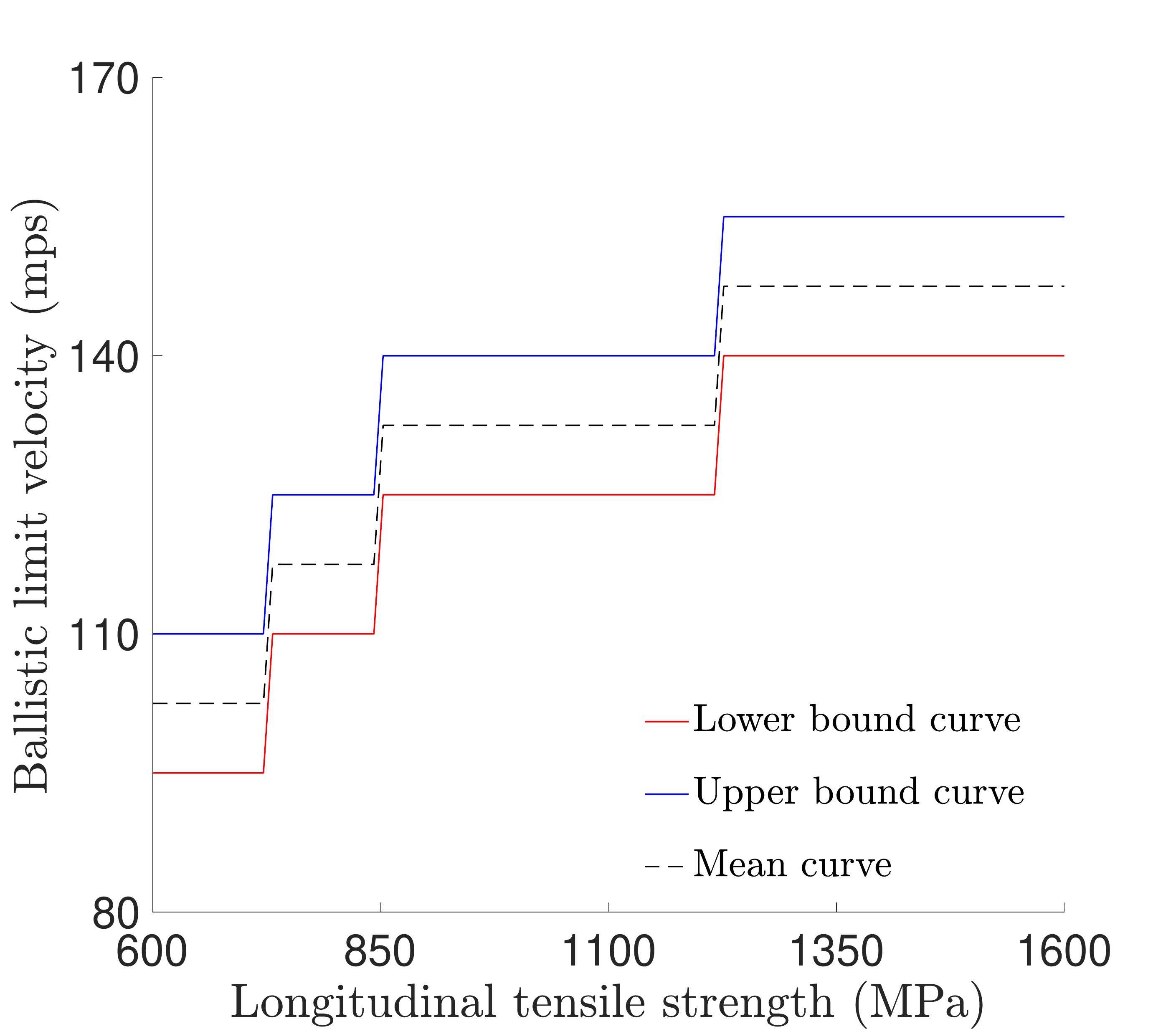}
\caption{}
\label{fig:2-D_uniform_samples_PVR_subfig6}
\end{subfigure}
\begin{subfigure}[b]{0.3\textwidth}
\centering
\includegraphics[width=\linewidth]{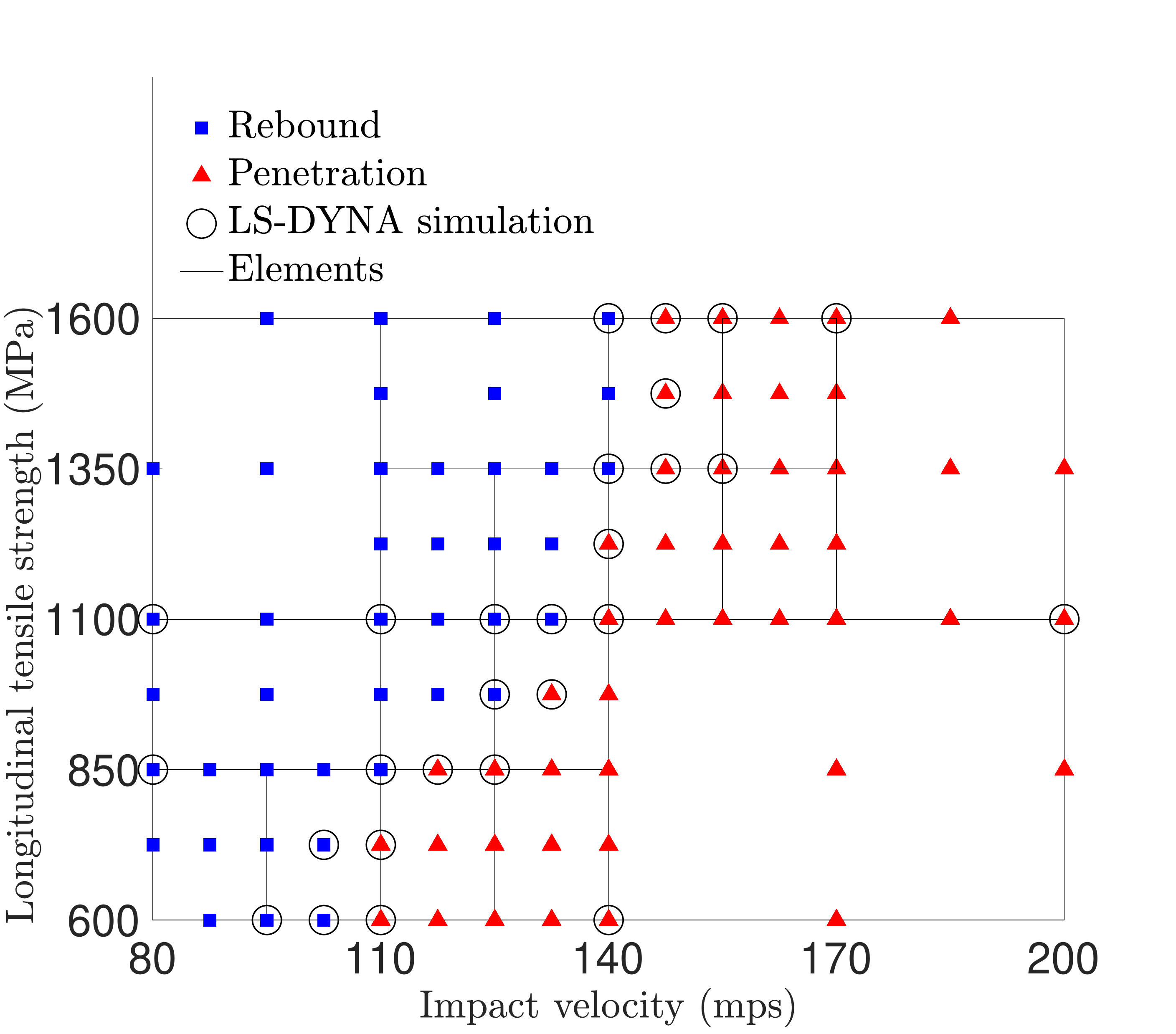}
\caption{}
\label{fig:2-D_uniform_samples_PVR_subfig7}
\end{subfigure}\quad 
\begin{subfigure}[b]{0.3\textwidth}
\centering
\includegraphics[width=\linewidth]{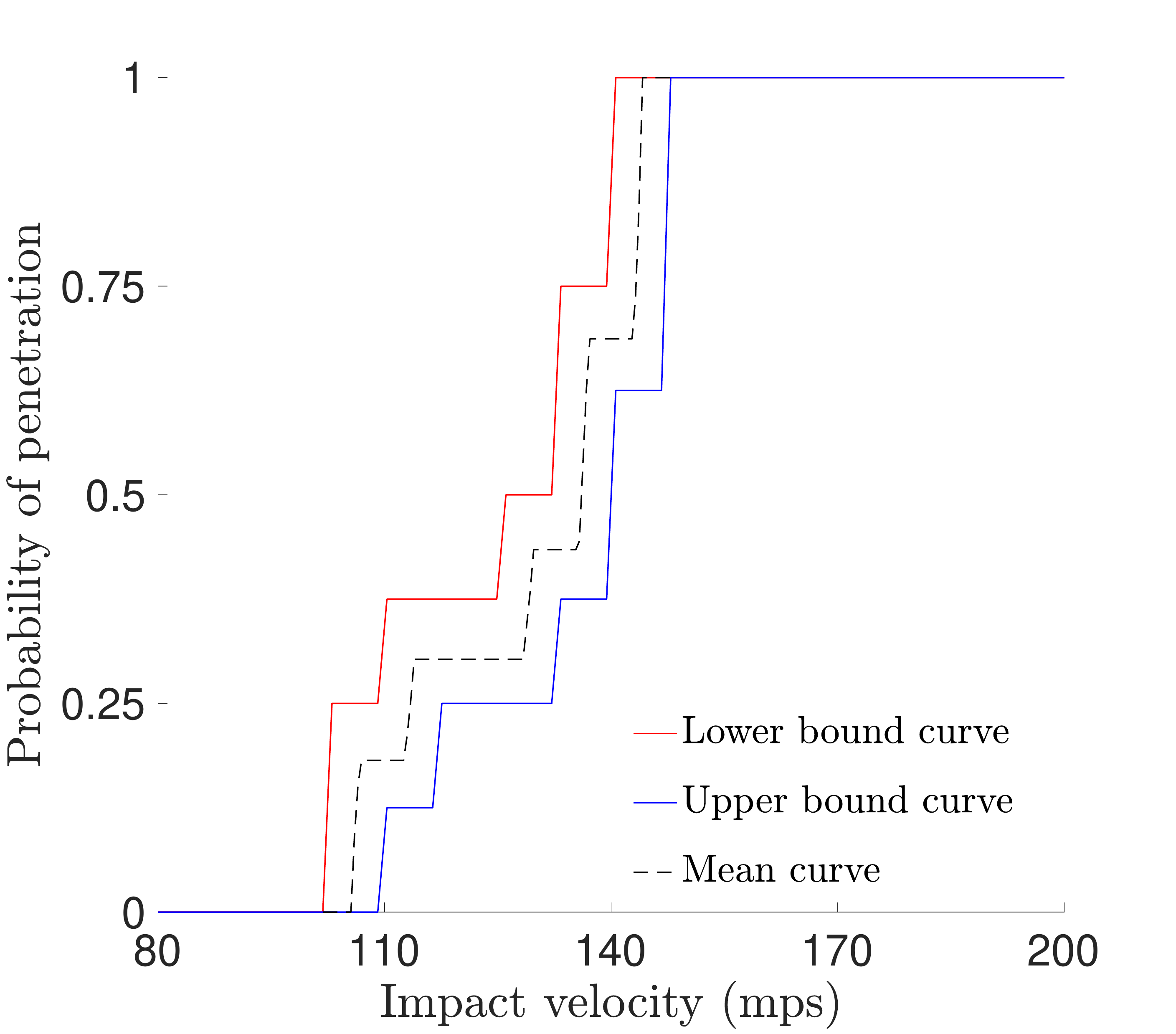}
\caption{}
\label{fig:2-D_uniform_samples_PVR_subfig8}
\end{subfigure}\quad 
\begin{subfigure}[b]{0.3\textwidth}
\centering
\includegraphics[width=\linewidth]{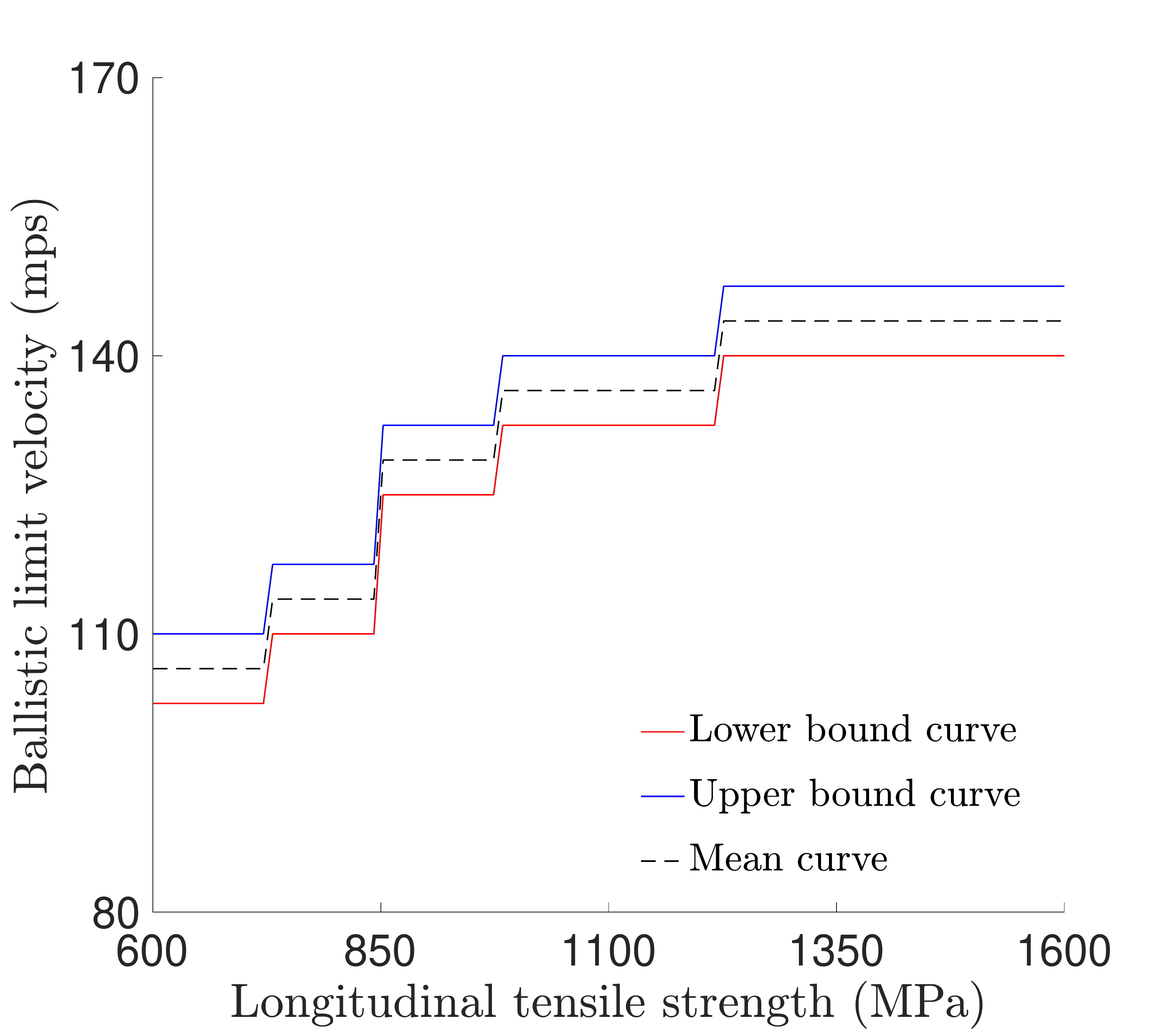}
\caption{}
\label{fig:2-D_uniform_samples_PVR_subfig9}
\end{subfigure}
\caption{(a, d, g): Adaptive sampling and domain decomposition in the $2$-dimensional input domain of impact velocity and longitudinal tensile strength with increase in the number of LS-DYNA simulations from (a) $16$ to (d) $19$ to (g) $27$; (b, e, h): Evolution in the lower, upper and mean PVR curves using (b) $16$, (e) $19$ and (h) $27$ LS-DYNA simulations; (c, f, i): Evolution of the lower, upper and mean ballistic limit velocity curves using (c) $16$, (f) $19$ and (i) $27$ LS-DYNA simulations; longitudinal tensile strength is assumed to follow a uniform distribution.}
\label{fig:2-D_uniform_samples_PVR}
\end{figure}
\begin{figure}
\centering
\begin{subfigure}[b]{0.3\textwidth}
\centering
\includegraphics[width=\linewidth]{Samples_ContinuumPW_Model3_iter4_2D.pdf}
\caption{}
\label{fig:Gaussian_to_Uniform_2D_subfig1}
\end{subfigure} \quad 
\begin{subfigure}[b]{0.3\textwidth}
\centering
\includegraphics[width=\linewidth]{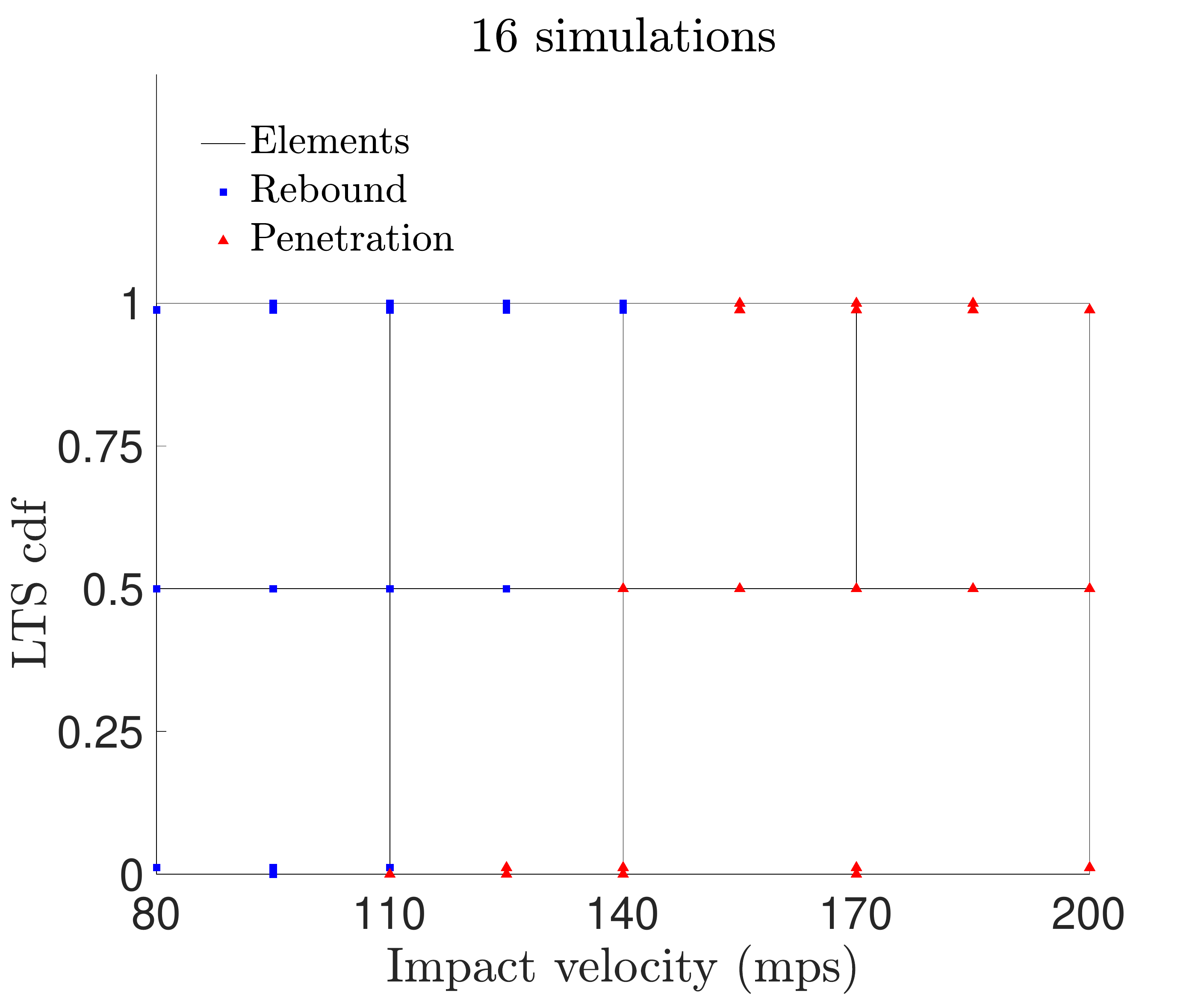}
\caption{}
\label{fig:Gaussian_to_Uniform_2D_subfig2}
\end{subfigure} \quad 
\begin{subfigure}[b]{0.3\textwidth}
\centering
\includegraphics[width=\linewidth]{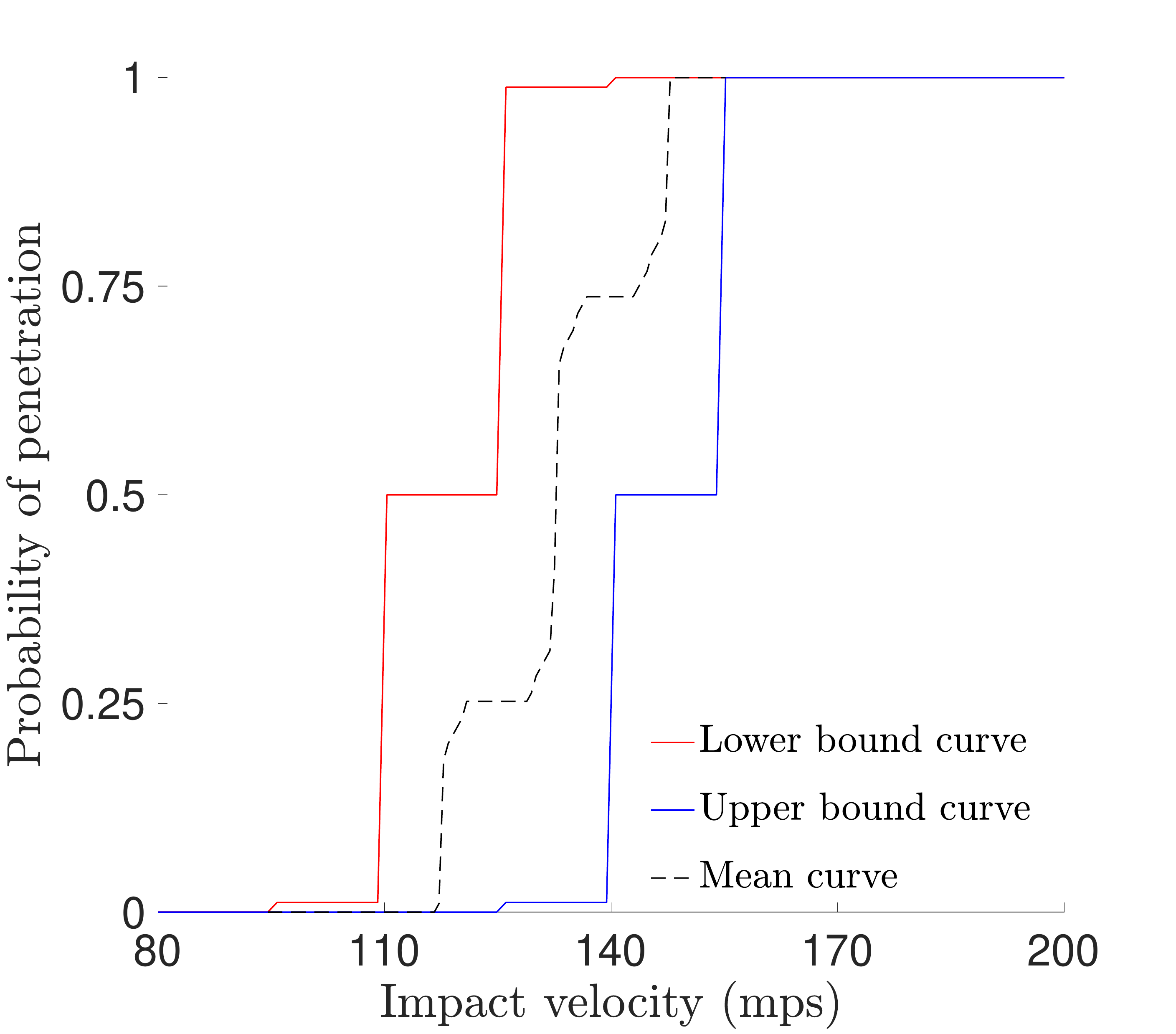}
\caption{}
\label{fig:Gaussian_to_Uniform_2D_subfig3}
\end{subfigure} 
\begin{subfigure}[b]{0.3\textwidth}
\centering
\includegraphics[width=\linewidth]{Samples_ContinuumPW_Model3_iter5_2D.pdf}
\caption{}
\label{fig:Gaussian_to_Uniform_2D_subfig4}
\end{subfigure}\quad 
\begin{subfigure}[b]{0.3\textwidth}
\centering
\includegraphics[width=\linewidth]{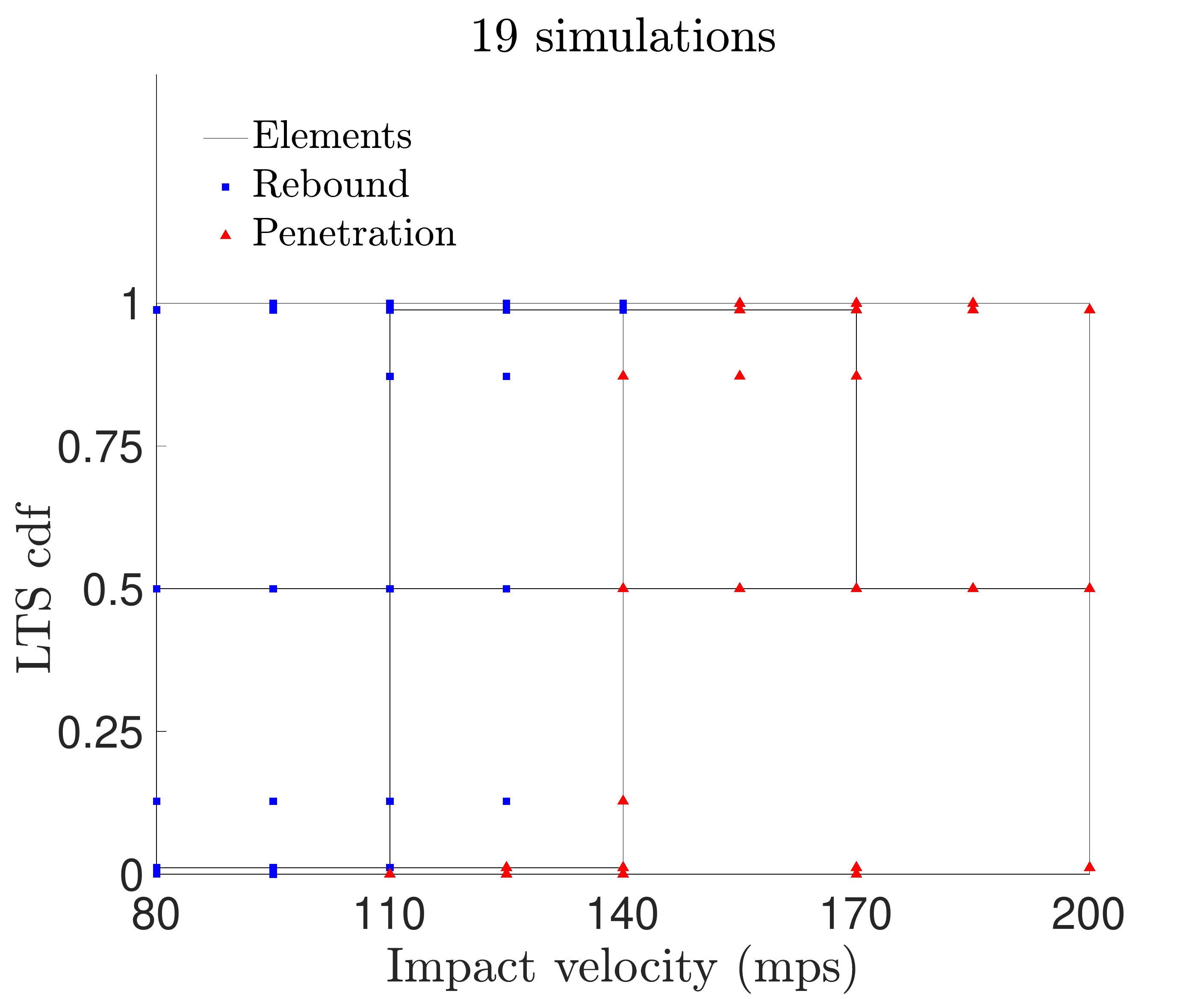}
\caption{}
\label{fig:Gaussian_to_Uniform_2D_subfig5}
\end{subfigure}\quad 
\begin{subfigure}[b]{0.3\textwidth}
\centering
\includegraphics[width=\linewidth]{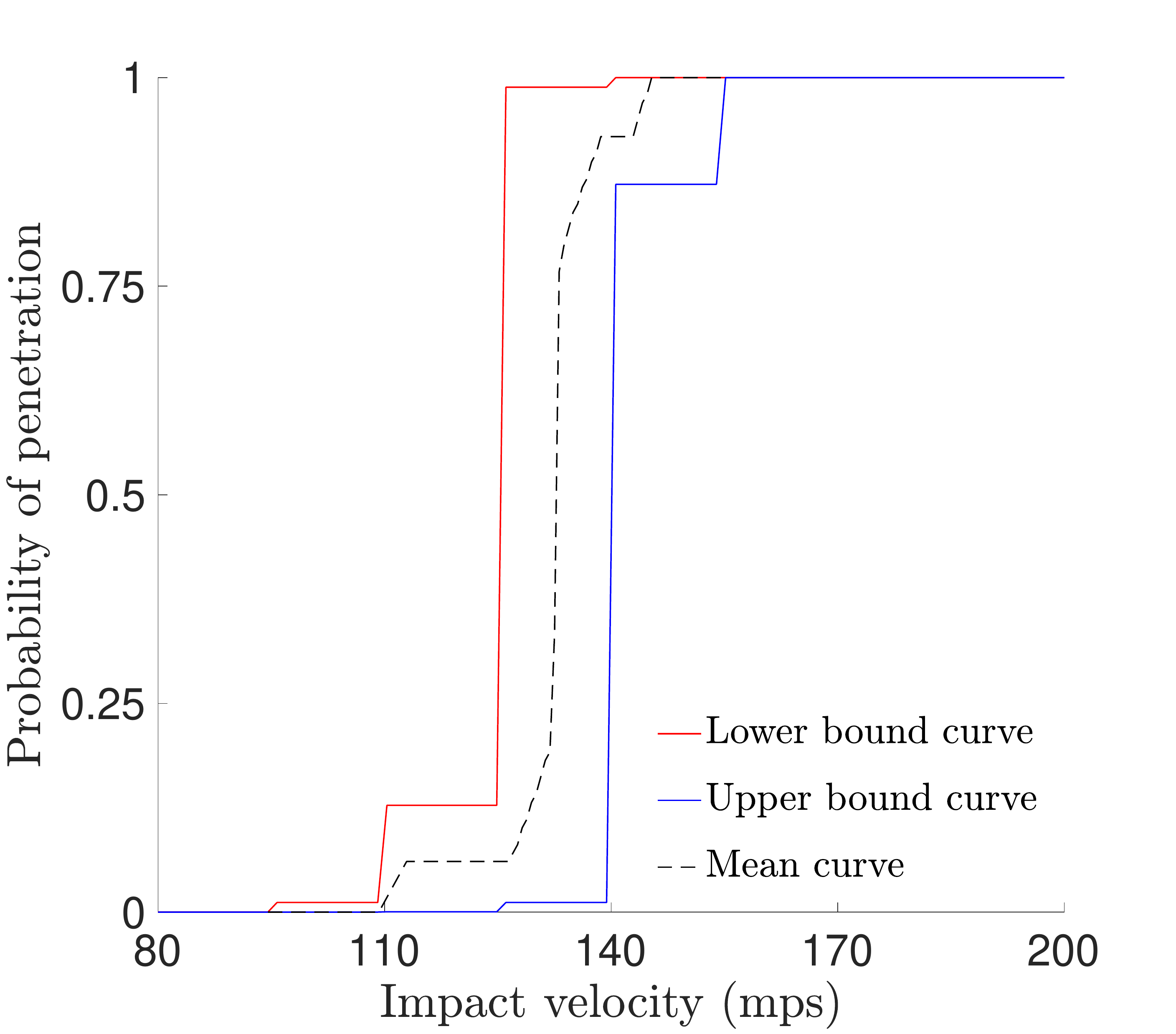}
\caption{}
\label{fig:Gaussian_to_Uniform_2D_subfig6}
\end{subfigure} 
\begin{subfigure}[b]{0.3\textwidth}
\centering
\includegraphics[width=\linewidth]{Samples_ContinuumPW_Model3_iter6_2D.pdf}
\caption{}
\label{fig:Gaussian_to_Uniform_2D_subfig7}
\end{subfigure}\quad 
\begin{subfigure}[b]{0.3\textwidth}
\centering
\includegraphics[width=\linewidth]{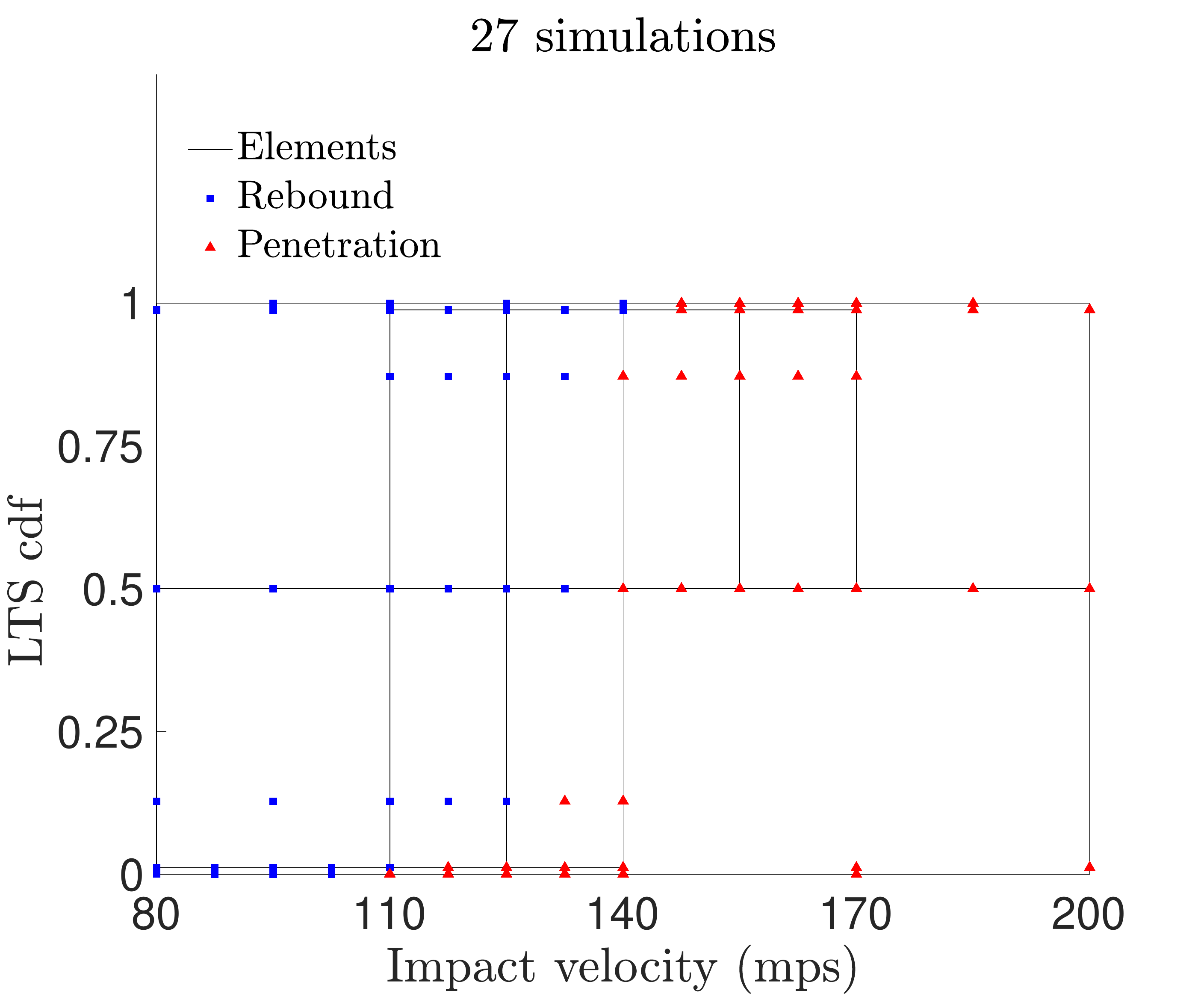}
\caption{}
\label{fig:Gaussian_to_Uniform_2D_subfig8}
\end{subfigure}\quad 
\begin{subfigure}[b]{0.3\textwidth}
\centering
\includegraphics[width=\linewidth]{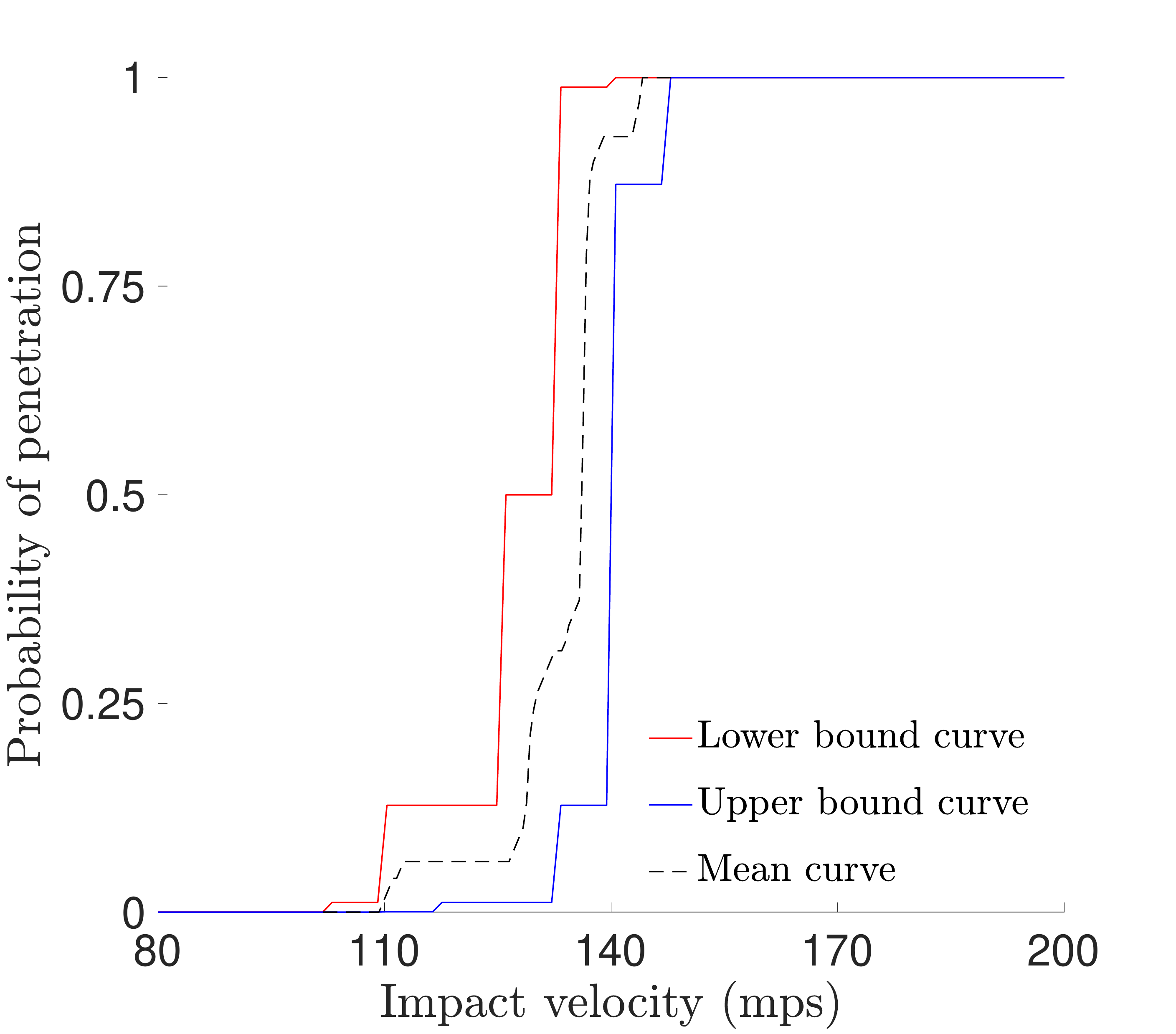}
\caption{}
\label{fig:Gaussian_to_Uniform_2D_subfig9}
\end{subfigure} 
\caption{(a, d, g): Adaptive sampling and domain decomposition in the $2$-dimensional input domain of impact velocity and normally distributed longitudinal tensile strength (LTS) with increase in the number of LS-DYNA simulations from (a) $16$ to (d) $19$ to (g) $27$; longitudinal tensile strength (LTS) is assumed to follow a normal distribution with $\mu=1100$ MPa and $\sigma=110$ MPa; (b, e, h): Transformation of samples and elements to an uniform domain involving (b) $16$, (e) $19$ and (h) $27$ LS-DYNA simulations; (c, f, i): Evolution of the lower, upper and mean PVR curves using (c) $16$, (f) $19$ and (i) $27$ LS-DYNA simulations where longitudinal tensile strength (LTS) is normally distributed.}
\label{fig:Gaussian_to_Uniform_2D}
\end{figure}
\begin{figure}
\centering
\includegraphics[width=0.4\textwidth]{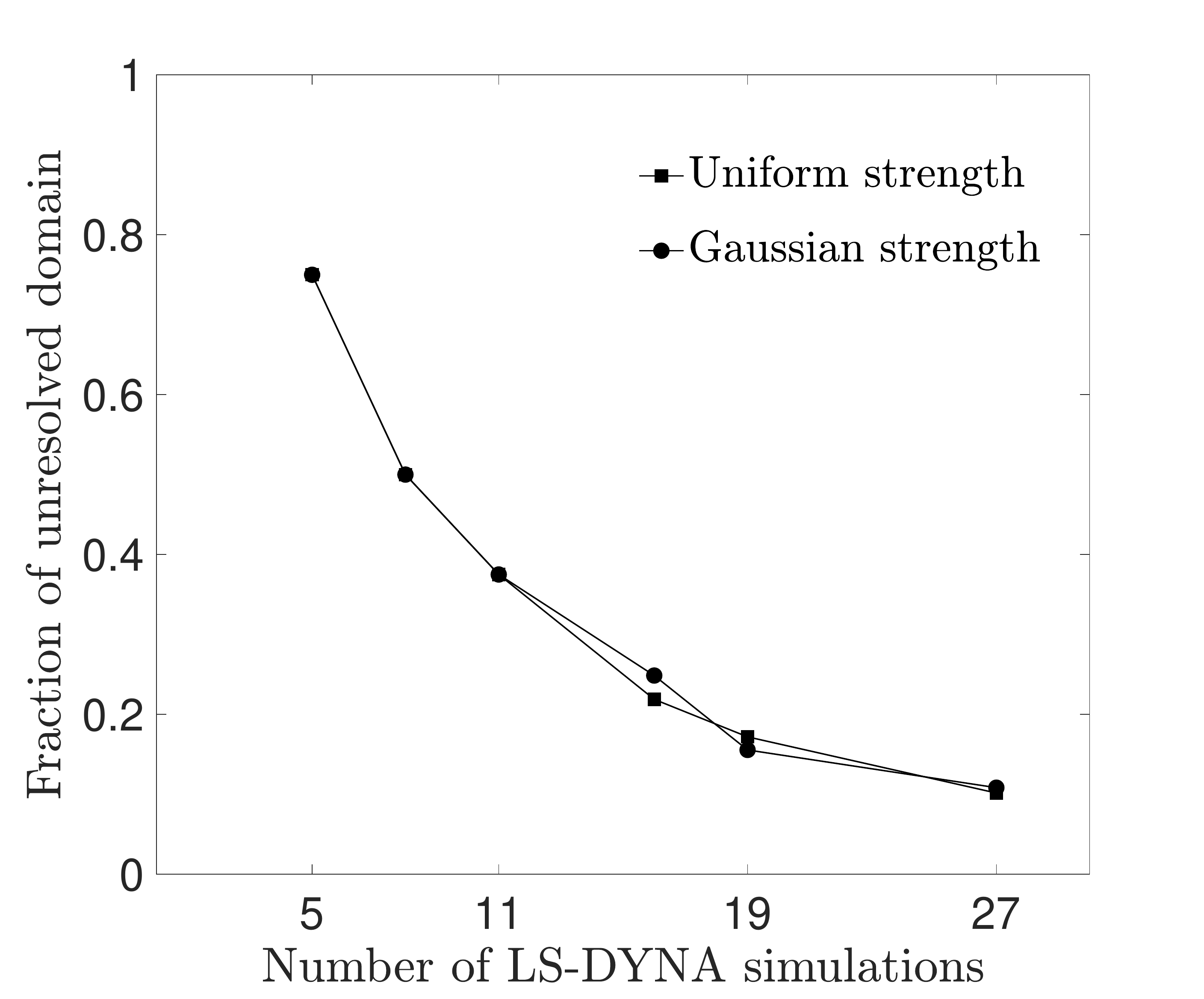}
\caption{Evolution of the fraction of unresolved domain for the $2$-dimensional problem where the impact velocity of the projectile and the longitudinal tensile strength of the plate are the variable parameters}
\label{fig:Fraction_of_unresolved_domain_2D}
\end{figure}
In this section, the effective longitudinal tensile strength (LTS) of the continuum level composite plate is considered as a variable parameter over the range $[600, 1600]$ MPa. The impact velocity variation is again considered over the range $[80, 200]$ m/s. From the methodology perspective, this is now a $2$-dimensional problem with impact velocity and the longitudinal tensile strength as the two parameters. It is of interest to generate the PVR curve by taking into account the variability of the tensile strength parameter, and assessing the variation of the ballistic limit with respect to the strength parameter variation. Figures \ref{fig:2-D_uniform_samples_PVR_subfig1}, \ref{fig:2-D_uniform_samples_PVR_subfig4} and \ref{fig:2-D_uniform_samples_PVR_subfig7} show the evolution of the sample in the input space using the domain decomposition method with increase in the number of iterations of the algorithm. The input domain can be roughly divided into three regions: `rebound' region, `penetration' region, and `transition' region separating the `penetration' and `rebound' region. It is of primary interest to resolve the different regions in the entire input domain in an efficient manner in order to estimate the relevant quantities of interest. The key to resolving the entire domain is to accurately locate the transition region. It is seen that the adaptive algorithm biases sampling towards the transition region. Figures \ref{fig:2-D_uniform_samples_PVR_subfig2}, \ref{fig:2-D_uniform_samples_PVR_subfig5} and \ref{fig:2-D_uniform_samples_PVR_subfig8} show the corresponding lower bound, upper bound and mean PVR curves assuming that the longitudinal tensile strength parameter follows a uniform distribution. It is observed that the area between the lower bound and upper bound PVR curves decreases with increase in the number of sample evaluations from $16$ in figure \ref{fig:2-D_uniform_samples_PVR_subfig2} to $27$ in figure \ref{fig:2-D_uniform_samples_PVR_subfig8}. The space between the upper bound and lower bound PVR curve gives a measure of the overall uncertainty of the PVR curve estimate and it is a function of the unresolved space in the input domain. Addition of more samples using the adaptive algorithm helps in resolving more space in the input domain which helps in closing the gap between the two PVR curves and reducing the overall uncertainty.
Figures \ref{fig:2-D_uniform_samples_PVR_subfig3}, \ref{fig:2-D_uniform_samples_PVR_subfig6} and \ref{fig:2-D_uniform_samples_PVR_subfig9} show the corresponding lower bound, upper bound and mean ballistic limit velocity curves. It is observed that the area between the lower bound and upper bound ballistic limit curves decreases with increase in the number of sample evaluations from $16$ in figure \ref{fig:2-D_uniform_samples_PVR_subfig3} to $27$ in figure \ref{fig:2-D_uniform_samples_PVR_subfig9}. As observed with the PVR curves, addition of more samples using the adaptive algorithm helps in resolving more of the input domain, which helps close the gap between the two ballistic limit curves. The PVR and the ballistic limit curves are calculated analytically as explained in section \ref{section: prediction step} where parameter $1$ is the impact velocity and parameter $2$ is the longitudinal tensile strength. The mean PVR curves shown as a black dotted line in figures \ref{fig:2-D_uniform_samples_PVR_subfig2}, \ref{fig:2-D_uniform_samples_PVR_subfig5} and \ref{fig:2-D_uniform_samples_PVR_subfig8} are obtained by implementing the k-nearest neighbor classification using Chebychev distance, assuming that the corresponding red and blue PVR curves are data with two different class labels. The mean PVR curves are basically the k-NN decision surfaces. The mean ballistic limit curves shown as a black dotted line in figures \ref{fig:2-D_uniform_samples_PVR_subfig3}, \ref{fig:2-D_uniform_samples_PVR_subfig6} and \ref{fig:2-D_uniform_samples_PVR_subfig9} are obtained by taking the average of the corresponding lower and upper bound ballistic limit curves.\\
\indent Figures \ref{fig:Gaussian_to_Uniform_2D_subfig1}, \ref{fig:Gaussian_to_Uniform_2D_subfig4} and \ref{fig:Gaussian_to_Uniform_2D_subfig7}, same as in figures \ref{fig:2-D_uniform_samples_PVR_subfig1}, \ref{fig:2-D_uniform_samples_PVR_subfig4} and \ref{fig:2-D_uniform_samples_PVR_subfig7}, show the evolution of samples in the input space using the domain decomposition method with increase in the number of iterations of the algorithm, but in this case, the longitudinal tensile strength parameter is assumed to be normally distributed with mean $\mu=1100$ MPa and standard deviation $\sigma=110$ MPa. Figures \ref{fig:Gaussian_to_Uniform_2D_subfig2}, \ref{fig:Gaussian_to_Uniform_2D_subfig5} and \ref{fig:Gaussian_to_Uniform_2D_subfig8} show the corresponding samples and elements after transformation of the tensile strength coordinates from the Gaussian (normal) domain to the uniform $[0, 1]$ domain. The transformation is performed using the `normcdf()' matlab \cite{MATLAB:2018} function which transforms the original sample coordinates in the Gaussian space to $[0,1]$-probability space using the mean and standard deviation values.
Figures \ref{fig:Gaussian_to_Uniform_2D_subfig3}, \ref{fig:Gaussian_to_Uniform_2D_subfig6} and \ref{fig:Gaussian_to_Uniform_2D_subfig9} show the corresponding lower bound, upper bound and mean PVR curves assuming that the longitudinal tensile strength parameter follows a gaussian (normal) distribution. As observed in the uniform distribution case, addition of more samples using the adaptive algorithm helps in resolving more space in the input domain which helps in closing the gap between the two PVR curves. The PVR curves are again calculated analytically as explained in section \ref{section: prediction step} where parameter $1$ is the impact velocity and parameter $2$ is the longitudinal tensile strength.\\
\indent Comparing the PVR curves in the uniformly distributed tensile strength case (uniform PVR curves) and the normally distributed tensile strength case (normal PVR curves), it is seen that the normal distribution has a stretching effect on the curves. It looks as if the probability of penetration (PoP) values greater than $0.5$ get magnified to values close to $1$, while PoP values less than $0.5$ get reduced to values close to $0$, when the underlying distribution of the strength changes from uniform to Gaussian. This happens because the strength coordinates of the samples as well as the elements under a transformation from the Gaussian domain to the uniform domain are biased towards the upper and lower bounds. 
Figure \ref{fig:Fraction_of_unresolved_domain_2D} shows the reduction in the fraction of the unresolved domain with increase in the number of LS-DYNA simulations. The fraction of the unresolved domain is measured by the total hypervolume of the unresolved (unclassified) elements of the finest resolution obtained after subdivision of the larger unresolved elements (remaining at the end of each iteration) along all dimensions. 

\begin{figure}
\centering
\begin{subfigure}[b]{0.4\textwidth}
\centering
\includegraphics[width=1.1\linewidth, height=4.5cm]{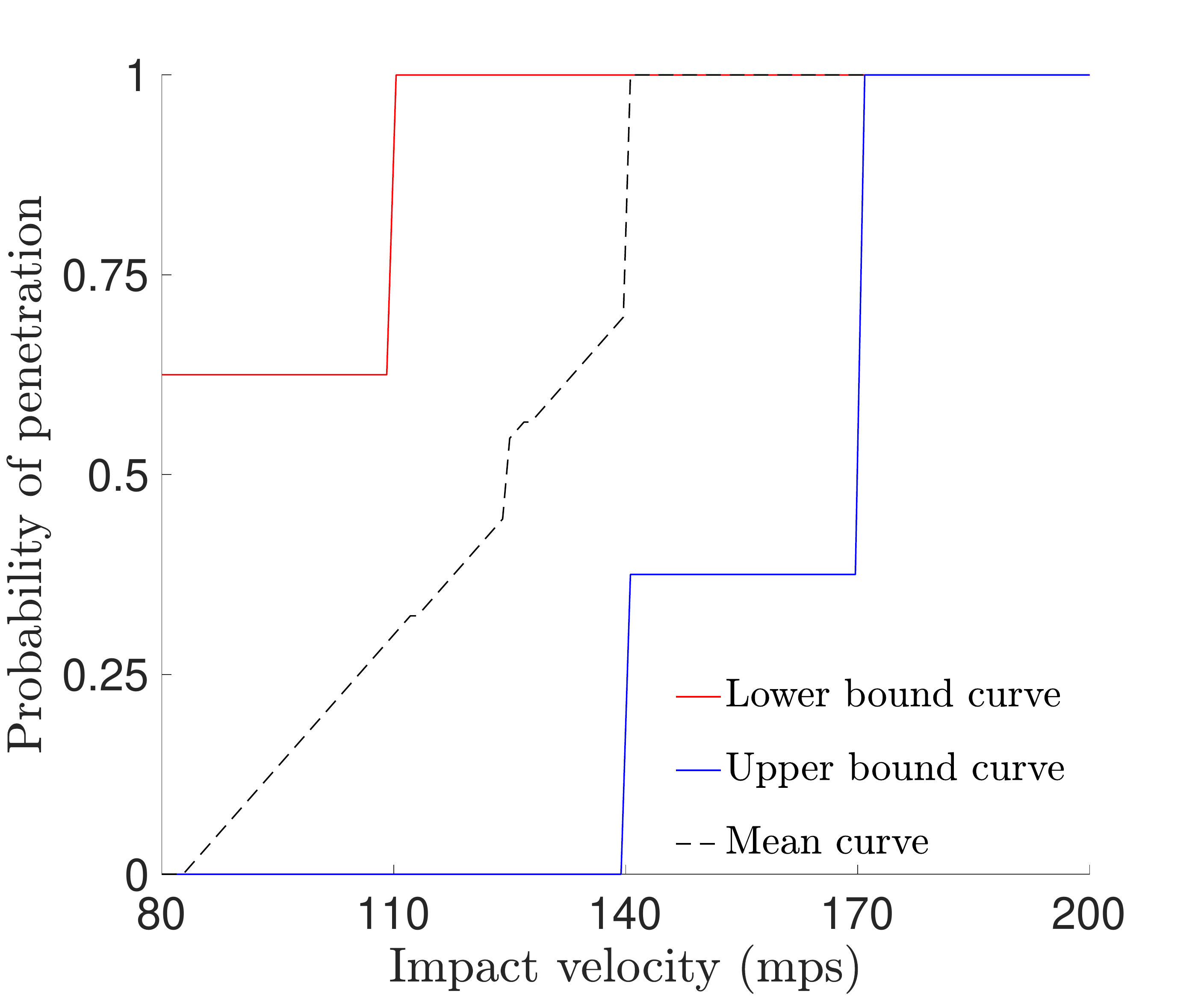}
\caption{}
\label{fig:3-D_uniform_PVR_curves_subfig1}
\end{subfigure} \quad 
\begin{subfigure}[b]{0.4\textwidth}
\centering
\includegraphics[width=1.1\linewidth, height=4.5cm]{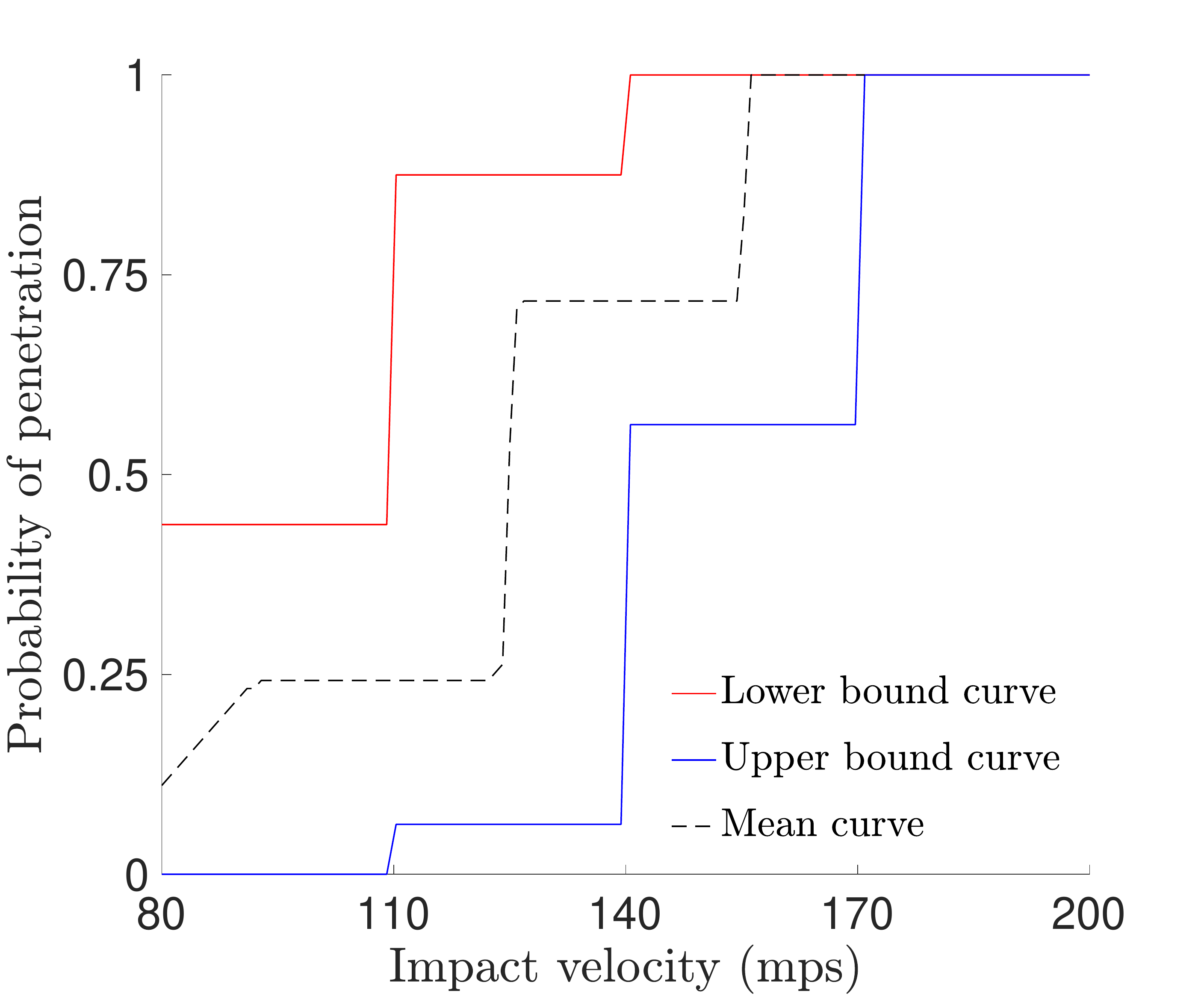}
\caption{}
\label{fig:3-D_uniform_PVR_curves_subfig2}
\end{subfigure} 
\begin{subfigure}[b]{0.4\textwidth}
\centering
\includegraphics[width=1.1\linewidth, height=4.5cm]{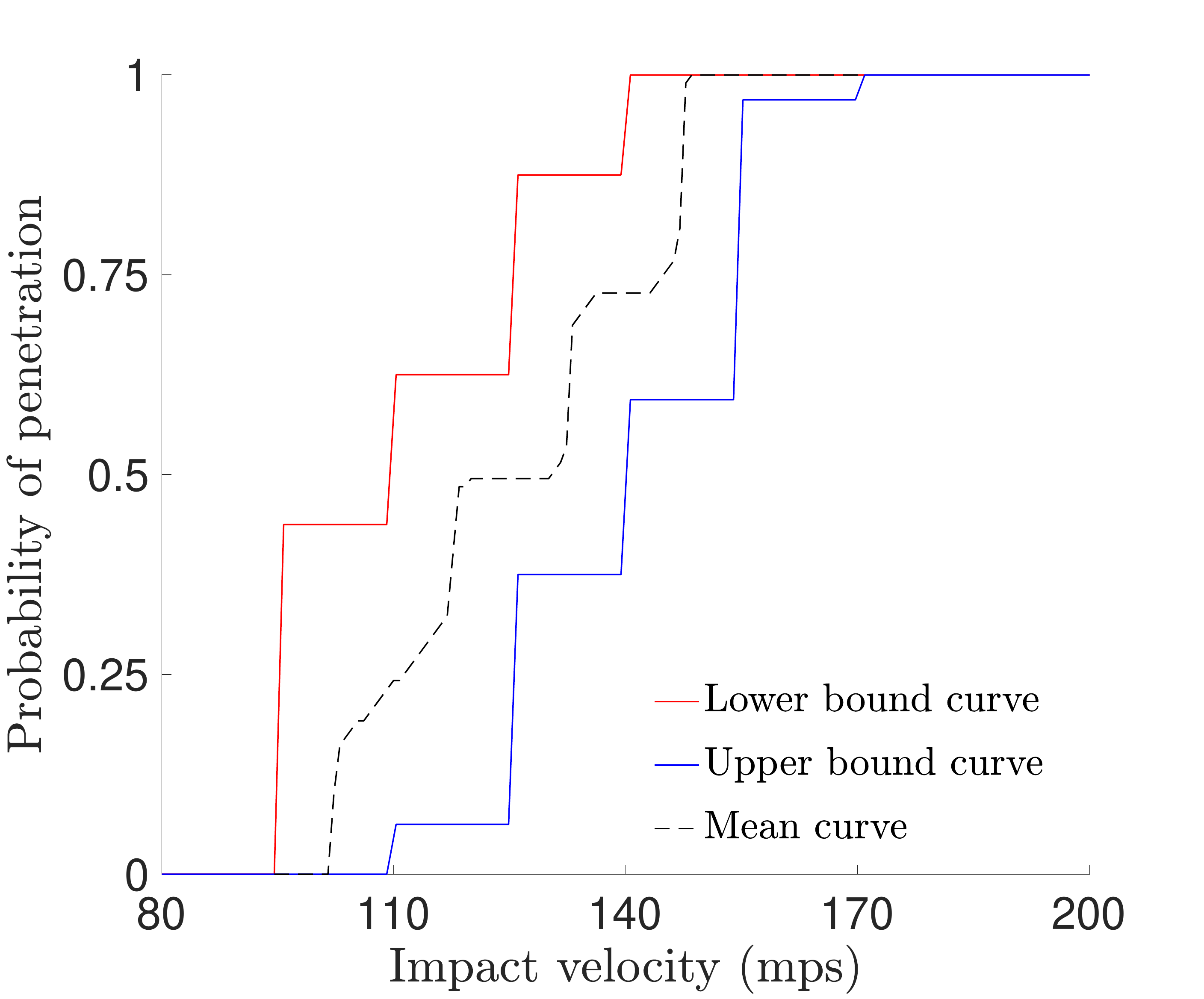}
\caption{}
\label{fig:3-D_uniform_PVR_curves_subfig3}
\end{subfigure}\quad 
\begin{subfigure}[b]{0.4\textwidth}
\centering
\includegraphics[width=1.1\linewidth, height=4.5cm]{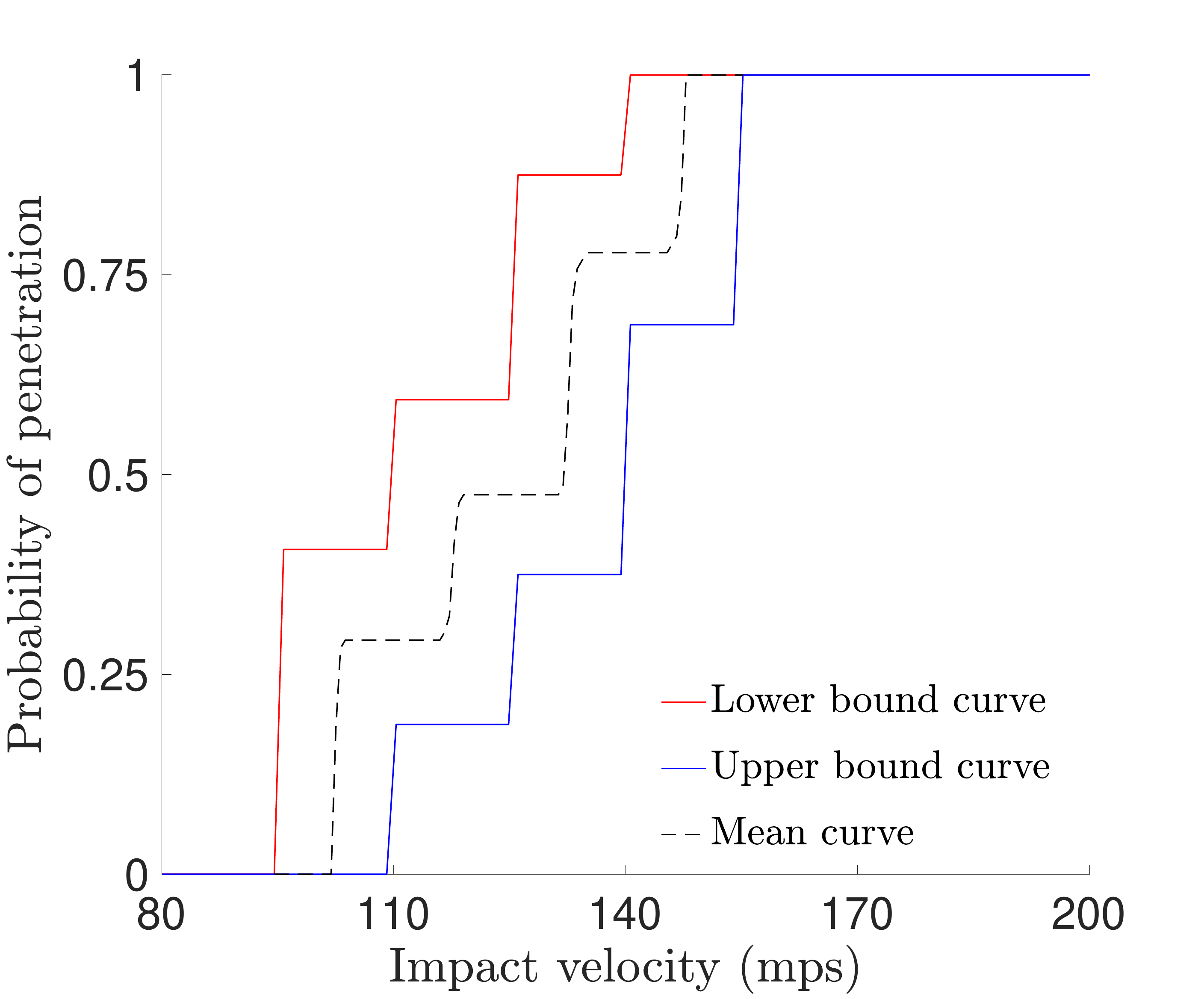}
\caption{}
\label{fig:3-D_uniform_PVR_curves_subfig4}
\end{subfigure}
\begin{subfigure}[b]{0.4\textwidth}
\centering
\includegraphics[width=1.1\linewidth, height=4.5cm]{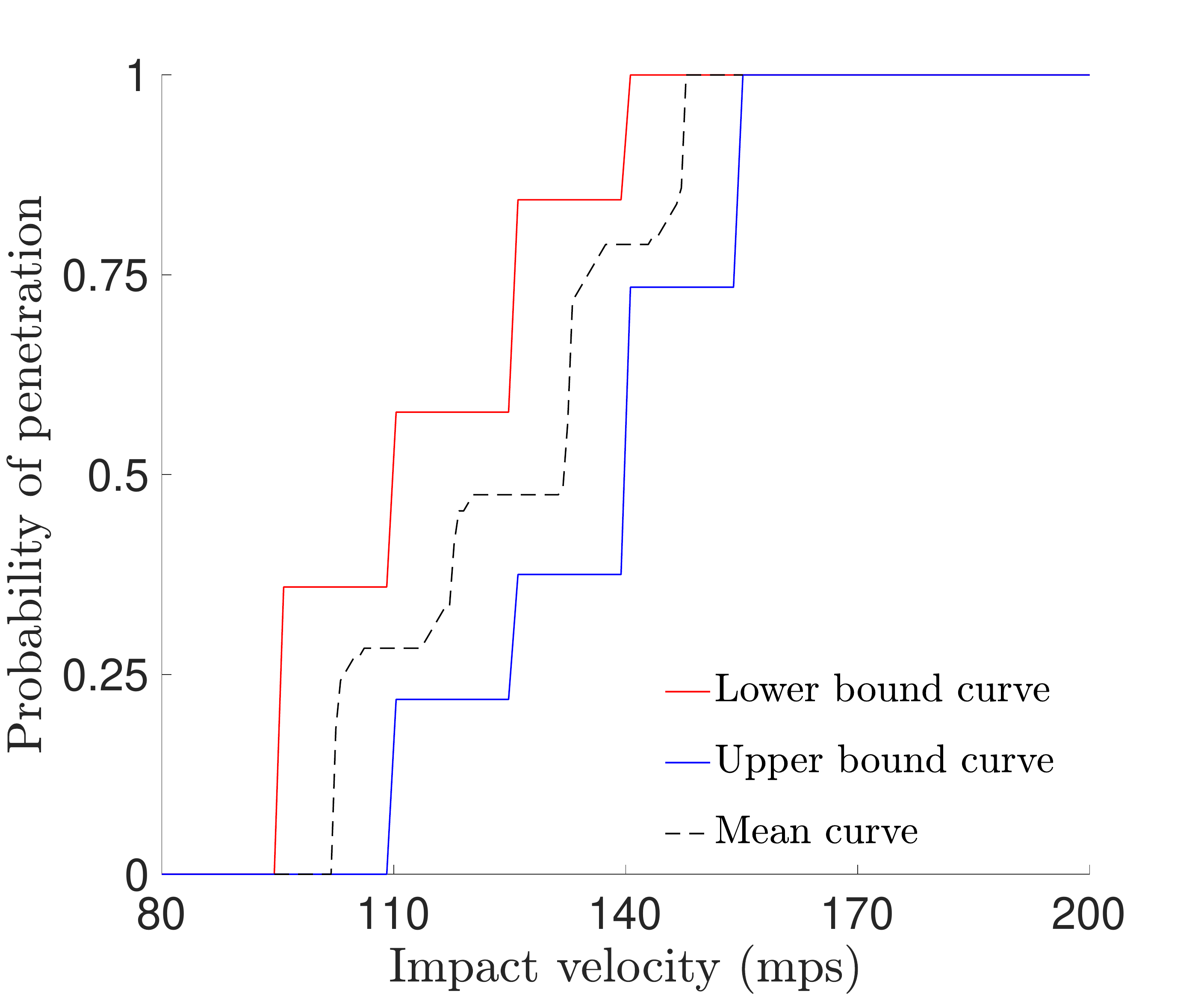}
\caption{}
\label{fig:3-D_uniform_PVR_curves_subfig5}
\end{subfigure}\quad 
\begin{subfigure}[b]{0.4\textwidth}
\centering
\includegraphics[width=1.1\linewidth, height=4.5cm]{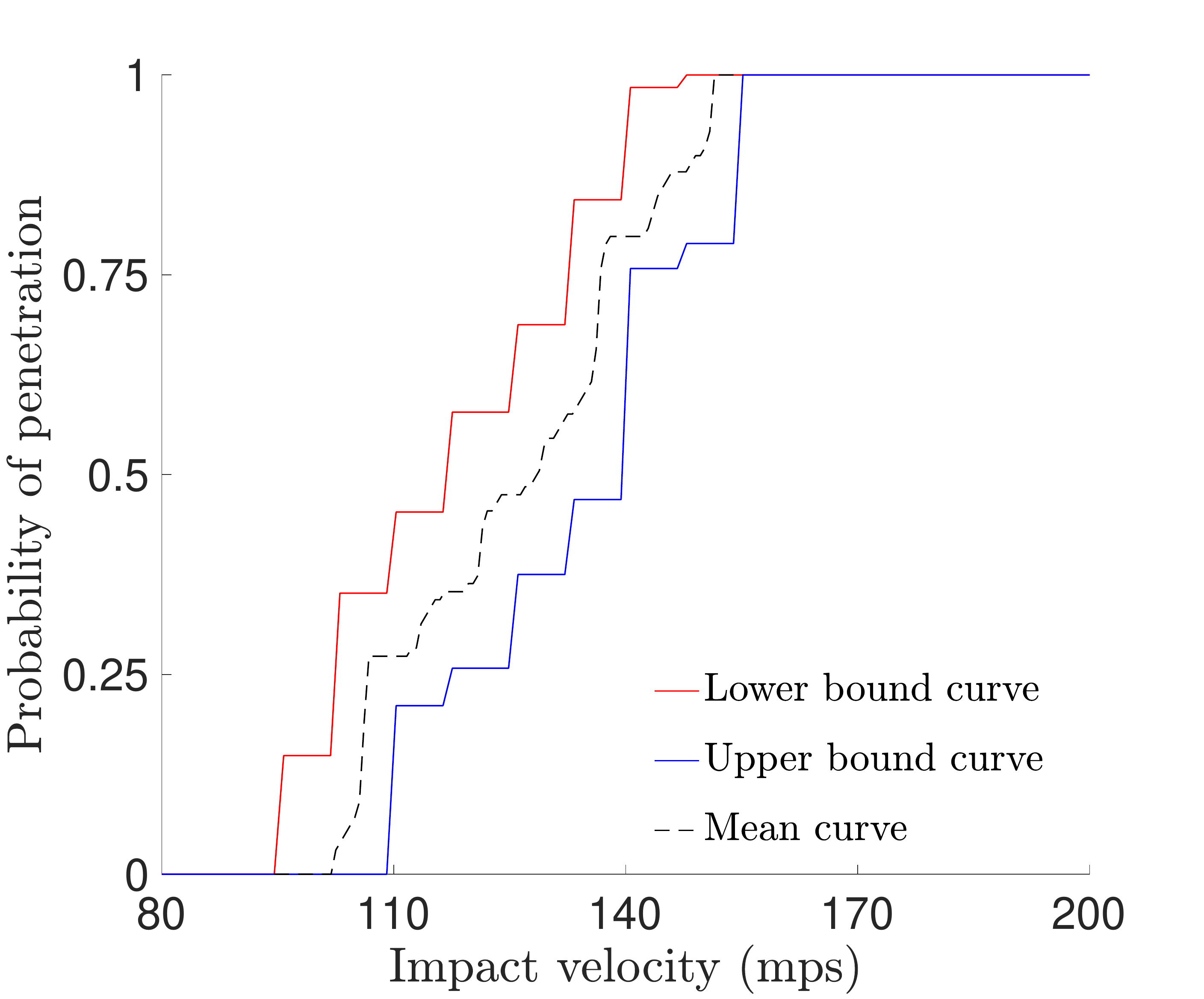}
\caption{}
\label{fig:3-D_uniform_PVR_curves_subfig6}
\end{subfigure}
\caption{Evolution of the lower, upper and mean PVR curves as a function of the impact velocity of the projectile using (a) 20, (b) 27, (c) 43, (d) 56, (e) 67 and (f) 104 LS-DYNA simulations for the 3-dimensional example case; the longitudinal tensile strength and the punch shear strength parameters are assumed to follow an uniform distribution in their respective ranges.}
\label{fig:3-D_uniform_PVR_curves}
\end{figure}
\begin{figure}
\centering
\begin{subfigure}[b]{0.4\textwidth}
\centering
\includegraphics[width=1.1\linewidth, height=4.5cm]{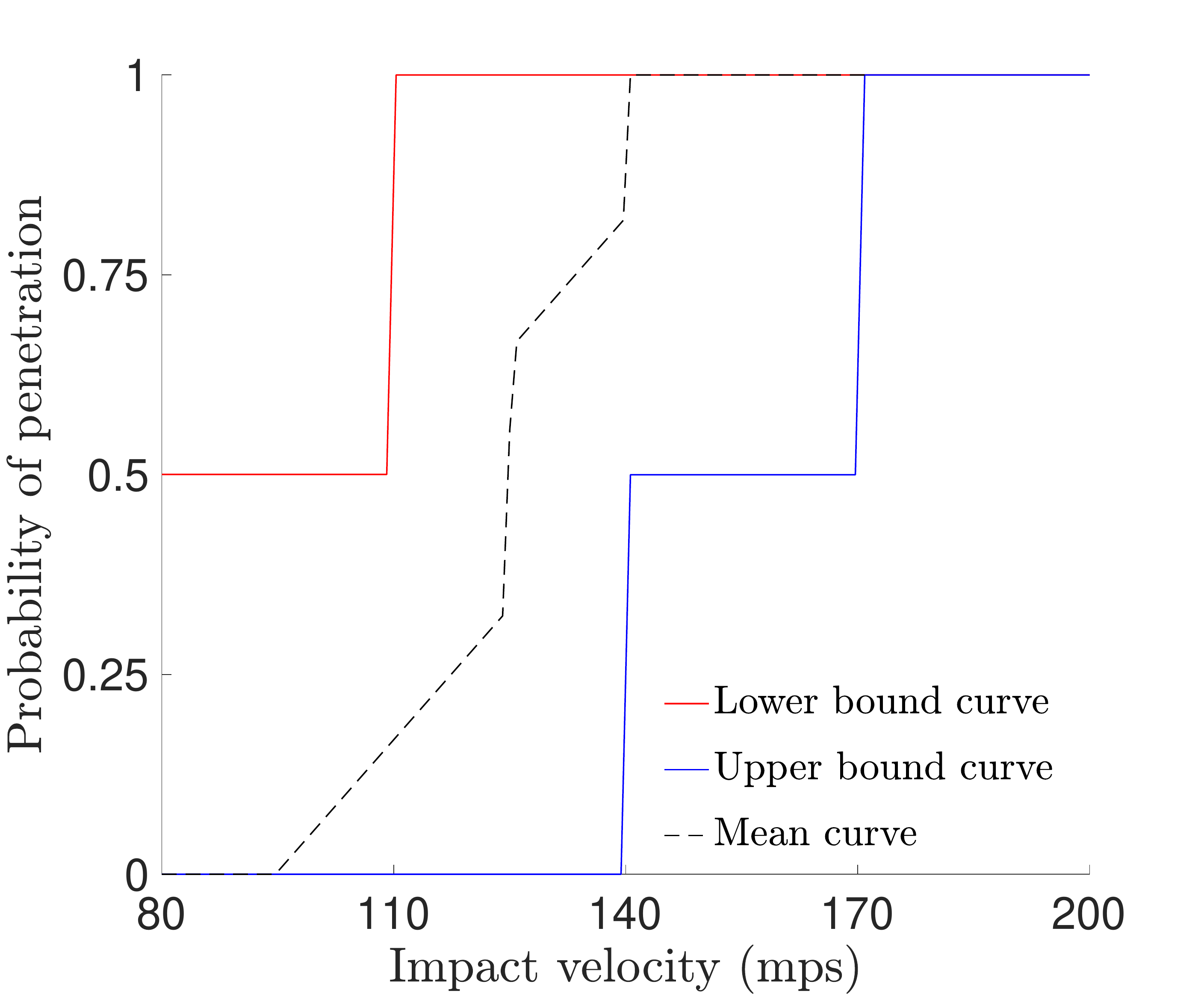}
\caption{}
\label{fig:3-D_gaussian_PVR_curves_subfig1}
\end{subfigure} \quad 
\begin{subfigure}[b]{0.4\textwidth}
\centering
\includegraphics[width=1.1\linewidth, height=4.5cm]{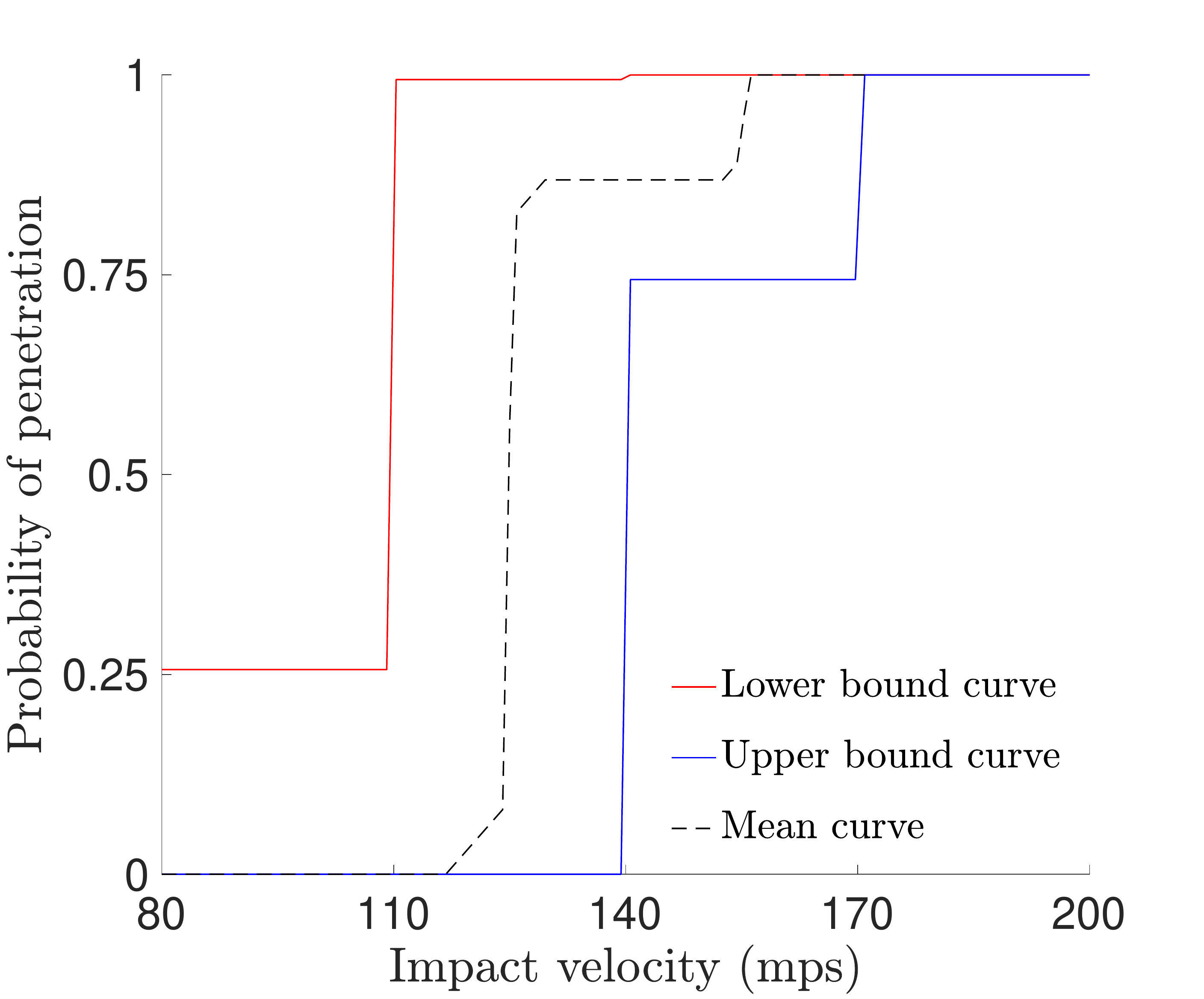}
\caption{}
\label{fig:3-D_gaussian_PVR_curves_subfig2}
\end{subfigure} 
\begin{subfigure}[b]{0.4\textwidth}
\centering
\includegraphics[width=1.1\linewidth, height=4.5cm]{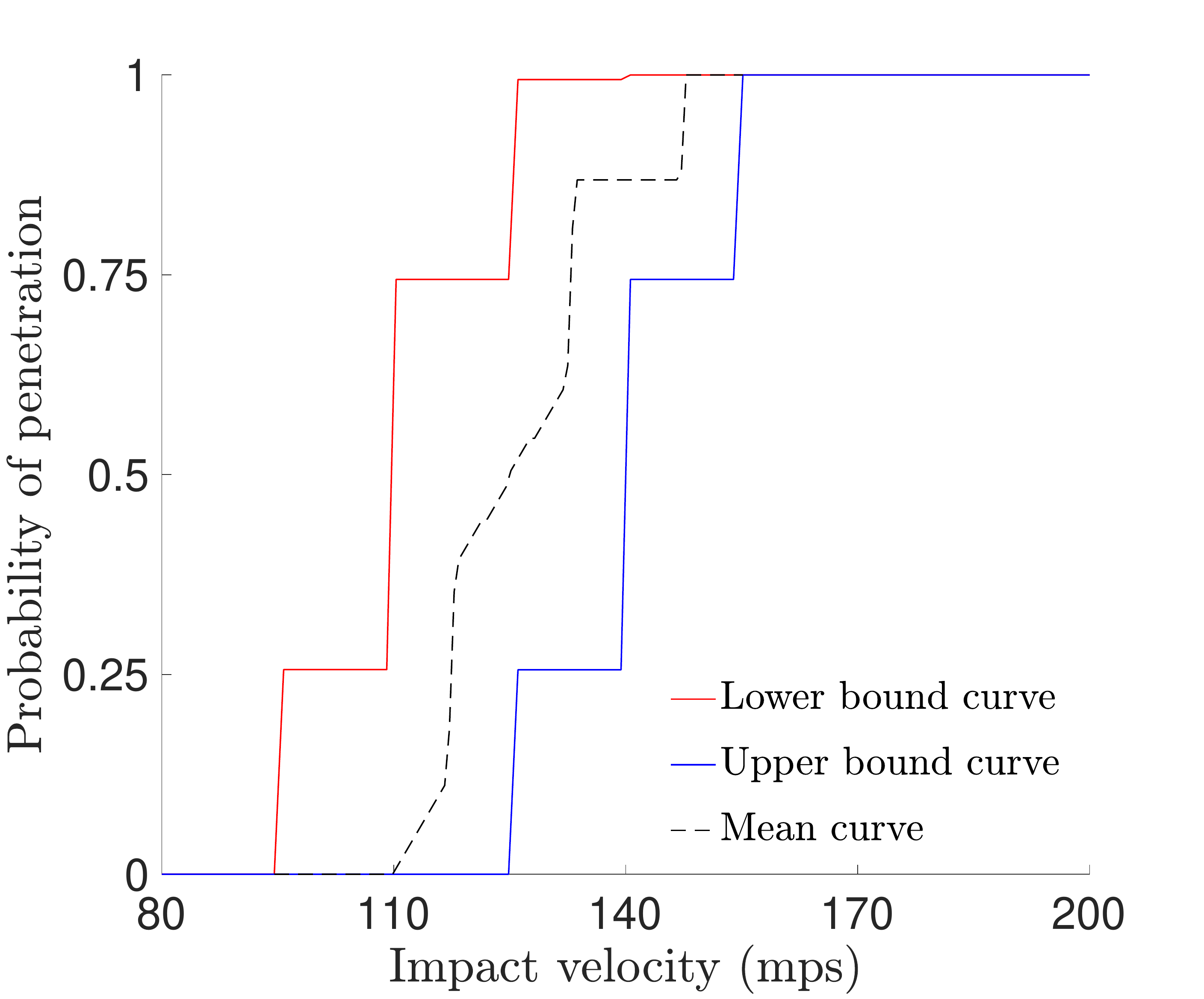}
\caption{}
\label{fig:3-D_gaussian_PVR_curves_subfig3}
\end{subfigure}\quad 
\begin{subfigure}[b]{0.4\textwidth}
\centering
\includegraphics[width=1.1\linewidth, height=4.5cm]{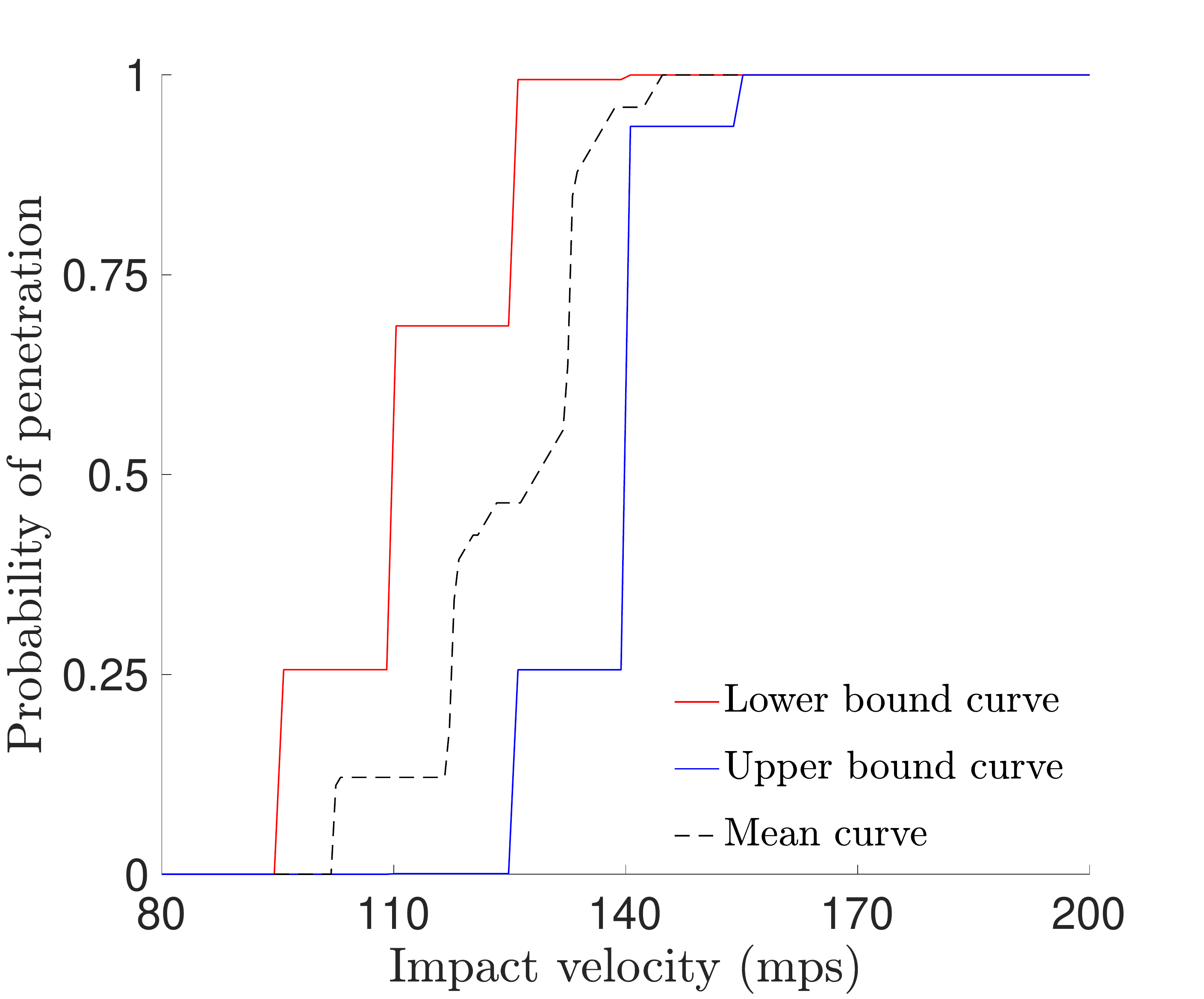}
\caption{}
\label{fig:3-D_gaussian_PVR_curves_subfig4}
\end{subfigure}
\begin{subfigure}[b]{0.4\textwidth}
\centering
\includegraphics[width=1.1\linewidth, height=4.5cm]{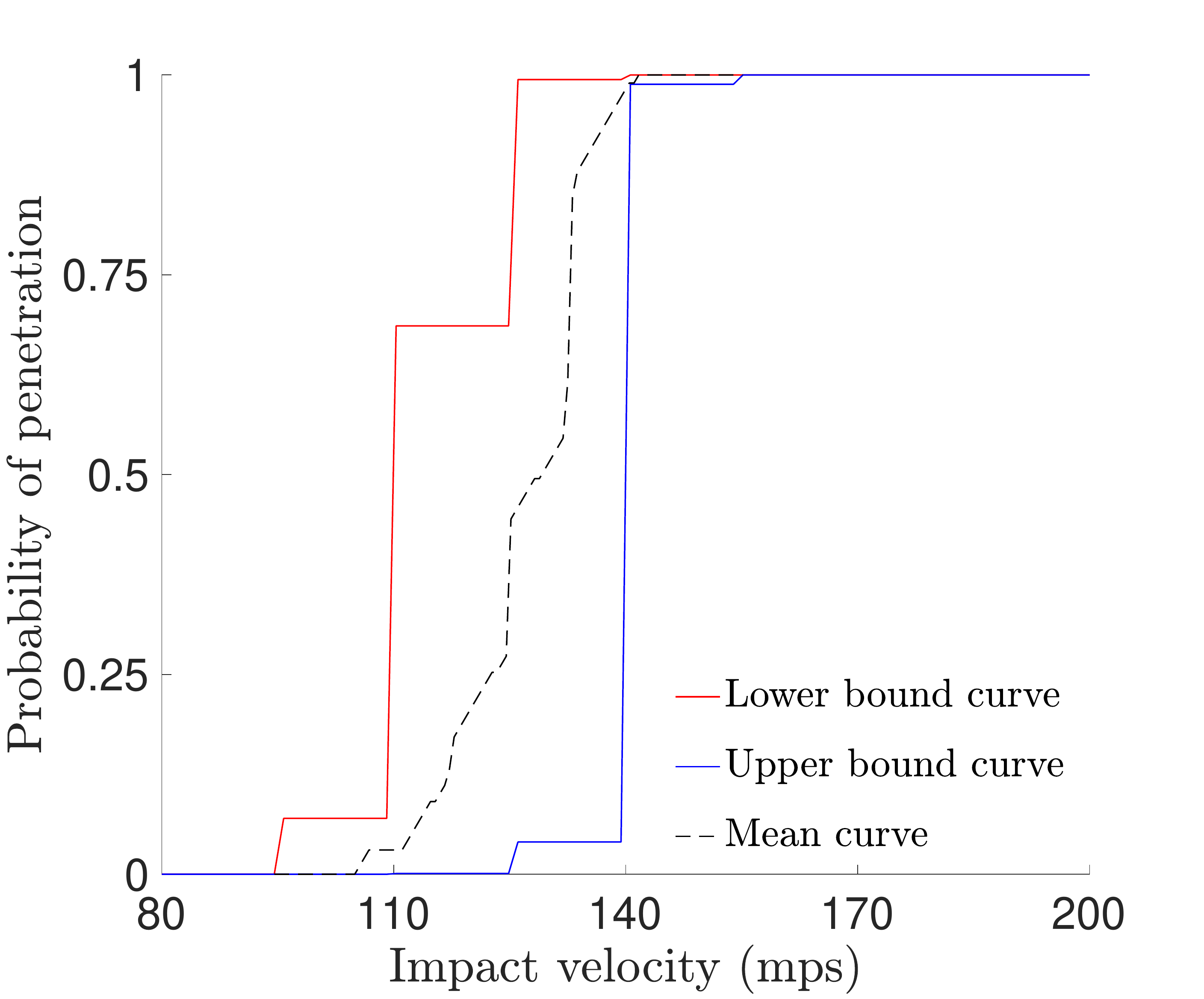}
\caption{}
\label{fig:3-D_gaussian_PVR_curves_subfig5}
\end{subfigure}\quad 
\begin{subfigure}[b]{0.4\textwidth}
\centering
\includegraphics[width=1.1\linewidth, height=4.5cm]{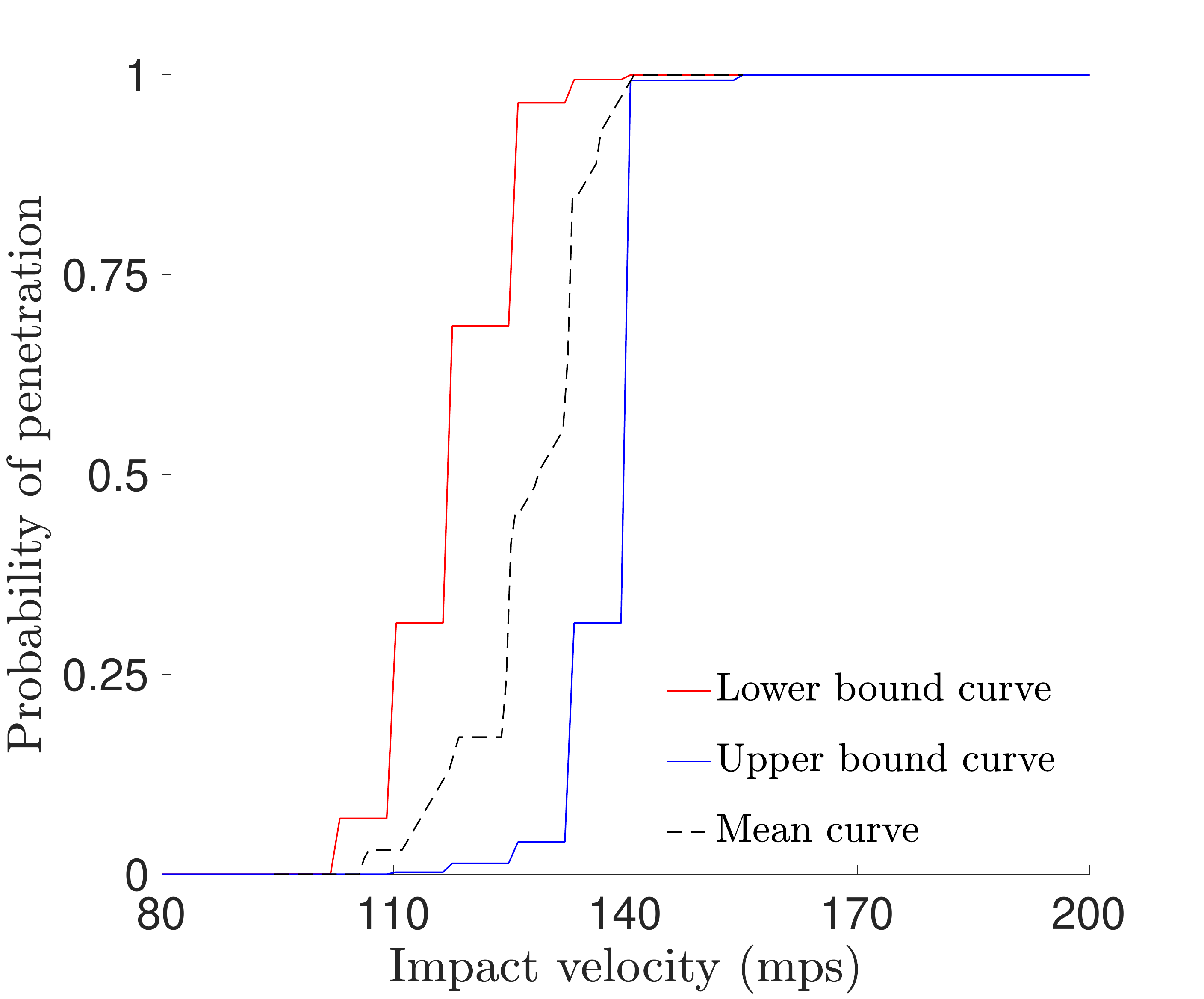}
\caption{}
\label{fig:3-D_gaussian_PVR_curves_subfig6}
\end{subfigure}
\caption{Evolution of the lower, upper and mean PVR curves as a function of the impact velocity of the projectile using (a) 20, (b) 27, (c) 43, (d) 56, (e) 67 and (f) 104 LS-DYNA simulations for the $3$-dimensional example case; the longitudinal tensile strength and the punch shear strength parameters are assumed to follow independent normal distributions with $\mu=1100$ MPa, $\sigma=110$ MPa and $\mu=300$ MPa, $\sigma=30$ MPa respectively.}
\label{fig:3-D_gaussian_PVR_curves}
\end{figure}
\subsection{3-dimensional results} \label{sec:results-3d}
In this section, in addition to the effective longitudinal tensile strength (LTS), the effective punch shear strength (PSS) of the continuum level composite plate is also considered as a variable parameter over the range $[100, 500]$ MPa. The effective longitudinal tensile strength (LTS) is again assumed to vary over the range $[600, 1600]$ MPa, and the impact velocity over the range $[80, 200]$ m/s. This is now a $3$-dimensional problem with impact velocity, the longitudinal tensile strength and the punch shear strength as the three variable parameters. The objective here is to generate the PVR curve by taking into account the variability of the tensile strength parameter as well as the punch shear strength, and assess the variation of the ballistic limit with respect to the variation in the strength parameters. 
Figure \ref{fig:3-D_uniform_PVR_curves} shows the evolution of the PVR curves with increase in the number of LS-DYNA simulations. As observed in the 2-dimensional case, addition of more samples using the adaptive algorithm helps in reducing the width between the lower bound and upper bound PVR curves. In this case, the longitudinal tensile strength and the punch shear strength parameters are assumed to follow an uniform distribution in $[600,1100]$ MPa and $[300,500]$ MPa respectively. Figure \ref{fig:3-D_gaussian_PVR_curves} shows the corresponding evolution of the PVR curves when the longitudinal tensile strength and the punch shear strength parameters are now assumed to follow independent normal distributions with mean $\mu=1100$ MPa and standard deviation $\sigma=110$ MPa for the longitudinal tensile strength and mean $\mu=300$ MPa and standard deviation $\sigma=30$ MPa for the punch shear strength. 
\begin{figure}
\centering
\begin{subfigure}[b]{0.4\textwidth}
\centering
\includegraphics[width=1.1\linewidth, height=4.5cm]{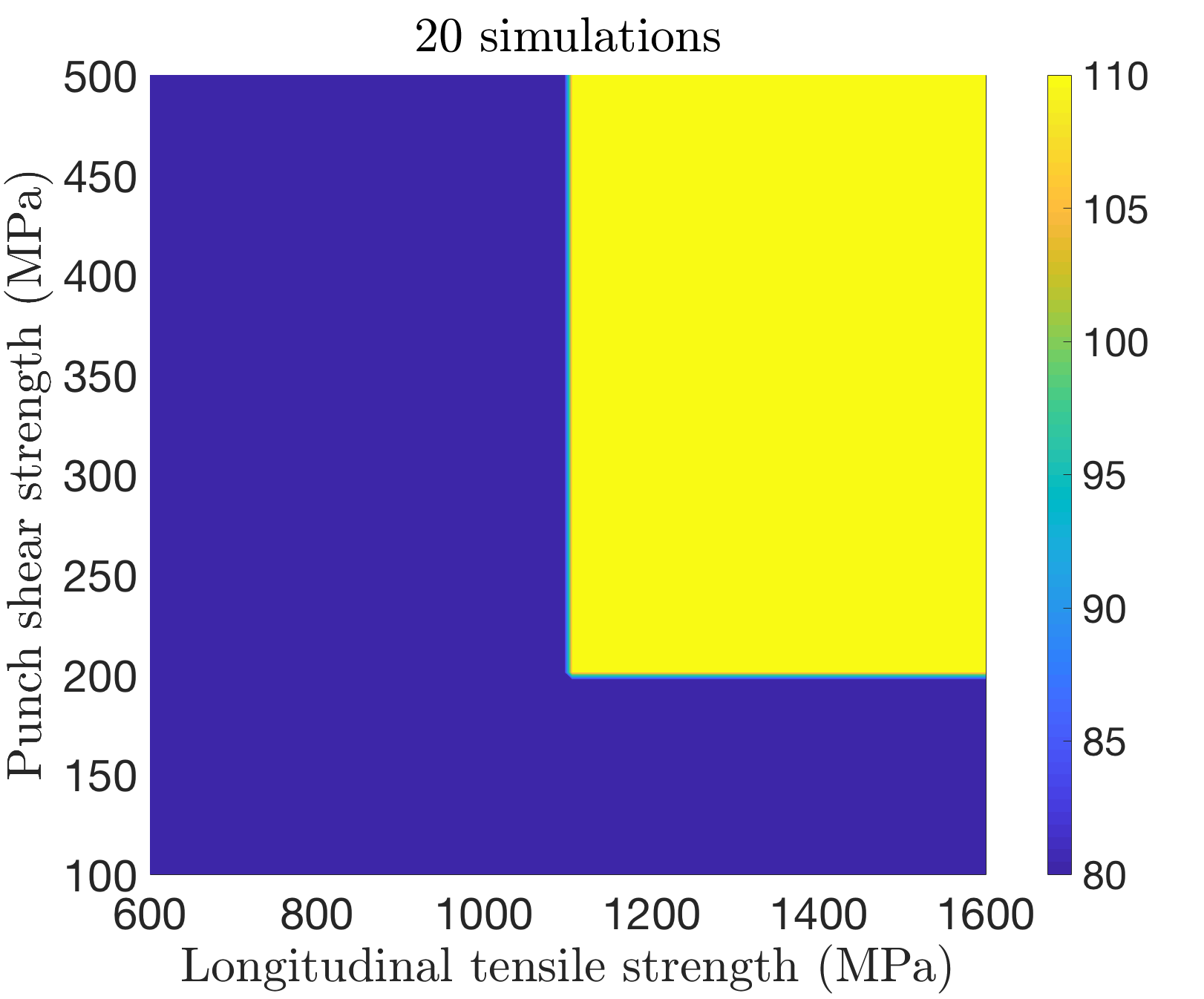}
\caption{}
\label{fig:3-D_ballistic_limit_LB_curves_subfig1}
\end{subfigure} \quad 
\begin{subfigure}[b]{0.4\textwidth}
\centering
\includegraphics[width=1.1\linewidth, height=4.5cm]{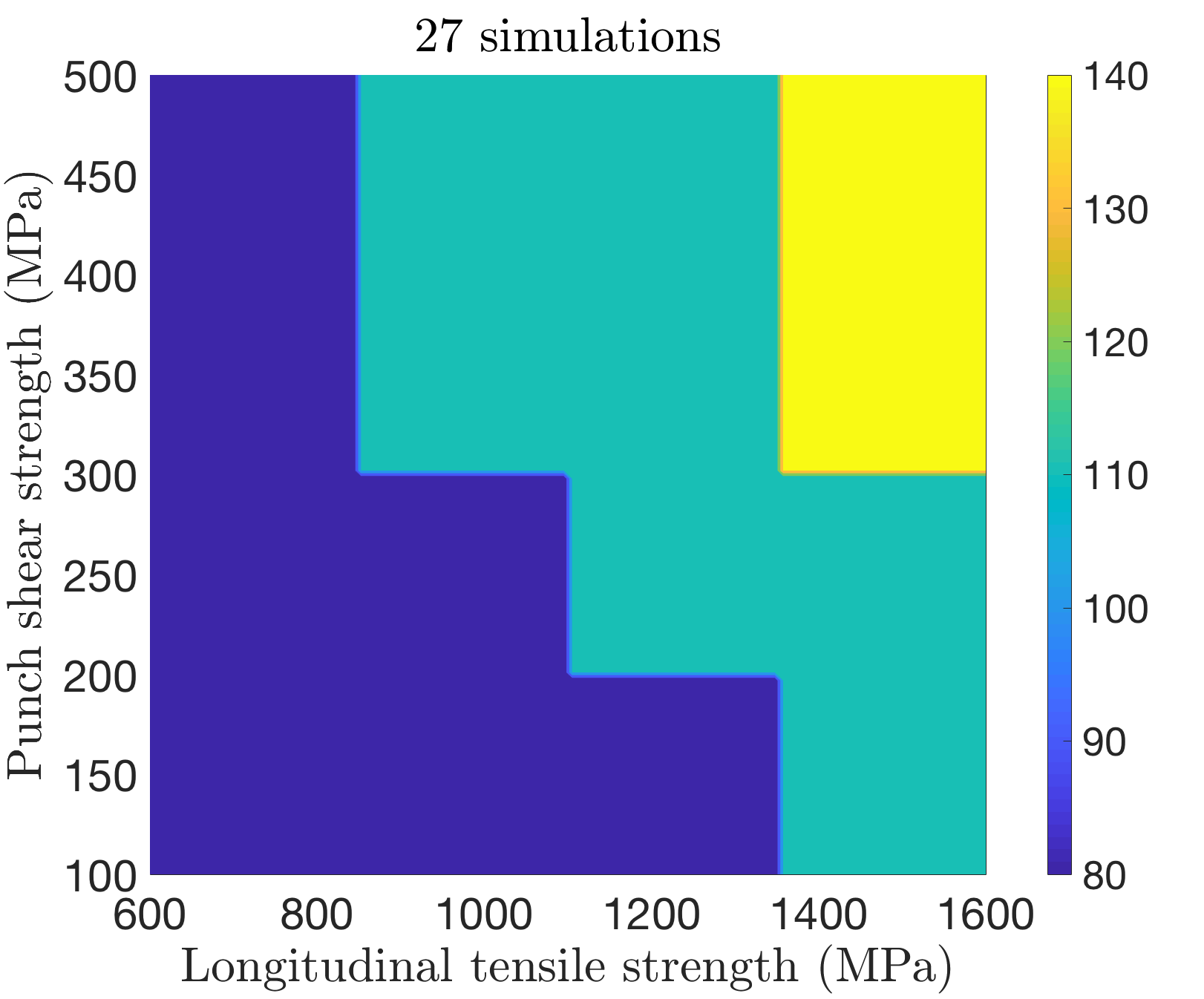} 
\caption{}
\label{fig:3-D_ballistic_limit_LB_curves_subfig2}
\end{subfigure} 
\begin{subfigure}[b]{0.4\textwidth}
\centering
\includegraphics[width=1.1\linewidth, height=4.5cm]{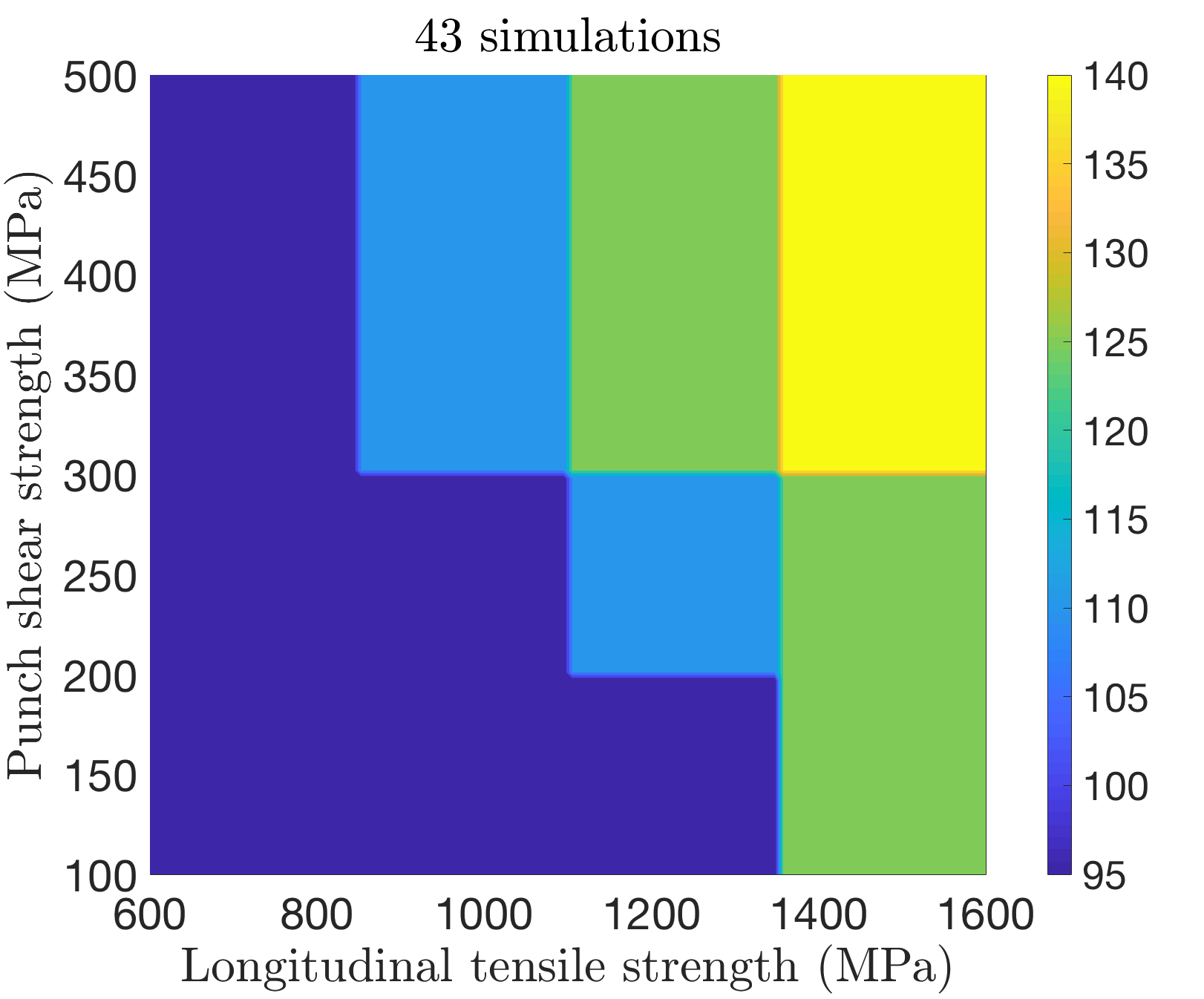} 
\caption{}
\label{fig:3-D_ballistic_limit_LB_curves_subfig3}
\end{subfigure}\quad 
\begin{subfigure}[b]{0.4\textwidth}
\centering
\includegraphics[width=1.1\linewidth, height=4.5cm]{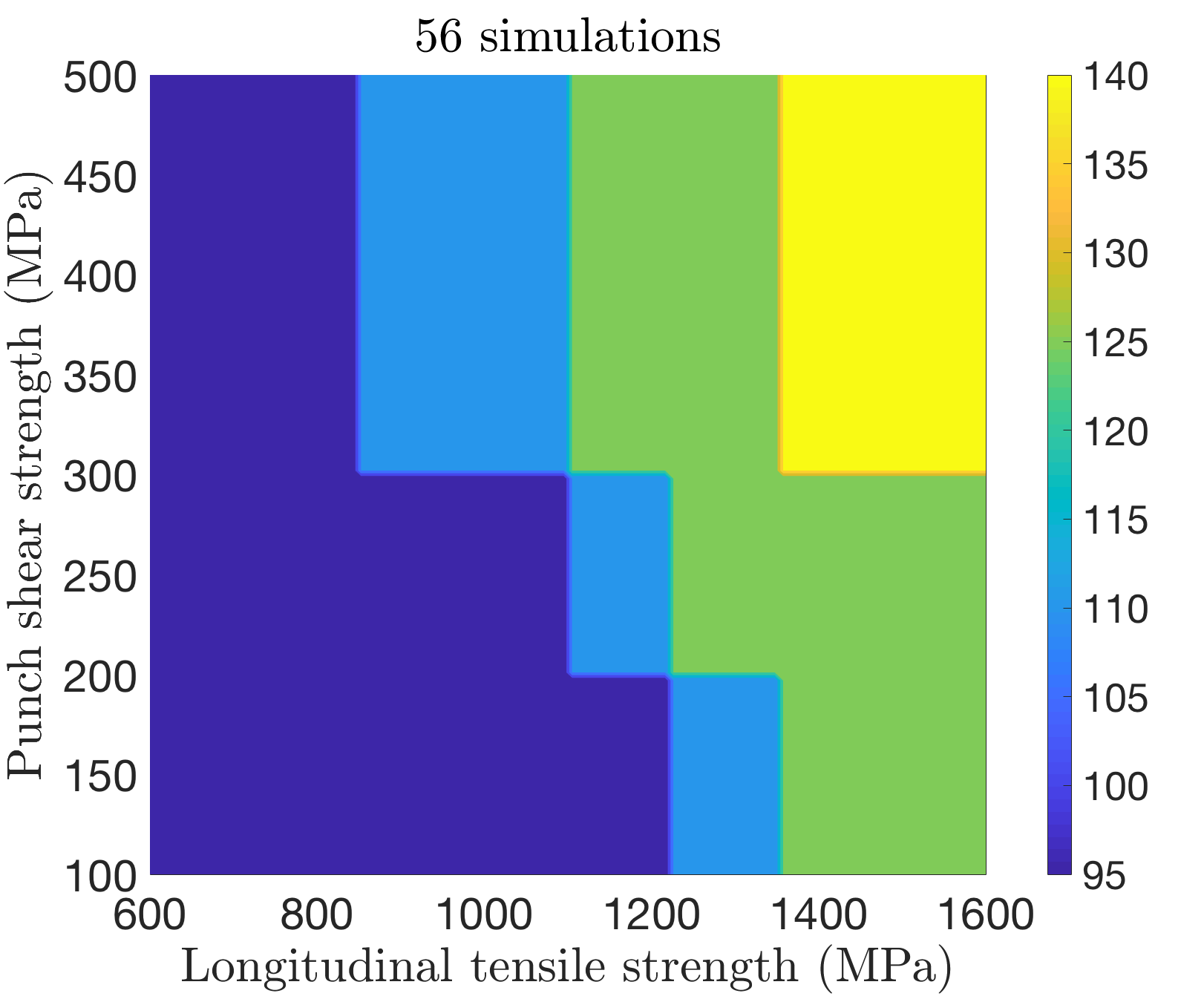}
\caption{}
\label{fig:3-D_ballistic_limit_LB_curves_subfig4}
\end{subfigure}
\begin{subfigure}[b]{0.4\textwidth}
\centering
\includegraphics[width=1.1\linewidth, height=4.5cm]{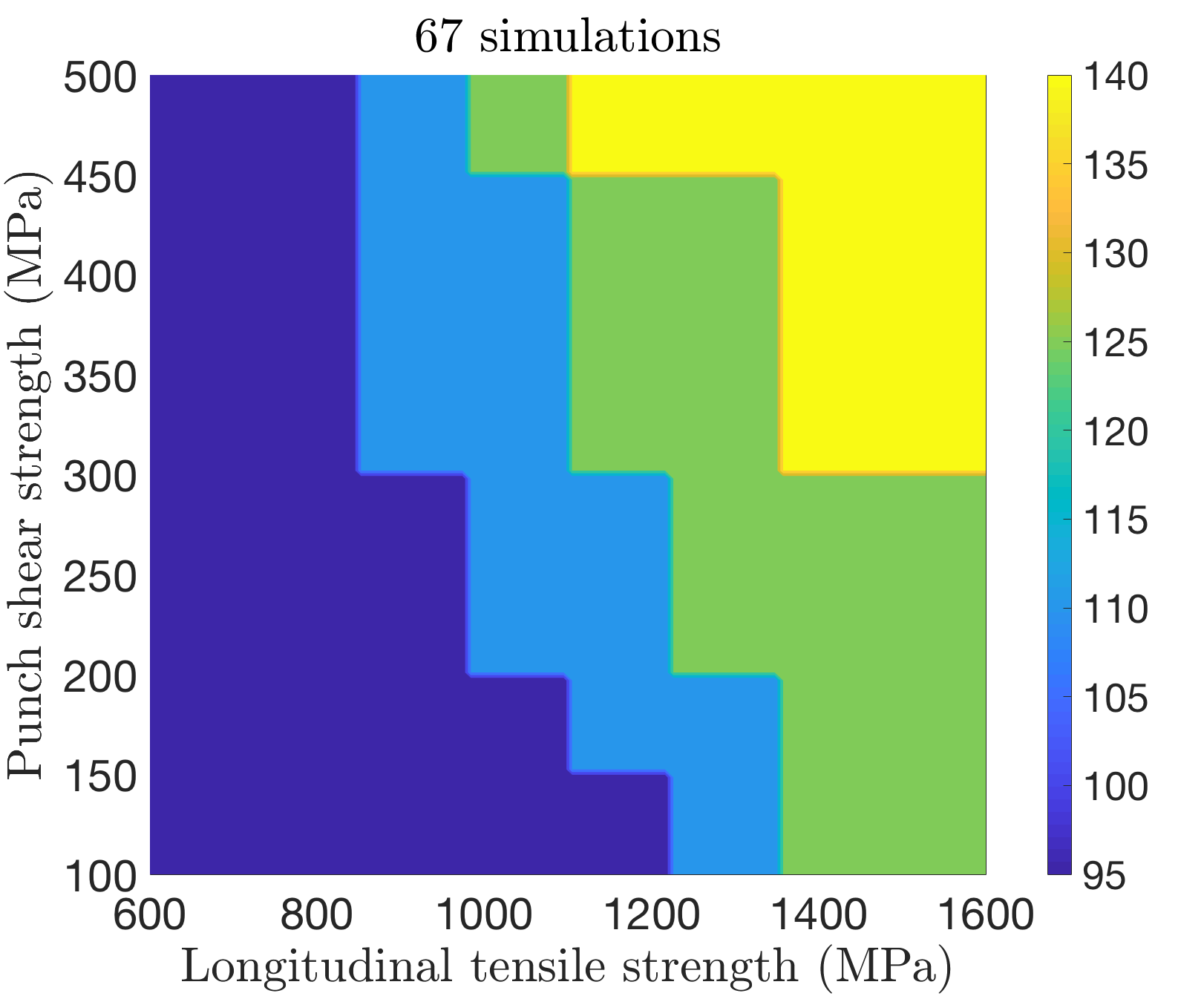}
\caption{}
\label{fig:3-D_ballistic_limit_LB_curves_subfig5}
\end{subfigure}\quad 
\begin{subfigure}[b]{0.4\textwidth}
\centering
\includegraphics[width=1.1\linewidth, height=4.5cm]{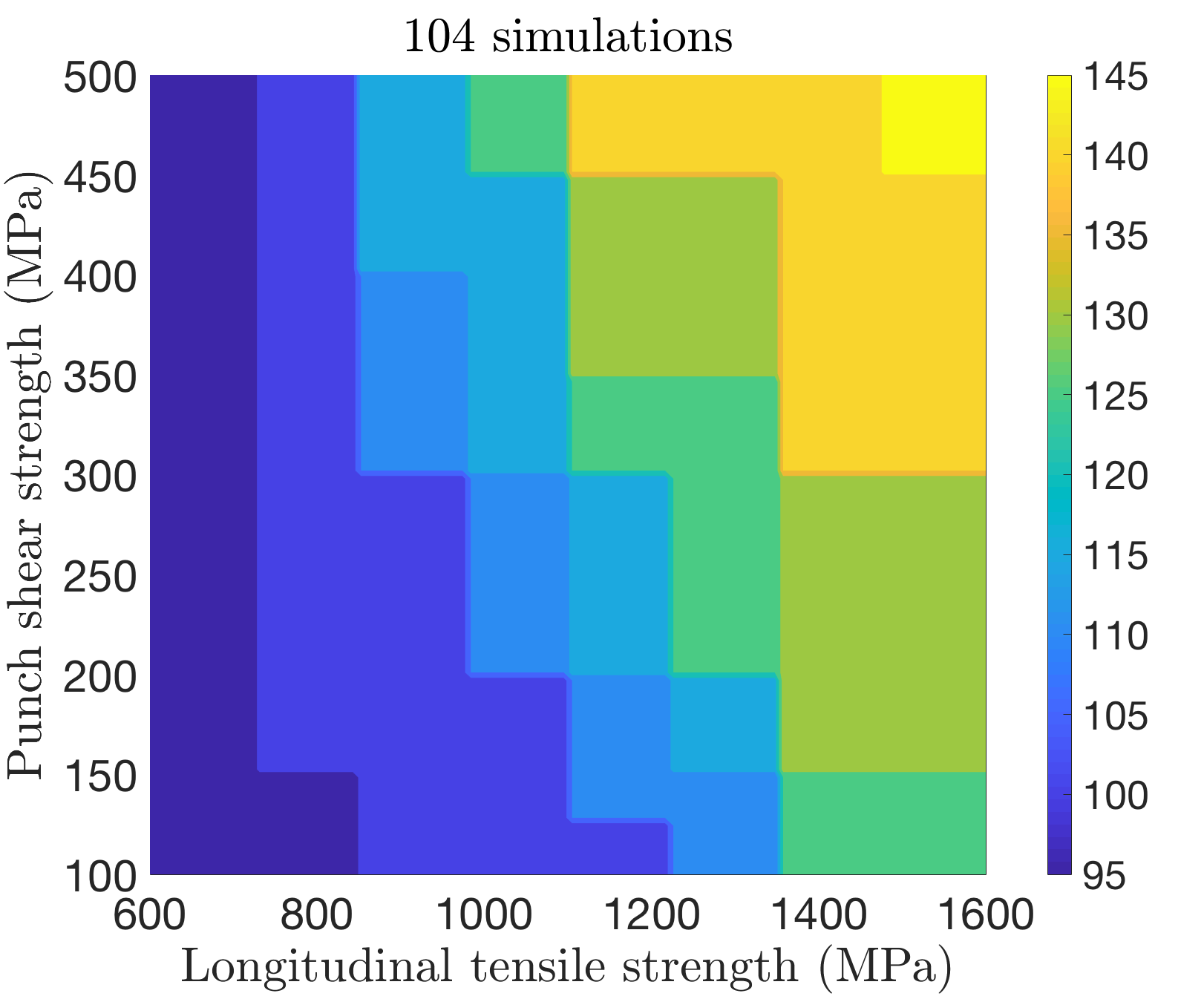}
\caption{}
\label{fig:3-D_ballistic_limit_LB_curves_subfig6}
\end{subfigure}
\caption{Evolution of the lower bound contour of the ballistic limit velocity as a function of the longitudinal tensile strength and punch shear strength using (a) 20, (b) 27, (c) 43, (d) 56, (e) 67 and (f) 104 LS-DYNA simulations for the $3$-dimensional example case.}
\label{fig:3-D_ballistic_limit_LB_curves}
\end{figure}
\begin{figure}
\centering
\begin{subfigure}[b]{0.4\textwidth}
\centering
\includegraphics[width=1.1\linewidth, height=4.5cm]{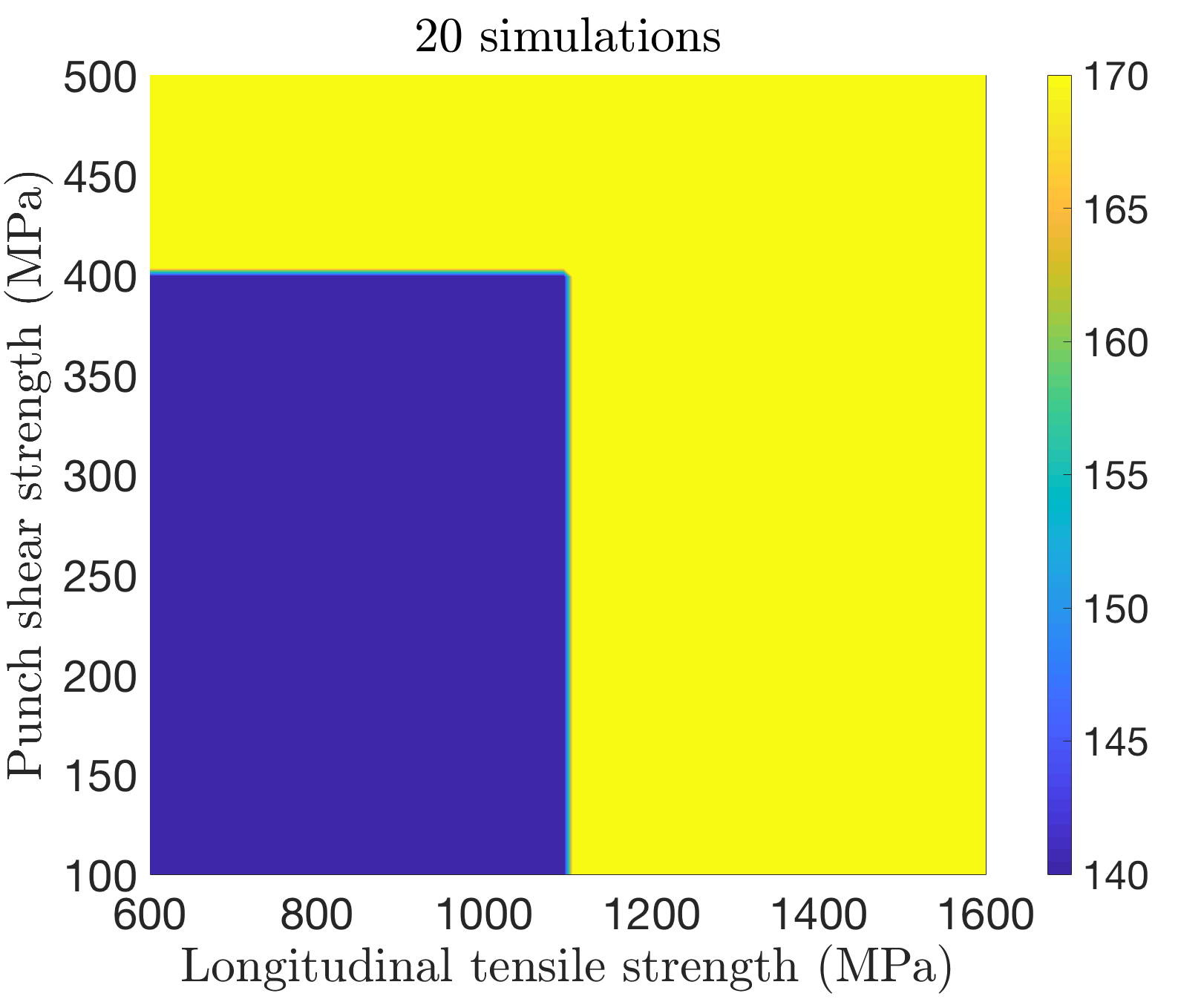}
\caption{}
\label{fig:3-D_ballistic_limit_UB_curves_subfig1}
\end{subfigure} \quad 
\begin{subfigure}[b]{0.4\textwidth}
\centering
\includegraphics[width=1.1\linewidth, height=4.5cm]{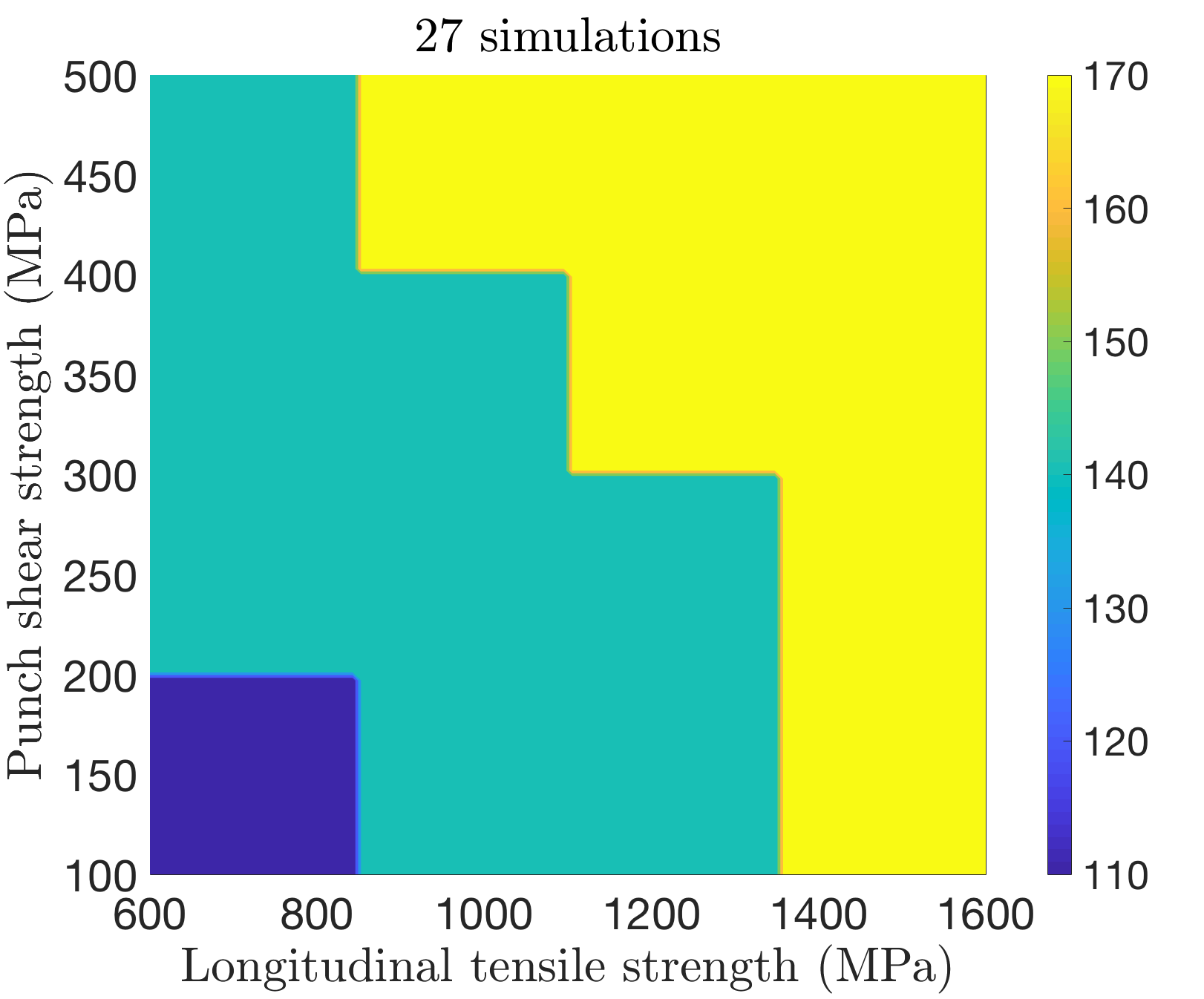} 
\caption{}
\label{fig:3-D_ballistic_limit_UB_curves_subfig2}
\end{subfigure} 
\begin{subfigure}[b]{0.4\textwidth}
\centering
\includegraphics[width=1.1\linewidth, height=4.5cm]{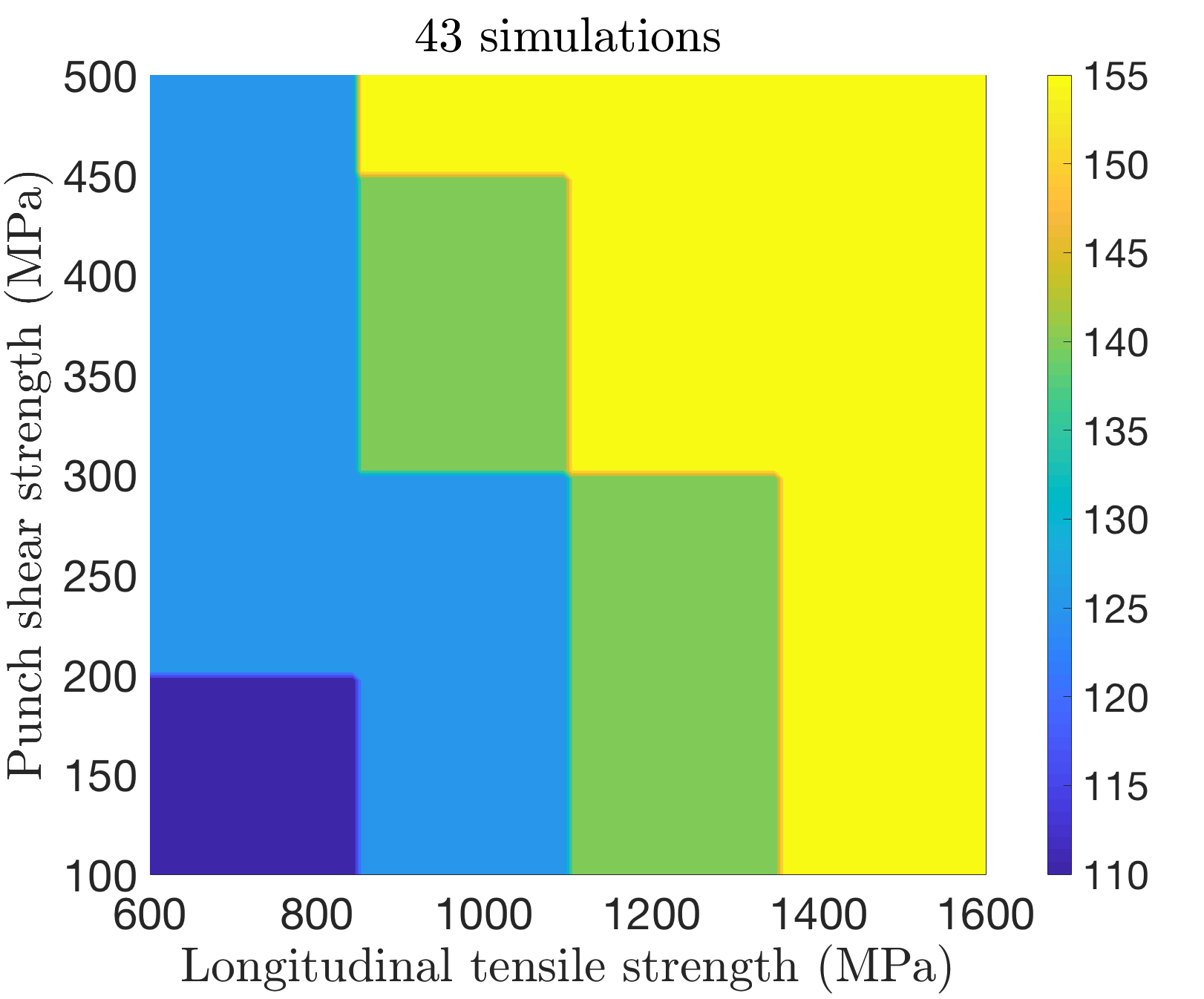} 
\caption{}
\label{fig:3-D_ballistic_limit_UB_curves_subfig3}
\end{subfigure}\quad 
\begin{subfigure}[b]{0.4\textwidth}
\centering
\includegraphics[width=1.1\linewidth, height=4.5cm]{BallisticLimit_LB_Contour_ContinuumPW_Model3_iter6_3D.png}
\caption{}
\label{fig:3-D_ballistic_limit_UB_curves_subfig4}
\end{subfigure}
\begin{subfigure}[b]{0.4\textwidth}
\centering
\includegraphics[width=1.1\linewidth, height=4.5cm]{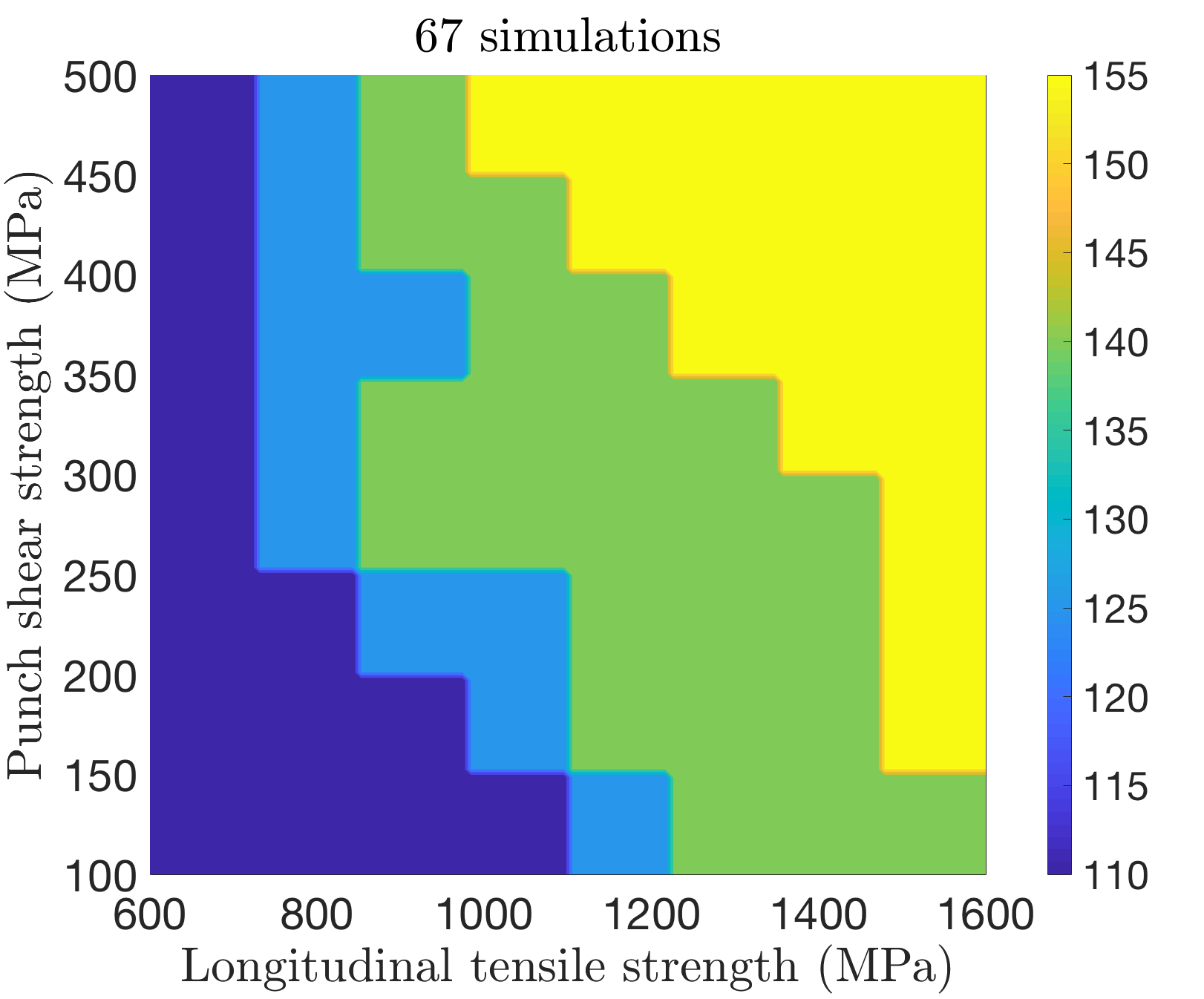}
\caption{}
\label{fig:3-D_ballistic_limit_UB_curves_subfig5}
\end{subfigure}\quad 
\begin{subfigure}[b]{0.4\textwidth}
\centering
\includegraphics[width=1.1\linewidth, height=4.5cm]{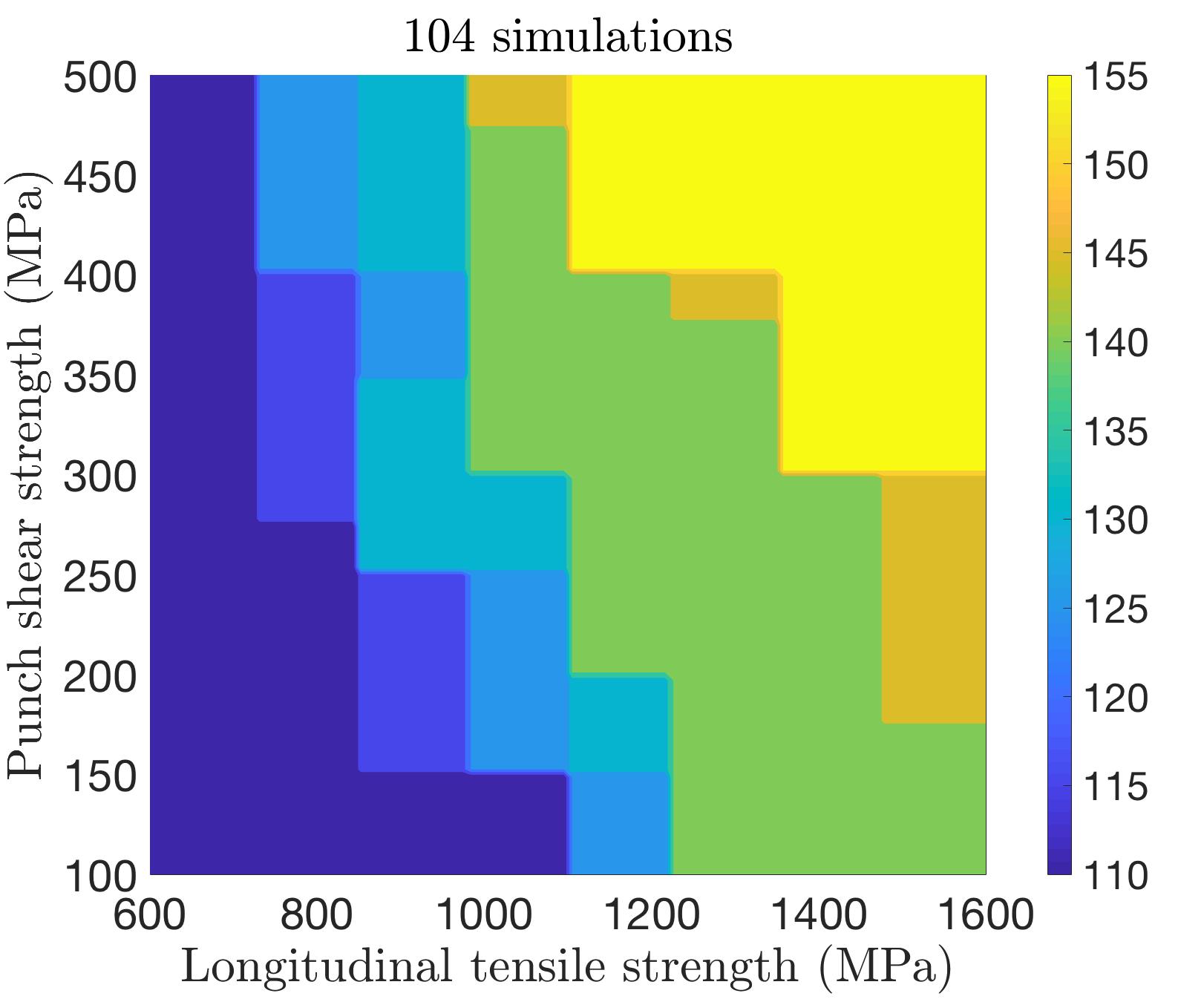}
\caption{}
\label{fig:3-D_ballistic_limit_UB_curves_subfig6}
\end{subfigure}
\caption{Evolution of the upper bound contour of the ballistic limit velocity as a function of the longitudinal tensile strength and punch shear strength using (a) 20, (b) 27, (c) 43, (d) 56, (e) 67 and (f) 104 LS-DYNA simulations for the $3$-dimensional example case.}
\label{fig:3-D_ballistic_limit_UB_curves}
\end{figure}
\begin{figure}
\centering
\begin{subfigure}[b]{0.4\textwidth}
\centering
\includegraphics[width=1.1\linewidth, height=4.5cm]{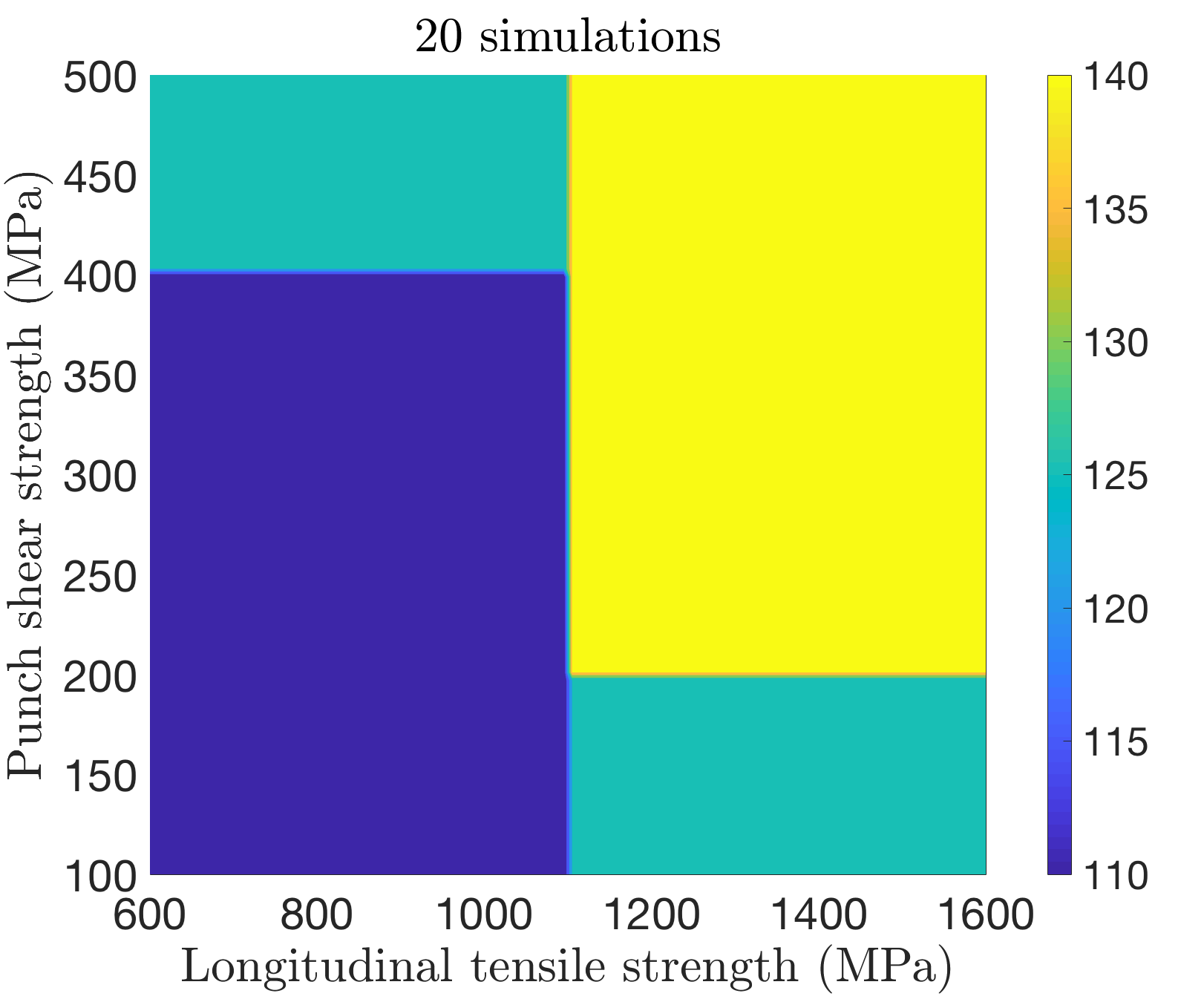}
\caption{}
\label{fig:3-D_ballistic_limit_mean_curves_subfig1}
\end{subfigure} \quad 
\begin{subfigure}[b]{0.4\textwidth}
\centering
\includegraphics[width=1.1\linewidth, height=4.5cm]{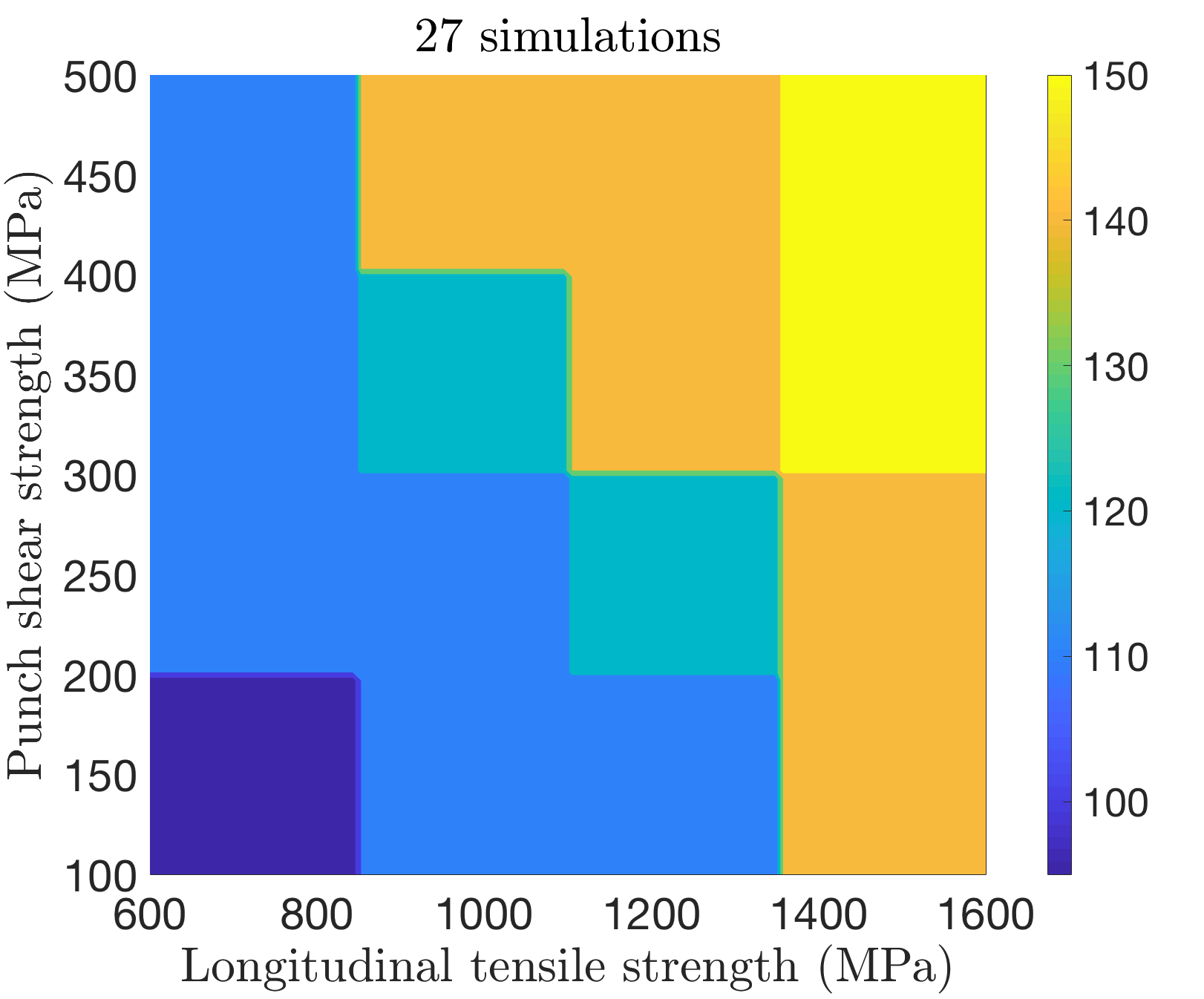} 
\caption{}
\label{fig:3-D_ballistic_limit_mean_curves_subfig2}
\end{subfigure} 
\begin{subfigure}[b]{0.4\textwidth}
\centering
\includegraphics[width=1.1\linewidth, height=4.5cm]{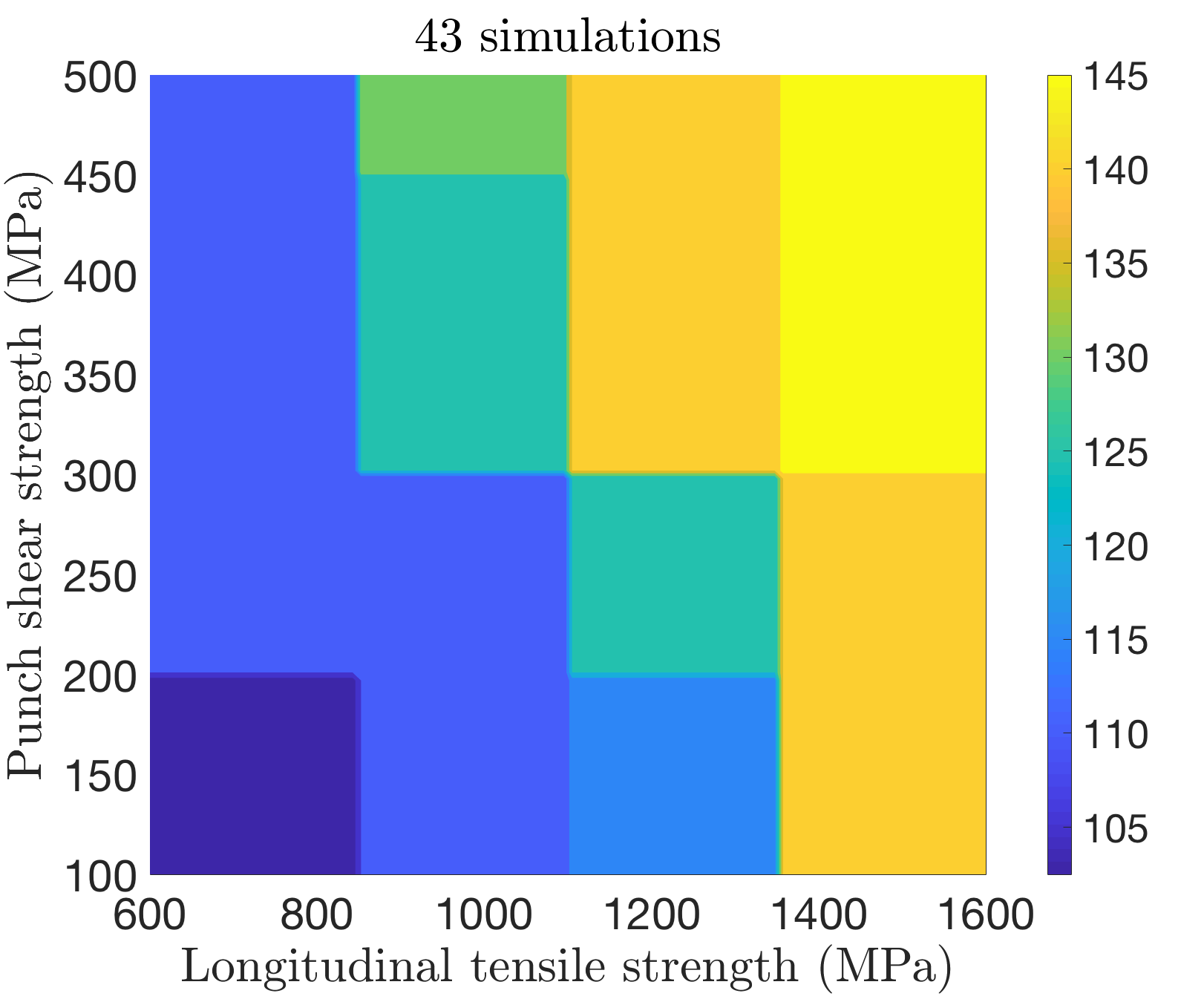} 
\caption{}
\label{fig:3-D_ballistic_limit_mean_curves_subfig3}
\end{subfigure}\quad 
\begin{subfigure}[b]{0.4\textwidth}
\centering
\includegraphics[width=1.1\linewidth, height=4.5cm]{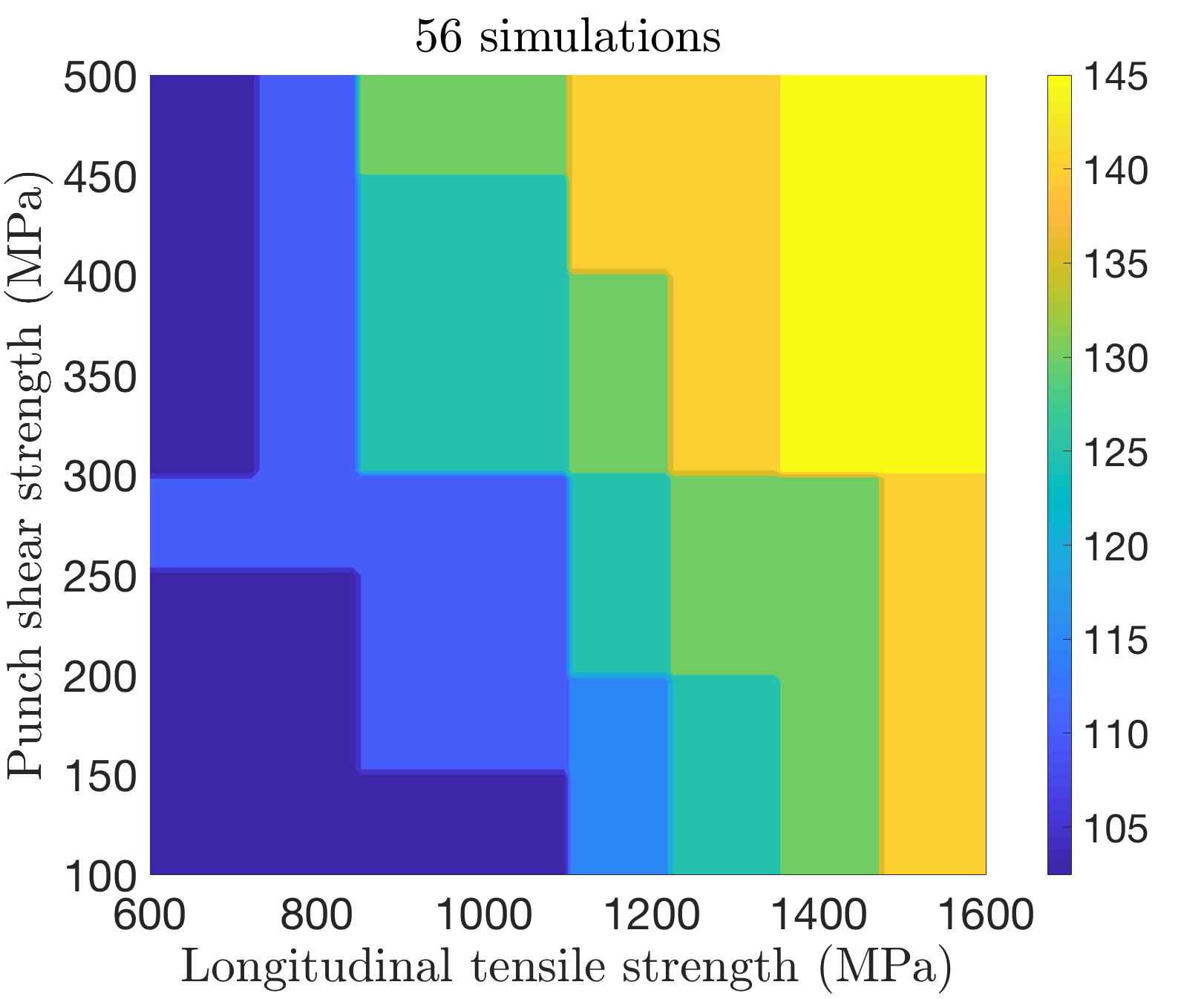}
\caption{}
\label{fig:3-D_ballistic_limit_mean_curves_subfig4}
\end{subfigure}
\begin{subfigure}[b]{0.4\textwidth}
\centering
\includegraphics[width=1.1\linewidth, height=4.5cm]{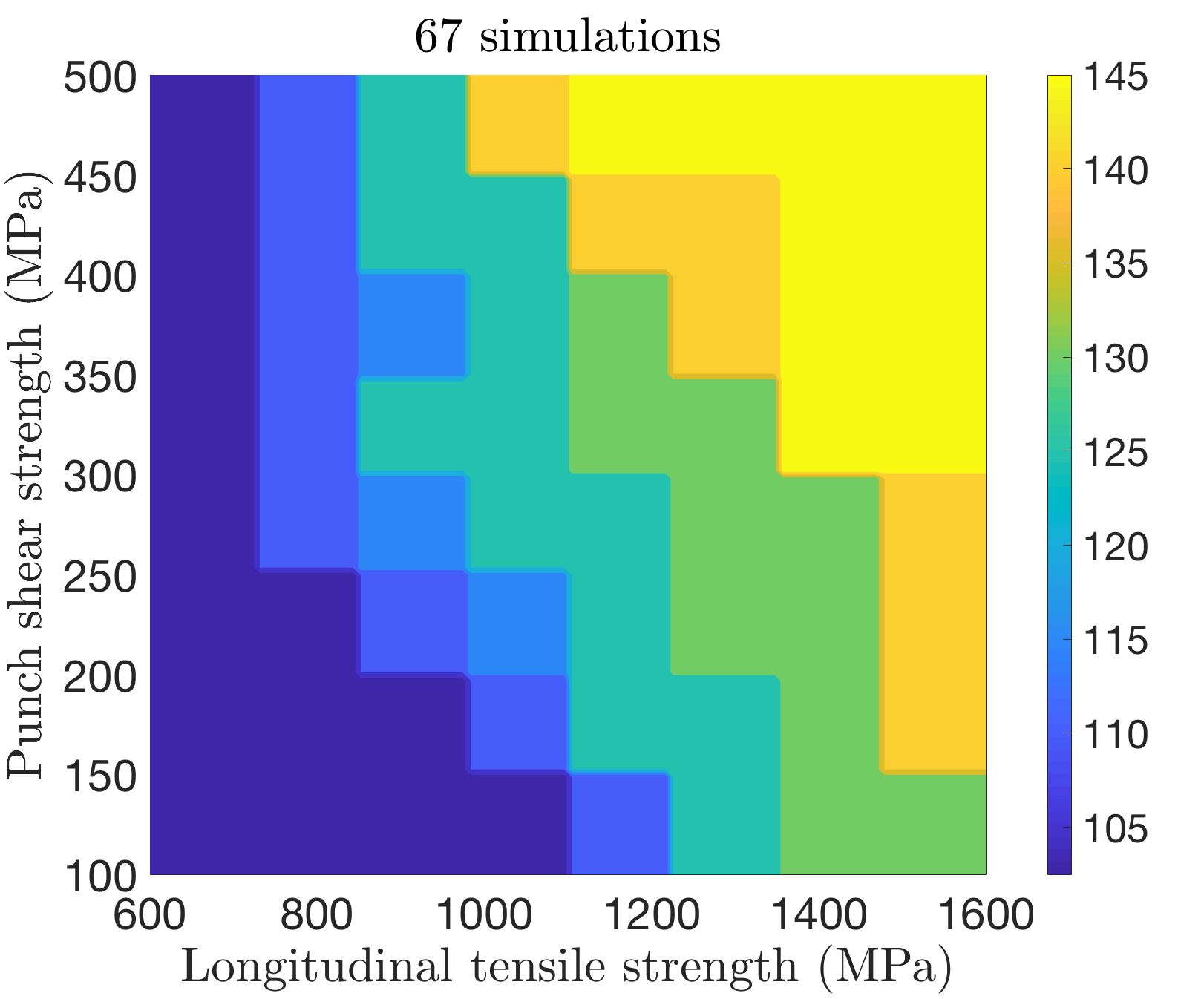}
\caption{}
\label{fig:3-D_ballistic_limit_mean_curves_subfig5}
\end{subfigure}\quad 
\begin{subfigure}[b]{0.4\textwidth}
\centering
\includegraphics[width=1.1\linewidth, height=4.5cm]{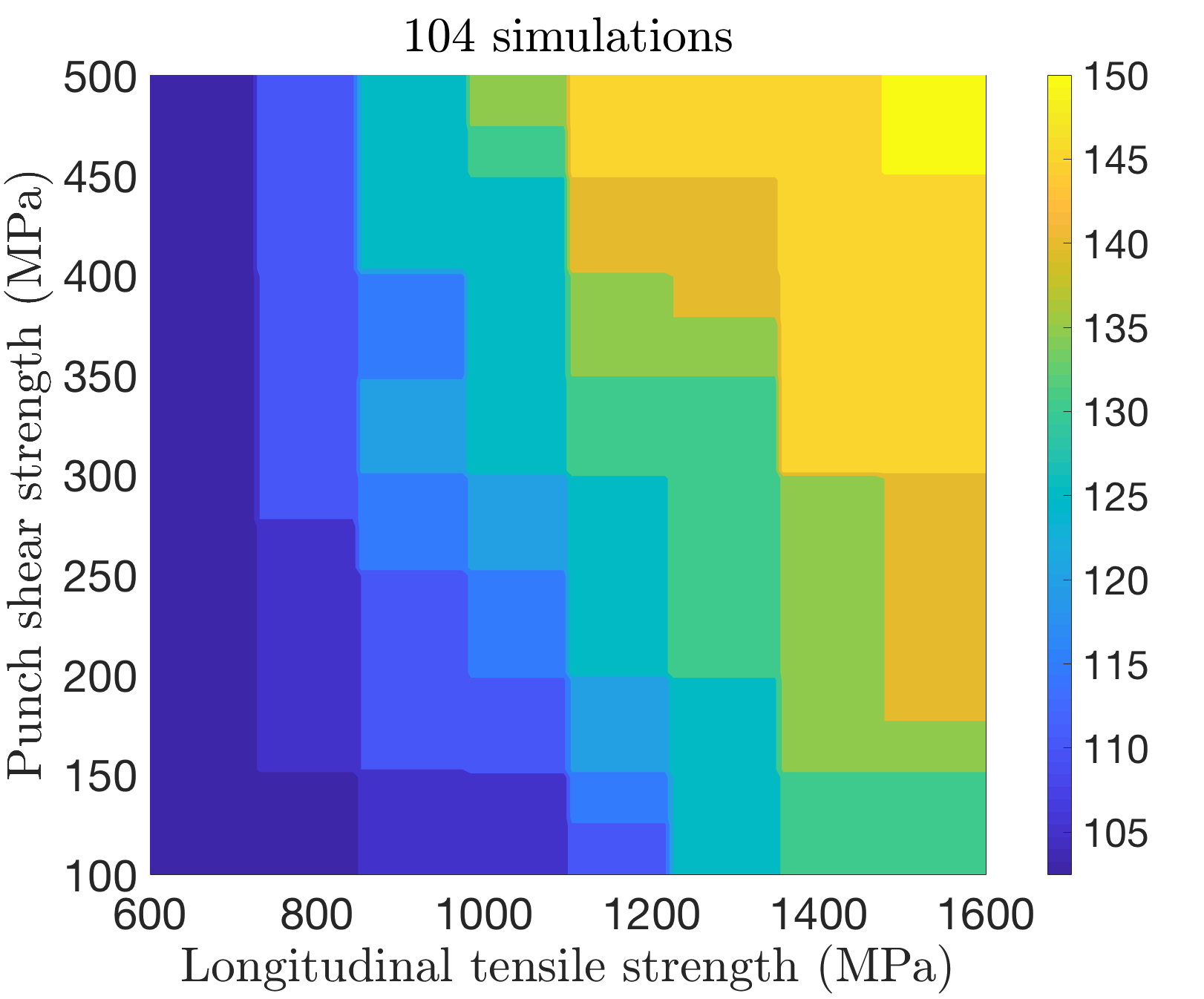}
\caption{}
\label{fig:3-D_ballistic_limit_mean_curves_subfig6}
\end{subfigure}
\caption{Evolution of the mean contour of the ballistic limit velocity as a function of the longitudinal tensile strength and punch shear strength using (a) 20, (b) 27, (c) 43, (d) 56, (e) 67 and (f) 104 LS-DYNA simulations for the $3$-dimensional example case.}
\label{fig:3-D_ballistic_limit_mean_curves}
\end{figure}

\begin{figure}
\centering
\includegraphics[width=0.4\textwidth]{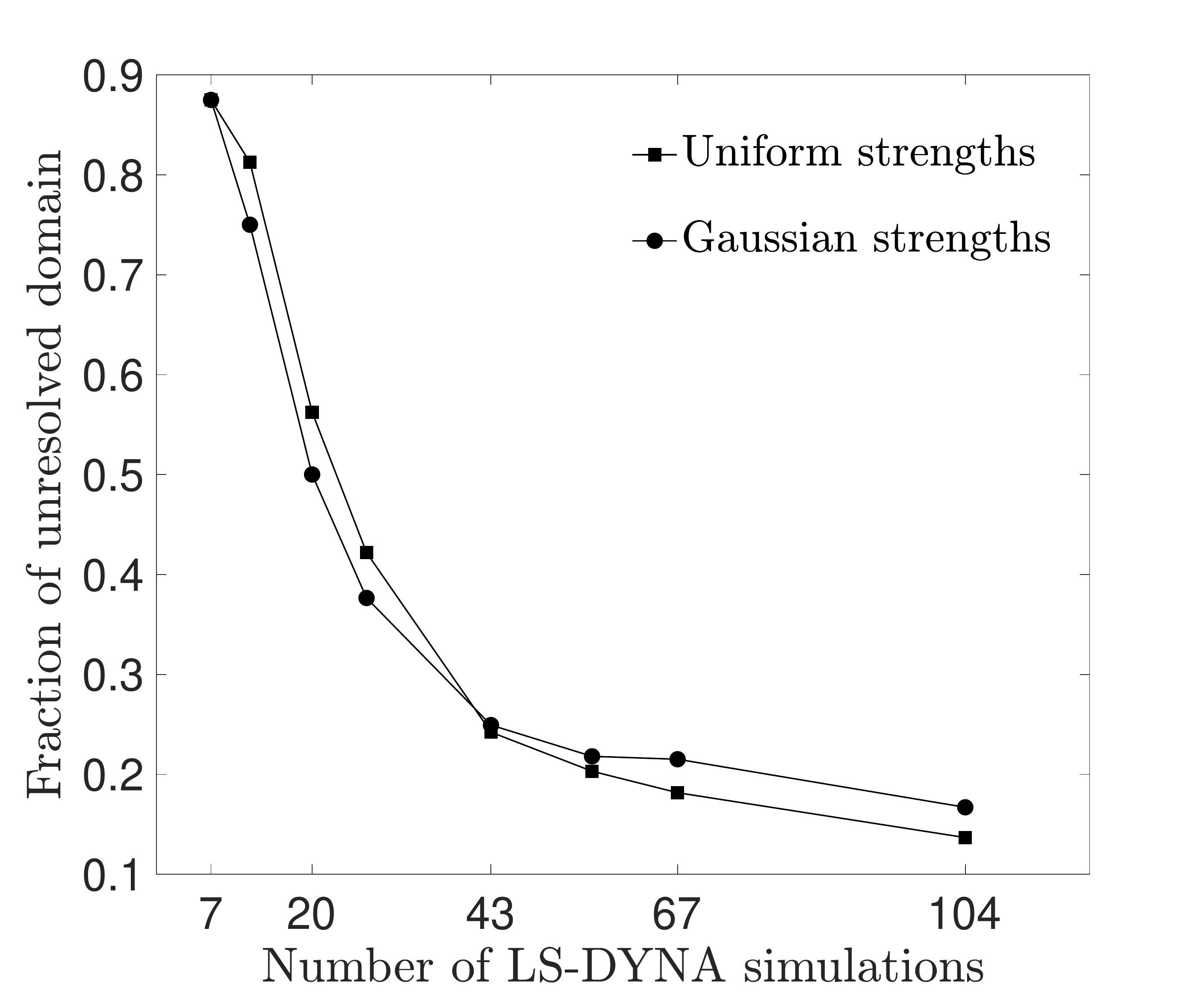}
\caption{Evolution of the fraction of unresolved domain for the $3$-dimensional problem where the impact velocity of the projectile, and the longitudinal tensile strength and punch shear strength of the plate are the variable parameters}
\label{fig:Fraction_of_unresolved_domain_3D}
\end{figure}

Next, results describing the variation of the ballistic limit velocity with respect to changes in both the strength parameters, the longitudinal tensile strength and the punch shear strength, are shown. Figure \ref{fig:3-D_ballistic_limit_LB_curves}(a-f) shows the evolution of the lower bound (LB) contour plot of the ballistic limit velocity with increase in the number of LS-DYNA simulations and figure \ref{fig:3-D_ballistic_limit_UB_curves}(a-f) shows the evolution of the upper bound (UB) contour plot of the ballistic limit velocity with increase in the number of LS-DYNA simulations. The evolution of the mean ballistic limit contour is shown in figure \ref{fig:3-D_ballistic_limit_mean_curves}(a-f) for increasing number of LS-DYNA simulations. As seen in all the three figures, more and more finer details about the ballistic limit variation get captured with increase in the number of simulations. It is also seen that the ballistic limit values (in the mean, LB or UB contours) has an increasing trend with respect to both the strength parameters, but the ballisitic limit is found to be more sensitive to changes in the longitudinal tensile strength than that of the punch shear strength over the given range of the strength parameters. This implies indirectly that the longitudinal tensile strength has a bigger effect on the PVR curve generation. A closer look at the ballistic limit values of the lower and upper bound contours reveals that the two contours converge closer to each other with increase in the number of LS-DYNA simulations, which signifies reduction in the uncertainty in the ballistic limit estimates. It is noted here that the new simulations should essentially be run at sample (parameter) values which correspond to important regions in the parameter space. This will help in extracting maximum information out of a new simulation. Figure \ref{fig:Fraction_of_unresolved_domain_3D} shows the reduction in the fraction of the unresolved domain with increase in the number of LS-DYNA simulations.
%
\section{Conclusions} \label{sec:conclusions_chap_composite_impact}
In this paper, an efficient classification surrogate algorithm has been developed to resolve the surface of separation between the regions with different discrete output labels in the input parameter domain. This study specifically considers the simulation model of a continuum level plain weave S-2 glass/SC-15 epoxy composite plate under ballistic impact with the goal of generating two important quantities of interest, the probabilistic velocity response (PVR) curve
as a function of the impact velocity, and the ballistic limit velocity prediction as a function of the model parameters. However, it is worth mentioning that the proposed methodology is general enough to be applied to other simulated models with discrete outputs.\\
\indent The state-of-the-art method for estimating the PVR curve works in the $1$-dimensional impact velocity space and all the different sources of variability are implicitly included. The method estimates the PVR curve based on $V_{50}$ mean and variance estimates by assuming a Gaussian distribution. The proposed methodology, on the other hand, explicitly accounts for the different sources of variability by including variable parameters associated with them. Although this involves more computational cost, it is a more robust approach with the capability of extracting much more reliable and useful information from a computational impact model, for example, ballistic limit velocity bound predictions at different model parameter values. The computational framework in this study efficiently characterizes the probabilistic response of a continuum plain weave composite plate with different sources of variability under projectile impact. As seen from the results, the methodology allows for an efficient determination of the lower and upper bounds of the PVR curve. It also captures the variation of the ballistic limit velocity bounds as a function of each of the different sources of variability, which also helps in assessing the sensitivity of the ballistic limit to these parameters. For the given simulation model and the input strength parameter ranges as discussed in section \ref{sec:results-3d}, the ballistic limit was found to be more sensitive to the longitudinal strength than the punch shear strength. The methodology is also shown to efficiently handle any non-uniform marginal distributions of strength parameters, like Gaussian distribution.\\
\indent This study demonstrates the implementation of the proposed algorithm to efficiently generate the PVR and ballistic limit curves corresponding to a continuum scale model. It is noted here that the proposed algorithm is model agnostic and the continuum model is used as an example model.
A direct extension of the present work with the continuum level composite plate model is to introduce more number of variable input parameters in the study and then assess the performance of the proposed algorithm for PVR curve generation. However, this model lacks information about the explicit woven architecture of the fiber yarns embedded in the matrix, and spatial variation of the corresponding properties. Thus, another potential future direction is to replace the continuum plate model with a meso-scale plate model which captures those local information. The meso-scale model is a more accurate representation of the plate under ballistic impact but more expensive to solve which can be a potential challenge for efficient PVR curve generation. \\
\indent Furthermore, there can be sources of variablity for which the assumption of monotonicity may not apply, for example, impact location in a meso-scale model (as penetration at the resin rich regions would be expected to be lower than at the tow cross-overs), which could also potentially reduce the efficiency of the current methodology. Thus, a third possible future direction is to further refine the proposed approach to efficiently handle such non-monotonic variables.
\section*{Acknowledgements}
Research was sponsored by the Army Research Laboratory and was accomplished under Cooperative Agreement Number W911NF-12-2-0023 and W911NF-12-2-0022. The views and conclusions contained in this document are those of the authors and should not be interpreted as representing the official policies, either expressed or implied, of the Army Research Laboratory or the U.S. Government. The U.S. Government is authorized to reproduce and distribute reprints for Government purposes notwithstanding any copyright notation herein.
\section*{References}
\bibliographystyle{ieeetr}
\bibliography{reference}

\end{document}